\documentclass[12pt]{article}
\usepackage[utf8]{inputenc}
\usepackage{geometry,amsmath,amssymb,bm}
\usepackage{natbib}
\usepackage{amsfonts}
\usepackage{latexsym}
\usepackage{graphicx}
\usepackage{amsmath}
\usepackage{color}
\usepackage{subfigure}
\usepackage{float}
\RequirePackage{jabbrv}
\newcommand{\beq}{\begin{equation}}
\newcommand{\bseqs}{\begin{subequations}}
\newcommand{\eseqs}{\end{subequations}}
\newcommand{\balign}{\begin{align}}
\newcommand{\ealign}{\end{align}}
\newcommand{\eeq}{\end{equation}}
\newcommand{\beql}{\begin{equation} \label}
\newcommand{\beqs}{\begin{eqnarray}}
\newcommand{\eeqs}{\end{eqnarray}}
\newcommand{\beas}{\begin{eqnarray*}}
\newcommand{\eeas}{\end{eqnarray*}}
\newcommand{\ber}{\begin{array}}
\newcommand{\eer}{\end{array}}
\newcommand{\becs}{\begin{cases}}
\newcommand{\eecs}{\end{cases}}
\newcommand{\leftm}{\left[\begin{array}}
\newcommand{\rightm}{\end{array}\right]}

\newcommand{\bfd}{{\mathbf d}}
\newcommand{\bfe}{{\mathbf e}}
\newcommand{\bff}{{\mathbf f}}

\newcommand{\bfn}{{\mathbf n}}

\newcommand{\bfu}{{\mathbf u}}
\newcommand{\bfv}{{\mathbf v}}

\newcommand{\bfx}{{\mathbf x}}
\newcommand{\bfy}{{\mathbf y}}
\newcommand{\bfz}{{\mathbf z}}

\newcommand{\bfC}{{\mathbf C}}

\newcommand{\bfG}{{\mathbf G}}

\newcommand{\bfK}{{\mathbf K}}

\newcommand{\bfP}{{\mathbf P}}
\newcommand{\bfQ}{{\mathbf Q}}
\newcommand{\bfR}{{\mathbf R}}
\newcommand{\bfS}{{\mathbf S}}
\newcommand{\bfT}{{\mathbf T}}

\newcommand{\bfV}{{\mathbf V}}

\newcommand{\bbm}{\begin{bmatrix}}
\newcommand{\ebm}{\end{bmatrix}}

 \usepackage{mathptmx}
 \usepackage{color}
\newtheorem{theorem}{\bf Theorem}[section]

\newtheorem{lemma}{\bf Lemma}[section]

\begin{document}

\title{Solutions to the generalized Eshelby conjecture for anisotropic media: Proofs of the weak version and counter-examples to the high-order and the strong versions}
\author{T. Y. Yuan$^{1,2}$, K. F. Huang$^{3,*}$ and J. X. Wang$^{1,2,*}$
\bigskip
\\
\small{$^{1}$State Key Laboratory for Turbulence and Complex System, College of Engineering,}
\\
\small{Peking University, Beijing 100871, P.R. China}
\smallskip
\\
\small{$^{2}$CAPT-HEDPS, and IFSA Collaborative Innovation Center of MoE, College of Engineering,}
\\
\small{Peking University, Beijing 100871, P.R. China}
\smallskip
\\
\small{$^{3}$Department of Mechanics and Aerospace Engineering,}
\\
\small{Southern University of Science and Technology, Shenzhen, Guangdong 518055, P.R. China}
\\
}
\date{}
\maketitle

\begin{abstract}
The Eshelby formalism for an inclusion in a solid is of significant theoretical and practical implications in mechanics and other fields of heterogeneous media.
Eshelby's finding that a uniform eigenstrain prescribed in a solitary ellipsoidal inclusion in an infinite isotropic medium results in a uniform elastic strain field in the inclusion  leads to the conjecture that the ellipsoid is the only inclusion that possesses the so-called {  Eshelby uniformity property}.
Previously, only the weak version of the conjecture has been proved for the isotropic medium, whereas the general validity of the conjecture for anisotropic media in three dimensions is yet to be explored.
In this work, firstly, we present proofs of the  weak version of the generalized Eshelby conjecture for anisotropic media that possess cubic, transversely isotropic, orthotropic, and monoclinic symmetries, which substantiates that only the ellipsoidal shape can transform {\it all} uniform eigenstrains into uniform elastic strain fields in a solitary inclusion in infinite media possessing these symmetries.
 Secondly, we prove that in these anisotropic media,  there exist non-ellipsoidal inclusions that can transform  particular polynomial eigenstrains of even degrees into  polynomial  elastic strain fields of the same even degrees in them.
These results constitute counter-examples, in the strong sense, to the generalized high-order Eshelby conjecture (inverse problem of {\it Eshelby's polynomial conservation theorem}) for polynomial eigenstrains in both anisotropic media and the isotropic medium (quadratic eigenstrain only). In addition, we also show that there are counter-examples to the  strong version of the generalized Eshelby conjecture for uniform eigenstrains in these anisotropic media. A sufficient condition for the existence of those counter-example inclusions is provided. These findings reveal striking richness of the uniformity between the eigenstrains and the correspondingly induced elastic strains in inclusions in anisotropic media beyond the canonical ellipsoidal inclusions.
 Since the strain fields in embedded and inherently anisotropic quantum dot crystals are effective tuning knobs of the quality of the emitted photons by the quantum dots, the results may have  implications in the technology of quantum information, in addition to in mechanics and materials science.
\end{abstract}

\thanks{
\begin{flushleft}
  \textbf{Subject Areas}: solid mechanics, applied mathematics, engineering\\
\end{flushleft}
\begin{flushleft}
\end{flushleft}
\textbf{Author for correspondence}: \\J. X. Wang: jxwang@pku.edu.cn;\quad K. F. Huang: huangkf@sustc.edu.cn
}

\section{Introduction}
Sixty years ago, Eshelby's seminal work on the elastic field within an ellipsoidal inclusion in an infinite medium
opened a fertile field in theories of heterogeneous materials \citep{Eshelby1957,Eshelby1959}.
Eshelby found that a uniform eigenstrain prescribed in the ellipsoidal inclusion induces a uniform elastic strain field in the inclusion.
The elegant Eshelby formalism inspires a wealth of investigations of various inclusion problems in multi-dimensions, multi-scales~\citep{Sharma2003,Sharma2004,Nanoinc2005,Lim2006,Sharma2007,Tian2007,Ma2018}, and multi-physical fields~\citep{Wang1992}, and it has been  widely applied to analyses and predictions of the behaviour of various composites.
Unfortunately, it is plausible that an inclusion with an arbitrary shape does not always possess the {  Eshelby uniformity property}
that transforms a uniform eigenstrain into a uniform elastic strain.
Hence, 
\cite{Eshelby1961} conjectured that ``{\it among closed surfaces the ellipsoid alone has this convenient
property.}" This conjecture is more specifically divided into the weak and strong versions~\citep{Kang2008}.
Eshelby's original assertion is commonly understood in the weak sense~\citep{Kang2008,Liu2008}, which means that the ellipsoid is the only inclusion that possesses the {  Eshelby uniformity property} for all uniform eigenstrains; however, the strong version requires that the ellipsoid is the only  inclusion that possesses the {  Eshelby uniformity property} for any single uniform eigenstrain~\citep{Kang2008,Liu2008}.

Many efforts have been devoted to investigations of the Eshelby conjectures \citep{Lubarda1998,Markenscoff1998,Mura1994,Mura1997,Ru1996,Sendeckyj1970,Vigdergauz2000} and the related quasi-Eshelby properties~\citep {Kawashita2001,Nozaki1997,Rodin1998,Xu2005,Zheng2006} in the context of isotropic elasticity.
\cite{Kang2008} pointed out that the weak and strong versions of the conjecture are also applicable to the inhomogeneity problem in an
infinite medium that is subjected to a remote strain field, in which case the induced field inside the inhomogeneity is caused by the remote loading.
Then, the solution of the Eshelby conjecture for the inhomogeneity problem is equivalent to that for the inclusion problem. Thus, they solved the inhomogeneity problem and proved the weak version for both the inclusion and inhomogeneity problems, through utilization of a theorem  in relation to the Newtonian potential of an ellipsoid.
In the same year, 
\cite{Liu2008} directly solved the inclusion problem and achieved the proof of the weak version by analyzing an obstacle problem with the variational method.
Further, a highly relevant proof was presented by 
\cite{Ammari2010} which reveals that the ellipsoid is the only  inclusion with the { Eshelby uniformity property}, when the three eigenvalues of the strain field induced by the remote loading are either identical or distinct.
In contrast to the isotropic case, there are few explorations of the conjecture in anisotropic media.
\cite{Markenscoff1997} found that the inclusions with the { Eshelby uniformity property} should be the ones without any planus surface.
For an inclusion in a two-dimensional domain, 
\cite{Xu2009} firstly verified that the strong version still holds for an anisotropic medium under plane and anti-plane eigenstrains using the Stroh formalism.

In addition to setting forth the classical conjecture regarding the uniform eigenstrain, 
\cite{Eshelby1961} also verified that the extraordinary {  Eshelby uniformity property} could be extended to the case when the eigenstrain is a polynomial of the coordinates of the interior points of the ellipsoidal inclusion.
Specifically, Eshelby stated that if the eigenstrain within an ellipsoidal inclusion is a polynomial of the coordinates of the points with degree $n$, resultantly, the induced elastic strain must be a polynomial with the same degree $n$, which is called  {\it Eshelby's polynomial conservation property}~\citep{CarmenandParnell2020} or {\it Eshelby's polynomial conservation theorem}~\citep{Rahman2002} in the subsequent research. Compared with the uniform eigenstrain,
polynomial eigenstrains are of more practical implications, besides theoretical significance.  For instance,  
\cite{Mura1987} pointed out that the equivalent inclusion method can be extended to nonuniform stress fields in inclusion problems by expanding the equivalent eigenstrains into polynomials of the coordinates.

 To study Eshelby's polynomial conservation theorem, the explicit expression of the induced elastic strain inside an ellipsoidal inclusion, when subjected to an eigenstrain of a polynomial form, is explicitly formulated by
\cite{Sendeckyj1967}
in the context of isotropy with the utilization of the theorem presented by \cite{Ferrers} and \cite{Dyson}.
\cite{Asaro1975} firstly studied the interior strain field of an anisotropic ellipsoidal inclusion subjected to an eigenstrain of a polynomial form. The same result was obtained by
\cite{Mura1978} with the exterior strain field additionally derived. The solutions derived by
\cite{Asaro1975}, and \cite{Mura1978}
are not entirely explicit; thus other researchers derived the explicit closed-form results for a spherical inhomogeneity \citep{Monchiet2011} and a cylindrical inhomogeneity \citep{MonchietanndBonnet2013}. Moreover,
\cite{Rahman2002} reported an explicit closed-form strain field inside an isotropic ellipsoid for a particular polynomial eigenstrain, and 
 \cite{NIE2007}, and \cite{NIE2011} derived the strain field of an elliptic inhomogeneity embedded in an orthotropic medium under linear and quadratic eigenstrains.
These studies all validate Eshelby's polynomial conservation theorem for both isotropic and anisotropic media.
Recently, 
\cite{Liu2013} also proposed a mathematically rigorous proof of Eshelby's polynomial conservation theorem for an ellipsoidal inclusion in an anisotropic medium via solving particular p-harmonic problems in arbitrary dimensions. \cite{CarmenandParnell2020} presented a new scheme  to evaluate the field inside an isolated elliptical inhomogeneity and further verified Eshelby's polynomial conservation theorem in two dimensions, via the approximation method  firstly established by \cite{Joyce2017} to deal with the Eshelby problem in the sense of Newtonian potentials. 
 \cite{Rashidinejad2019} further proved that Eshelby's polynomial conservation theorem remains valid even when multi-field effects are considered. They found that magneto-electro-elastic ellipsoidal inclusions retain Eshelby's polynomial conservation property, but pointed a limitation of this striking property, which requires that the anisotropy is rectilinear.

Although Eshelby's polynomial conservation theorem has been proved for the  ellipsoidal inclusion problem in the context of linearly elastic isotropy and rectilinear anisotropy, conversely, the inverse problem, namely, whether the ellipsoid is the only shape  that possesses Eshelby's polynomial conservation property  for {\it any single} polynomial eigenstrain, is not explored. The answer to this question depends on the proof or disproof of the conjecture that {\it no inclusion other than an ellipsoid transforms  a polynomial eigenstrain into a polynomial elastic strain field of the same degree in it (high-order Eshelby conjecture)}.  Note that the conventional Eshelby conjecture on the uniform field is a special case of the high-order Eshelby conjecture when the degree of the polynomial is zero. Though there are studies dealing with the non-ellipsoidal or non-elliptical inclusions like polygons \citep{Duong2001,Lee2015,Yue2015} with polynomial eigenstrains prescribed,  the results only show that the considered non-ellipsoidal or non-elliptical inclusions do not exhibit Eshelby's polynomial conservation property, which neither falsifies nor substantiates the high-order Eshelby conjecture.

Today, the emerging technology of quantum information may open a new area of applications of the Eshelby formalism for anisotropic inclusions. In this regard, it has been revealed that the strain fields in embedded quantum dots (inclusions) have remarkable implications in strain engineering of the dots, and thus are regarded as viable ``tuning knobs" of the behaviour of the emitted photons \citep{Chen2016,Stepanov2016,Trotta2015,Trotta2016}. An anisotropic-strain field in quantum dots can tune the energy of the emitted polarization-entangled photons while un-affecting the degree of entanglement \citep{Trotta2015,Trotta2016}; biaxial stresses in quantum dots eliminate the fine structure splitting (FSS), which is essential to the generation of high-fidelity entangled photon pairs for quantum communication and other applications~\citep{Chen2016,Wang2012}. Thus, strain modulation in quantum dots may help to realize ``{\it the perfect source of entangled photons}"~\citep{Chen2016,Lu2014,Trotta2015,Trotta2016}.  For laser applications, a uniform biaxial strain in colloidal quantum dots successfully decreases the band-edge degeneracy and photoluminescence linewidth, and the biaxially-strained quantum dots enable continuous-wave lasing compared to the hydrostatically-strained ones \citep{Fan2017}, which manifests the pivotal role of the state and uniformity of the strain in quantum dots. The strain fields in quantum dots are caused by the lattice mismatch between the crystals of the dots and its surroundings (shell or matrix) \citep{Cassette2012,Fan2017,Jing2016,Smith2009,Veilleux2010,Zhang2010,Zhao2019}, which is a type of eigenstrain in the Eshelby formalism for inclusions~\citep{Downes1997,Gosling1995}, or by external force \citep{Chen2016,Trotta2015,Trotta2016}. In addition, an eigenstrain can also be induced by variation of the temperature when the coefficients of thermal expansion of a dot and its surrounding are different. It is worth noting that the crystals making quantum dots are generally anisotropic; for example, InAs and GaAs are cubic crystals, and CdSe has the wurtzite structure (hexagonal, transversely isotropic) \citep{Choi2009,Fan2017,Jing2016}, and the dots can have various shapes, for example, sphere, ellipsoid, prolate/oblate spheroid, multi-facet, cylinder, truncated-pyramid, and even branches \citep{Cassette2012,Choi2009,Efros1996,Fan2017,Smith2009,Steindl2019}. Thus, with the material anisotropy bringing a new dimension, the combination of the material symmetry and the eigenstrain expands the domain of the classical Eshelby conjecture, which raises the question whether the ellipsoid is {\it still the only inclusion shape} that possesses the {  Eshelby uniformity property}, for any combinations of the elastic tensors of a given material symmetry and the uniform eigenstrains. This question is still open -- 
\cite{Kang2008}, and \cite{Parnell2016}
 have explicitly
stated that the Eshelby conjecture in the context of  three-dimensional anisotropic elasticity has not been proved either
in the weak or strong sense.

For generally anisotropic materials, the interior Eshelby tensor is directly determined by the stiffness tensor of the material and the shape of the inclusion~\citep{Mura1987}.
Therefore, the {  Eshelby uniformity property} is a special transformation of a uniform eigenstrain into a uniform elastic strain in an inclusion by
the stiffness tensor of the material and the shape of the inclusion. Then, the weak and strong versions of the generalized Eshelby conjecture for generally anisotropic media can be posed as follows:

 (1) {\it Generalized weak version: An ellipsoid alone transforms all combinations of elastic tensors of a given material symmetry and uniform eigenstrains into uniform elastic strain fields in it.}

(2) {\it Generalized strong version: No inclusion other than an ellipsoid transforms a combination of an elastic tensor of a given material symmetry and a uniform eigenstrain into a uniform elastic strain field in it.}

Noting that there may be different ways of statements of these versions,  for example,
 those in the literature  \citep{Kang2008,Liu2008,Xu2009}, we choose to follow the essence of the original statement made by~\cite{Eshelby1961} to express the weak version,
and use a way of exclusion to express the strong version. The mathematical expressions of these conjectures will be given in the next section. It is noted that the weak and strong versions of the high-order conjecture concerning polynomial eigenstrains can be posed by replacing `uniform eigenstrains' with `polynomial eigenstrains' in the above statements.
That the generalized strong version is true means that the generalized weak version is true.

In this work, we will explore these versions of the generalized Eshelby conjecture for both uniform and polynomial eigenstrains.  After mathematically defining the weak and strong versions of the conjecture in Section 2, we will provide proofs of the  weak version of the conjecture for four anisotropic materials, namely, cubic, transversely isotropic, orthotropic, and monoclinic ones in Section 3. In  Section 4 and Section 5, with the help of the variational method proposed by  
\cite{Liu2008}, we prove the existence of non-ellipsoidal inclusions that possess Eshelby's polynomial conservation property in these anisotropic media and also the isotropic medium, when the eigenstrain is expressed in the form of a quadratic polynomial. Then, more counter-examples are constructed to extend the proof of the invalidity of the strong version of the high-order Eshelby conjecture for quadratic eigenstrains to that for polynomial eigenstrains of any even degree. We also show that the counter-example given by 
\cite{Liu2008} concerning the strong version of the Eshelby conjecture in the isotropic medium can be utilized to disprove the  strong version of the generalized Eshelby conjecture for uniform eigenstrains in the anisotropic media. A sufficient condition for the existence of counter-examples for polynomial eigenstrains of any even degree in these anisotropic media and the proof of its validity are also provided.

We stress that in this work, we only consider {  one-component inclusions which are connected and bounded}, and the boundary of the studied inclusion region is required to be Lipschitz continuous, as is restricted in the previous work~\citep{Ammari2010,Kang2008,Liu2008}.

\section{Formulation of the generalized weak and strong versions}

\subsection{Conjecture for uniform eigenstrains}

Let  $\Omega\subset \mathbb{R}^{3}$ denote the inclusion domain embedded in an infinite medium in the three-dimensional Euclidean space $\mathbb{R}^{3}$.
The equilibrium equation for the three-dimensional inclusion problem  of linear elasticity
can be expressed as
\begin{align}\label{1.1}
\begin{split}
\boldsymbol{\nabla} {\rm{\cdot}}\left[\bfC:\left( {\boldsymbol{\nabla}\otimes \bfu\left(\bfx\right) - \chi_\Omega \left( \bfx \right)\boldsymbol{{\varepsilon ^*}}(\bfx)} \right) \right] = 0 \;\;\mathrm{in} \;\;\;\mathbb{R}^{3}
\end{split}
\end{align}
where $\boldsymbol{\nabla}$ is the gradient operator; $\bfC$ is the fourth-order anisotropic elastic tensor; $\bfu(\bfx)$ is the displacement vector; $\boldsymbol{\varepsilon^*}(\bfx)$ is the eigenstrain which is a second-order tensor; and $\chi_\Omega \left( \bfx \right)$ is the characteristic function.
It is known that \eqref{1.1} admits a solution~\citep{Mura1987}:
\begin{align}\label{1.2}
\begin{split}
{\bfu}\left( \bfx \right) = \mathop \int_{\Omega} \boldsymbol{\nabla}\otimes \bfG\left( {\bfx - \bfy} \right):\left( {\bfC:{\boldsymbol{\varepsilon ^*}}\left( {\bfy} \right)} \right)\;d\bfy.
\end{split}
\end{align}
Here, $\bfG(\bfx)$ is the Green function, a second-order tensor solved by the equation
\beas
\begin{split}
  \boldsymbol{\nabla} {\rm{\cdot}}[\bfC: \boldsymbol{\nabla}\otimes \bfG\left(\bfx\right)] = \delta \left( \bfx \right)\boldsymbol{{\tilde I}},
\end{split}
\eeas
where $\delta \left( {\bfx} \right)$ is the Dirac function, and $\boldsymbol{{\tilde I}}$ is the second-order identity tensor.

The inclusion with the {  Eshelby uniformity property} is such a domain $\Omega^E$ which makes
\begin{align}\label{4.4}
\begin{split}
\bfu\left( \bfx \right) =\int_{\Omega^E} \boldsymbol{\nabla}\otimes \bfG\left(  {\bfx - \bfy} \right):\left( {\bfC:{\boldsymbol{\varepsilon ^*}}} \right)\;d\bfy = \mathcal{L}(\bfx),\  \bfx\in \Omega^E.
\end{split}
\end{align}
Here, $\mathcal{L}(\bfx)$ denotes a linear vector function of $\bfx$, and $\boldsymbol{\varepsilon ^*}$ is a uniform eigenstrain (independent of $\bfy$)
prescribed in the inclusion.

Define a mapping $F$ that maps the Cartesian product of the set $\{\boldsymbol{\varepsilon ^*}\}$ of uniform eigenstrains  and the set $\{\bfC\}$  of elastic tensors of a given material symmetry into the set $\{\Omega^E\}$ of corresponding configurations which possess the {  Eshelby uniformity property}, i.e.,
\begin{align}\label{4.3}
\begin{split}
F \;:\;\;\{\boldsymbol{\varepsilon ^*}\}\times \{\bfC\} \rightarrow\{\Omega^E\} ,\;(\boldsymbol{\varepsilon ^*},\bfC) \mapsto F(\boldsymbol{\varepsilon ^*},\bfC) \subset {\mathbb{R}^3}
\end{split}
\end{align}
with \eqref{4.4} regarded as the implicit expression of $F$.
Note that $F$  is  a multiple-valued function of the two variables, i.e., the uniform eigenstrain and the elastic tensor, for ellipsoids have been shown to possess the {  Eshelby uniformity property} for any uniform eigenstrain regardless of the material symmetry 
 through the Fourier analysis \citep{Mura1987}; in other words,
\begin{align}\label{1.3}
\begin{split}
\forall (\boldsymbol{\varepsilon ^*},\bfC)\in \{(\boldsymbol{\varepsilon ^*},\bfC)\},\;\;\left\{ E \right\}\subseteq\ \{F(\boldsymbol{\varepsilon ^*},\bfC)\},
\end{split}
\end{align}
where $\left\{ E \right\}$ represents the set of ellipsoids, and $\{F(\boldsymbol{\varepsilon ^*},\bfC)\}$  represents the image set of the combination $(\boldsymbol{\varepsilon ^*},\bfC)$ of a  uniform eigenstrain $\boldsymbol{\varepsilon ^*}$ and an elastic tensor $\bfC$ of a given material symmetry.

Define $\left\{ {\left\{ {F\left( {{\boldsymbol{\varepsilon ^*},\bfC}} \right)} \right\}} \right\}$ ${:=} \left\{ {\left\{ {F\left( {{\boldsymbol{\varepsilon ^*},\bfC}} \right)} \right\}|{\boldsymbol{\varepsilon ^*}} \in \left\{ {{\boldsymbol{\varepsilon ^*}}} \right\}}\;,\bfC \in \left\{ {\bfC} \right\} \right\}$ as the class of the image sets $\{F(\boldsymbol{\varepsilon ^*},\bfC)\}$.  Then, the two versions of the generalized Eshelby conjecture are expressed as
\begin{subequations}
\begin{align}\label{weak}
\mathrm{Generalized \;\; weak\;\; version:} \;\;\cap \{\{F(\boldsymbol{\varepsilon ^*}&,\bfC)\}\}= \{ E\};
\end{align}
\begin{align}\label{strong}
\mathrm{Generalized \;\;strong\;\; version:}  {\;\;\mathrm{For\; a \;given}\;} (\boldsymbol{\varepsilon ^*},\bfC),\;\;\{F(\boldsymbol{\varepsilon ^*},\bfC)\} = \{ E\}
\end{align}
\end{subequations}
with
\begin{align}\label{10.1a}
\begin{split}
&\cap \{\{F(\boldsymbol{\varepsilon ^*},\bfC)\}\}=\{\;\xi\;\;|\;\;\forall\ \{F(\boldsymbol{\varepsilon ^*},\bfC)\} \in \{\{F(\boldsymbol{\varepsilon ^*},\bfC)\}\}, \xi \in \{F(\boldsymbol{\varepsilon ^*},\bfC)\}\}. \\
\end{split}
\end{align}

\subsection{Conjecture for polynomial eigenstrains (High-order conjecture)}

The generalized weak and strong versions of the high-order Eshelby conjecture can be formulated by following the same procedure from \eqref{4.3} to \eqref{10.1a}, which leads to the same expression as those in \eqref{weak} and \eqref{strong}.

 The only differences are that ${\boldsymbol{\varepsilon ^*}}$ no longer denotes uniform eigenstrains, but denotes polynomial eigenstrains of degree $n$ with $n$ a non-negative integer, and in this case, the implicit expression of $F$ is no longer defined by \eqref{4.4} which corresponds to the Eshelby's uniformity property, but defined by
\beas
\begin{split}
\bfu\left( \bfx \right) =\int_{\Omega^E} \boldsymbol{\nabla}\otimes \bfG\left(  {\bfx - \bfy} \right):\left( {\bfC:{\boldsymbol{\varepsilon ^*}}(\bfy)} \right)\;d\bfy = \mathcal{P}^{(n+1)}(\bfx),\  \bfx\in \Omega^E
\end{split}
\eeas
where $\mathcal{P}^{(n+1)}(\bfx)$ represents a polynomial vector function of $\bfx$ of degree $n+1$, which corresponds to Eshelby's polynomial conservation property.

\section{Proofs of the generalized weak version for uniform eigenstrains}

For brevity, we will follow the Voigt notation  of the anisotropic elastic tensor $\bfC$. The Voigt notation expresses $C_{mnkl}\;(m,n,k,l=1,2,3)$, which has the symmetries $C_{mnkl}=C_{klmn}=C_{klnm}$, as $C_{ij}\;(i,j=1,2,3,4,5,6)$ with subscript $i$ representing $mn$ and subscript $j$ representing $kl$, according to the following rule:
\beas
\begin{split}
&i(\text{or}\;j)=1\;\text{corresponding to}\;mn(\text{or}\;kl)=11;\quad i(\text{or}\;j)=2\;\text{corresponding to}\;mn(\text{or}\;kl)=22;\\
&i(\text{or}\;j)=3\;\text{corresponding to}\;mn(\text{or}\;kl)=33;\quad i(\text{or}\;j)=4\;\text{corresponding to}\;mn(\text{or}\;kl)=23;\\
&i(\text{or}\;j)=5\;\text{corresponding to}\;mn(\text{or}\;kl)=13;\quad i(\text{or}\;j)=6\;\text{corresponding to}\;mn(\text{or}\;kl)=12.\\
\end{split}
\eeas

{Then we use the Fourier form of the Eshelby formalism to facilitate mathematical manipulations.}
The Green function in \eqref{4.4} can be expressed as \citep{Mura1987}
 \beas
\begin{split}
G_{mn}(\bfx)=\frac{1}{(2\pi)^3}\int_{\mathbb{R}^3}L_{mn}(\boldsymbol{\xi})e^{-\mathrm{i}\boldsymbol{\xi}\cdot \bfx} d\boldsymbol{\xi},
\end{split}
\eeas
 substitution of which into \eqref{1.2} yields
 \begin{align}\label{Fourier}
\begin{split}
u_m(\bfx)=\frac{-\mathrm{i}}{(2\pi)^3}\int_{\mathbb{R}^3}L_{mn}(\boldsymbol{\xi})\sigma ^*_{ns}\xi_s\int_{\Omega}e^{-\mathrm{i}\boldsymbol{\xi}\cdot( \bfx - \bfy)}d\bfy d\boldsymbol{\xi},
\end{split}
\end{align}
where the Einstein summation convention is applied,
with $\mathrm{i}=\sqrt{-1}$ the imaginary unit, $\sigma ^*_{ns}=C_{nspq}\varepsilon^*_{pq}$ the uniform eigenstress, and
 \begin{align}\label{L}
\begin{split}
L_{mn}(\boldsymbol{\xi})=(C_{mknl}\xi_k\xi_l)^{-1}\;(m,n,k,l=1,2,3).
\end{split}
\end{align}

Then, we present the following theorem:
\begin{theorem}\label{cubictheorem}
 {Let $\Omega\subset \mathbb{R}^{3}$ be a { one-component connected bounded open domain} with a Lipschitz boundary}. There exist combinations $(\boldsymbol{\overline{\varepsilon}^{(1)}},{\bfC})$ and $(\boldsymbol{\overline{\varepsilon}}^{(2)},{\bfC})$,  where  $\boldsymbol{\overline{\varepsilon}^{(1)}}$ and $\boldsymbol{\overline{\varepsilon}^{(2)}}$
are { two linearly independent} uniform eigenstrains, and ${\bfC}$ is the elastic tensor of {an anisotropic material with one of the cubic, transversely isotropic, orthotropic, and monoclinic symmetries}, such that {$\bfu(\bfx)$ in \eqref{Fourier} is linear inside $\Omega$} for
$(\boldsymbol{\overline{\varepsilon}^{(1)}},{\bfC})$ and $(\boldsymbol{\overline{\varepsilon}}^{(2)},{\bfC})$ simultaneously,
if and only if $\Omega$ is of ellipsoidal shape.
\end{theorem}

 Theorem \ref{cubictheorem} means
\begin{align}\label{newstrategy}
\begin{split}
 &\exists\; \boldsymbol{\overline{\varepsilon}^{(1)}},\boldsymbol{\overline{\varepsilon}^{(2)}}\in \{\boldsymbol{\varepsilon^*}\},\;{\bfC}\in \{{\bfC^{\;\text{an}}}\} \quad s.t.\;\;\{\{F(\boldsymbol{\overline{\varepsilon}^{(1)}}, {\bfC})\}\} \cap \{\{F(\boldsymbol{\overline{\varepsilon}^{(2)}},{\bfC})\}\}=\{ E\},
\end{split}
\end{align}
where {$\{\bfC^{\;\text{an}}\}$} denotes the set of the elastic tensors of {one of the anisotropic materials mentioned in Theorem \ref{cubictheorem}}.

{
It is straightforward to see
\beqs\label{assist4bf}
\begin{split}
\cap \{\{F(\boldsymbol{\varepsilon ^*},\bfC)\}\}\subseteq \{\{F(\boldsymbol{\overline{\varepsilon}^{(1)}}, {\bfC})\}\} \cap \{\{F(\boldsymbol{\overline{\varepsilon}^{(2)}}, {\bfC})\}\},
\end{split}
\eeqs
 substitution of which into \eqref{newstrategy} yields
 \beqs\label{assist3}
 \cap \{\{F(\boldsymbol{\varepsilon ^*},\bfC)\}\}\subseteq \left\{ E \right\}.
  \eeqs
Further, since \eqref{1.3} indicates
\beqs\label{assist4}
 \left\{ E \right\}\subseteq \cap \{\{F(\boldsymbol{\varepsilon ^*},\bfC)\}\},
  \eeqs
then combining \eqref{assist4} with \eqref{assist3} leads to \eqref{weak}.}

Then we turn to prove Theorem \ref{cubictheorem}.




\subsection*{{ Proof of Theorem \ref{cubictheorem}}}

We consider four anisotropic materials in sequence according to the number of the independent elastic parameters.

\subsubsection*{{(1) Cubic material}}

For a cubic material, let the Cartesian coordinate system $\bfx= \left( {{x_1},{x_2},{x_3}} \right)$ be set in $\mathbb{R}^{3}$ with the three coordinate axes coinciding with the 4-fold axes of rotational symmetry of the infinite homogeneous cubic material. In addition, the origin is placed within the inclusion $\Omega$.  Hence, the three independent elastic parameters of the cubic material are ${C_{11}},\;{C_{12}}$ and ${C_{44}}$. 
The positive definiteness of the strain energy requires that
 $\forall \boldsymbol{\varepsilon} \ne \textbf{0}$, $C_{ijmn}\varepsilon_{ij}\varepsilon_{mn}>0$, which means
\begin{align}\label{121.3}
\begin{split}
&{C_{11}} > 0,\;{C_{44}} > 0,\;{C_{11}}>{C_{12}}>-\frac{1}{3}{C_{11}}.
\end{split}
\end{align}

We choose
\begin{align}\label{cubiceigenstrain}
\begin{split}
\boldsymbol{\overline{\varepsilon}^{(1)}}=\begin{bmatrix} 0&0&0\\0&0&0\\0&0&\overline{\varepsilon}^{(1)}_{33}\end{bmatrix},\;\boldsymbol{\overline{\varepsilon}^{(2)}}=\begin{bmatrix} 0&0&0\\0&0&\overline{\varepsilon}^{(2)}_{23}\\0&\overline{\varepsilon}^{(2)}_{23}&0\end{bmatrix},
\end{split}
\end{align}
where $\overline{\varepsilon}^{(1)}_{33},\overline{\varepsilon}^{(2)}_{23}\neq 0$ are two real constants, and thus the corresponding eigenstresses $\boldsymbol{\overline{\sigma}^{(i)}}=\bfC:\boldsymbol{\overline{\varepsilon}^{(i)}}\;(i=1,2)$ are
\begin{align}\label{cubiceigenstrain2}
\begin{split}
\boldsymbol{\overline{\sigma}^{(1)}}=\overline{\varepsilon}^{(1)}_{33}\begin{bmatrix} C_{12}&0&0\\0&C_{12}&0\\0&0&C_{11}\end{bmatrix},\;\boldsymbol{\overline{\sigma}^{(2)}}=2\overline{\varepsilon}^{(2)}_{23}\begin{bmatrix} 0&0&0\\0&0&C_{44}\\0&C_{44}&0\end{bmatrix}.
\end{split}
\end{align}
Note that the superscript (1) will always represent the field quantity that results from $\boldsymbol{\overline{\varepsilon}^{(1)}}$, and the superscript (2) will always represent the field quantity that results from $\boldsymbol{\overline{\varepsilon}^{(2)}}$ in the following derivations.

As for ${\bfC}$, we require
\begin{align}\label{cubiconstrain}
\begin{split}
C_{12}+C_{44}=0.
\end{split}
\end{align}
Note that although under the condition \eqref{cubiconstrain}, a cubic material only possesses two independent elastic parameters, it prevents the cubic material from degenerating into an isotropic material since for an isotropic material, $C_{12}+C_{44}=\frac{C_{11}+C_{12}}{2}=\kappa+\frac{1}{3}\mu>0$, where $\kappa$ and $\mu$ are the bulk modulus and shear modulus, respectively.

By substituting \eqref{cubiconstrain} and $\boldsymbol{\overline{\sigma}^{(1)}}$ and $\boldsymbol{\overline{\sigma}^{(2)}}$ in \eqref{cubiceigenstrain2} into \eqref{Fourier}, we obtain
\begin{align}\label{121.7}
\begin{split}
&u_1^{(1)}(\bfx)=\frac{-\mathrm{i}\;\overline{\varepsilon}^{(1)}_{33}}{(2\pi)^3}\int_{\mathbb{R}^3}\frac{C_{12}\xi_1}{C_{11} \xi_1^2 + C_{44} (\xi_2^2 + \xi_3^2)}\int_{\Omega}e^{-\mathrm{i}\boldsymbol{\xi}\cdot( \bfx - \bfy)}d\bfy d\boldsymbol{\xi},\\
&u_2^{(1)}(\bfx)=\frac{-\mathrm{i}\;\overline{\varepsilon}^{(1)}_{33}}{(2\pi)^3}\int_{\mathbb{R}^3}\frac{C_{12}\xi_2}{C_{11} \xi_2^2 + C_{44} (\xi_1^2 + \xi_3^2)}\int_{\Omega}e^{-\mathrm{i}\boldsymbol{\xi}\cdot( \bfx - \bfy)}d\bfy d\boldsymbol{\xi},\\
&u_3^{(1)}(\bfx)=\frac{-\mathrm{i}\;\overline{\varepsilon}^{(1)}_{33}}{(2\pi)^3}\int_{\mathbb{R}^3}\frac{C_{11}\xi_3}{C_{11} \xi_3^2 + C_{44} (\xi_1^2 + \xi_2^2)}\int_{\Omega}e^{-\mathrm{i}\boldsymbol{\xi}\cdot( \bfx - \bfy)}d\bfy d\boldsymbol{\xi}.\\
\end{split}
\end{align}
and
\begin{align}\label{121.8}
\begin{split}
&u_1^{(2)}(\bfx)=0,\\
&u_2^{(2)}(\bfx)=\frac{-\mathrm{i}\;\overline{\varepsilon}^{(2)}_{23}}{(2\pi)^3}\int_{\mathbb{R}^3}\frac{2C_{44}\xi_3}{C_{11} \xi_2^2 + C_{44} (\xi_1^2 + \xi_3^2)}\int_{\Omega}e^{-\mathrm{i}\boldsymbol{\xi}\cdot( \bfx - \bfy)}d\bfy d\boldsymbol{\xi},\\
&u_3^{(2)}(\bfx)=\frac{-\mathrm{i}\;\overline{\varepsilon}^{(2)}_{23}}{(2\pi)^3}\int_{\mathbb{R}^3}\frac{2C_{44}\xi_2}{C_{11} \xi_3^2 + C_{44} (\xi_1^2 + \xi_2^2)}\int_{\Omega}e^{-\mathrm{i}\boldsymbol{\xi}\cdot( \bfx - \bfy)}d\bfy d\boldsymbol{\xi}.\\
\end{split}
\end{align}
Then we require that the right-hand sides of \eqref{121.7} and \eqref{121.8} should be linear with respect to $\bfx$.
Given this, by combining $\eqref{121.7}_3$ and $\eqref{121.8}_3$, we see
\begin{align}\label{cubiclinear}
\begin{split}
&\frac{\partial}{\partial x_2} U(\bfx) =\mathcal{L}_1(\bfx);\;\quad\frac{\partial}{\partial x_3} U(\bfx)=\mathcal{L}_2(\bfx),\;\bfx\in\Omega,
\end{split}
\end{align}
where $\mathcal{L}_i(\bfx)\;(i=1,2)$ denote two linear scalar functions of $\bfx$, and
\begin{align}
\begin{split}\label{NewtonianpotentialFourier}
U(\bfx){:=}\frac{1}{(2\pi)^3}\int_{\mathbb{R}^3}\frac{1}{(\xi_1^2+\xi_2^2+\frac{1}{t^2}\xi_3^2)}\int_{\Omega}e^{-\mathrm{i}\boldsymbol{\xi}\cdot( \bfx - \bfy)}d\bfy d\boldsymbol{\xi},
\end{split}
\end{align}
with $t{:=}\sqrt{\frac{C_{44}}{C_{11}}}$.
It is noted that by using the Fourier transformation, \eqref{NewtonianpotentialFourier} is a solution to
\begin{align}\label{laplacian}
\begin{split}
\Delta^t U(\bfx)=\chi_{\Omega}(\bfx)\;\;\text{in}\;\;\mathbb{R}^3,
\end{split}
\end{align}
with the boundary conditions
\begin{align}\label{laplacianboundary}
\begin{split}
&\left.\frac{\partial}{\partial x_2}U(\bfx)\right|_{|\bfx|\rightarrow +\infty}=\left.\frac{1}{\overline{\varepsilon}_{23}^{(2)}}u_3^{(2)}(\bfx)\right|_{|\bfx|\rightarrow +\infty};\quad \left.\frac{\partial}{\partial x_3}U(\bfx)\right|_{|\bfx|\rightarrow +\infty}=\left.\frac{C_{44}}{C_{11}\overline{\varepsilon}_{33}^{(1)}}u_3^{(1)}(\bfx)\right|_{|\bfx|\rightarrow +\infty},
\end{split}
\end{align}
where
\beas
\begin{split}
\Delta^t {:=}\frac{\partial^2}{\partial x_1^2}+\frac{\partial^2}{\partial x_2^2}+\frac{\partial^2}{t^2\partial x_3^2}.
\end{split}
\eeas

To continue our derivation, we prove the following lemma:
\begin{lemma}\label{lemma1}
{The solution of the partial differential equation \eqref{laplacian} under the condition \eqref{laplacianboundary} is}
\begin{align}\label{laplaciansolution}
\begin{split}
U(\bfx)=-t\int_\Omega\frac{1}{4\pi R_t(\bfx-\bfy)}d\bfy+\hat{\mathcal{L}}(x_1),
\end{split}
\end{align}
where $R_t(\bfx-\bfy){:=}\sqrt{(x_1-y_1)^2+(x_2-y_2)^2+t^2(x_3-y_3)^2}$, and $\hat{\mathcal{L}}(x_1)$ denotes a linear scalar function of $x_1$.
\end{lemma}

\subsubsection*{Proof of Lemma \ref{lemma1}}


Recall that $u_3^{(1)}$ and $u_3^{(2)}$ in \eqref{laplacianboundary} are the displacements along the $x_3$-axis due to two different eigentrains $\overline{\boldsymbol{\varepsilon}}^{(1)}$ and $\overline{\boldsymbol{\varepsilon}}^{(2)}$, respectively. According to the work of \cite{Tanuma2007}, the Green function utilized to obtain the solution $\bfu$ of \eqref{1.2} can be given as
\begin{align}\label{Ganiso}
\begin{split}
\bfG(\bfx)=\frac{1}{8\pi^2|\bfx|}\int_{-\pi}^{\pi}\bfT^{-1}(\theta)d\theta.
\end{split}
\end{align}
 Here  $T_{ik}(\theta)=C_{ijkl}n_jn_l$,  where $\bfn(\theta)=-\bfe_1\sin\theta+\bfe_2\cos\theta$ with $\bfe_1$ and $\bfe_2$ being two unit vectors satisfying $\bfe_1\cdot\bfe_2=0$ and $\bfe_1\times\bfe_2=\frac{\bfx}{|\bfx|}$. {  The symmetric matrix $\bfT$ is positive definite \citep{Tanuma2007} so that $\bfT$ is invertible for any $\bfx$ and $\theta$. Thus, $\int_{-\pi}^{\pi}\bfT^{-1}(\theta)d\theta$, which only relies on the direction $\frac{\bfx}{|\bfx|}\in S^2$ of $\bfx$, is bounded.
Given this, by substituting \eqref{Ganiso} back into \eqref{1.2}, it is straightforward to verify that}
\begin{align}\label{SIboundary}
\begin{split}
{\bfu}\left( \bfx \right)|_{|\bfx|\rightarrow +\infty}\rightarrow 0.
\end{split}
\end{align}

By introducing transformations
\begin{align}\label{cotrans}
\bfx' {:=} \left( {\begin{array}{*{20}{c}}
  1&0&0 \\
  0&1&0 \\
  0&0&{t}
\end{array}} \right) \cdot \bfx,\ \ \ \  \
\bfy' {:=} \left( {\begin{array}{*{20}{c}}
  1&0&0 \\
  0&1&0 \\
  0&0&{t}
\end{array}} \right) \cdot \bfy,
\end{align}
and then substituting \eqref{cotrans} into \eqref{laplacian} along with substituting \eqref{cotrans} and \eqref{SIboundary} into \eqref{laplacianboundary}, we get
\begin{align}\label{laplaciantrans}
\begin{split}
\left\{ {\begin{array}{*{20}{c}}
  {\Delta_{x'} U(\bfx')=\chi_{\Omega'}(\bfx')\;\;\text{in}\;\;\mathbb{R}^3}, \\
  {\frac{\partial U(\bfx')}{\partial x_2'}\rightarrow 0\;\;\text{at infinity}}, \\
  {\frac{\partial U(\bfx')}{\partial x_3'}\rightarrow 0\;\;\text{at infinity}},
\end{array}\;\;\;\;\;\;\;\;\;\;\;\;\;\;\;\;\;\;\;\;\;\;\;\;\;\;\;\;\;\;}  \right.\;\;\;
\end{split}
\end{align}
where $\Delta_{x'}$ denotes the Laplacian operator with respect to $\bfx'$ and
\begin{align}\label{7.2}
\begin{split}
\Omega' {:=} \left\{ {\bfy'\left| {\left( {\begin{array}{*{20}{c}}
  1&0&0 \\
  0&1&0 \\
  0&0&{\frac{1}{{{t}}}}
\end{array}} \right)\cdot\bfy'} \right. \in \Omega } \right\}.
\end{split}
\end{align}

Therefore, our aim is to prove that {the solution of \eqref{laplaciantrans} must be expressed as}
\begin{align}\label{laplaciansolutiontrans}
\begin{split}
U(\bfx')=-\int_{\Omega'}\frac{1}{4\pi|\bfx'-\bfy'|}d\bfy'+\hat{\mathcal{L}}(x_1'),
\end{split}
\end{align}
 {which substantiates that the solution of \eqref{laplacian} under the condition \eqref{laplacianboundary} is $\eqref{laplaciansolution}$.  We will achieve our goal by contradiction.}

 {Firstly, it is easy to verify that \eqref{laplaciansolutiontrans} solves \eqref{laplaciantrans}.
 Then we assume that there is another solution  $U'(\bfx')$ of \eqref{laplaciantrans}, which can not be expressed as \eqref{laplaciansolutiontrans}.}
Let $U^*(\bfx'):=$ $U(\bfx')-U'(\bfx')$, and then we know $U^*(\bfx')$ shall be a harmonic function satisfying
\begin{align}\label{Ustar}
\begin{split}
\left\{ {\begin{array}{*{20}{c}}
  {\Delta_{x'} U^*(\bfx')=0\;\;\text{in}\;\;\mathbb{R}^3}, \\
  {\frac{\partial U^*(\bfx')}{\partial x_2'}\rightarrow 0\;\;\text{at infinity}}, \\
  {\frac{\partial U^*(\bfx')}{\partial x_3'}\rightarrow 0\;\;\text{at infinity}}.
\end{array}\;\;\;\;\;\;\;\;\;\;\;\;\;\;\;\;\;\;\;\;\;\;\;\;\;\;\;\;\;\;}  \right.\;\;\;
\end{split}
\end{align}
Since  harmonic functions are analytic~\citep{Han2012},  $U^*(\bfx')\in C^{\infty}(\mathbb{R}^3)$, and thus $\frac{\partial U^*(\bfx')}{\partial x_2'}\in C^{\infty}({\mathbb{R}^3})$ and $\frac{\partial U^*(\bfx')}{\partial x_3'}\in C^{\infty}({\mathbb{R}^3})$.

Then, due to the boundary condition  in \eqref{Ustar} and the analyticity of $\frac{\partial U^*(\bfx')}{\partial x_2'} $ and $\frac{\partial U^*(\bfx')}{\partial x_3'}$,  $\frac{\partial U^*(\bfx')}{\partial x_2'} $ and $\frac{\partial U^*(\bfx')}{\partial x_3'}$ should be bounded in ${\mathbb{R}^3}$.
Owing to the Liouville theorem,  which stipulates that {\it any harmonic function in $\mathbb{R}^3$ bounded from above or
below is constant}~\citep{Han2012}, and the boundary condition, which indicates that $\frac{\partial U^*(\bfx')}{\partial x_2'}$ and $\frac{\partial U^*(\bfx')}{\partial x_3'}$ will tend to $0$ at infinity, we conclude that
\beas
\begin{split}
\frac{\partial U^*(\bfx')}{\partial x_2'}\equiv 0,\;\frac{\partial U^*(\bfx')}{\partial x_3'}\equiv 0 \;\;\;{\text{in}}\;\; \mathbb{R}^3,
\end{split}
\eeas
which implies
\begin{align}\label{phi1}
\begin{split}
U^*(\bfx')=\varphi(x_1'),
\end{split}
\end{align}
where $\varphi(x_1')$ is an unknown function. Further, by substituting \eqref{phi1} into \eqref{Ustar}, we have

\begin{align}\label{Ustarfinal}
\begin{split}
\frac{d^2\varphi(x_1')}{dx_1'^2}=0.
\end{split}
\end{align}
{Substituting (\ref{Ustarfinal}) back into \eqref{phi1} exhibits that $U^*(\bfx')$ is a linear scalar function only dependent on $x_1'$. Therefore, $U'(\bfx')=U(\bfx')-U^*(\bfx')$ can be expressed as \eqref{laplaciansolutiontrans} since the linear scalar function $U^*(\bfx')$ is contained in the expression \eqref{laplaciansolutiontrans} of $U(\bfx')$, which contradicts the previous  assumption that $U'(\bfx')$ can not be expressed as \eqref{laplaciansolutiontrans}. Thus the solution of \eqref{laplaciantrans} must be expressed as \eqref{laplaciansolutiontrans}.}
 The proof is completed.

Then based on Lemma \ref{lemma1}, substituting \eqref{laplaciansolution} into \eqref{cubiclinear} yields
\begin{align}
\begin{split}
&\frac{\partial}{\partial x_2} \int_\Omega \frac{1}{{{R_t}\left(\bfx-\bfy\right)}}d\bfy=\tilde{\mathcal{L}}_1(\bfx);\quad\frac{\partial}{\partial x_3} \int_\Omega\frac{1}{{{R_t}\left(\bfx-\bfy\right)}}d\bfy=\tilde{\mathcal{L}}_2(\bfx),\;\bfx\in\Omega,
\end{split}
\end{align}
where $\tilde{\mathcal{L}}_i(\bfx)\;(i=1,2)$ denote linear scalar functions of $\bfx$,
which yield
\begin{align}\label{cubicsolution}
\begin{split}
-\int_\Omega \frac{1}{4\pi{{R_t}\left(\bfx-\bfy\right)}}d\bfy=q(\bfx)+\psi(x_1),\;\bfx\in\Omega
\end{split}
\end{align}
where $q(\bfx)$ denotes a quadratic function of $\bfx$, and $\psi(x_1)$ denotes an unknown function.

Finally, by substituting \eqref{cotrans} into \eqref{cubicsolution}, we obtain
\begin{align}\label{121.9}
\begin{split}
{N_{\Omega '}}\left( {\bfx}' \right) =&-\int_{\Omega'}\frac{1}{{4\pi | \bfx'-\bfy' |}}d\bfy'={{t}\left[ {q}\left( {{x_1}',{x_2}',\frac{{{x_3}'}}{{{t}}}} \right) + {\psi}\left( {{{x_1}'}} \right) \right]},\;\bfx'\in\Omega'.
\end{split}
\end{align}
  ${N_{\Omega'}}\left( \bfx' \right)$ denotes the Newtonian potential induced by the inclusion $\Omega'$ that is transformed from the original inclusion $\Omega$ via \eqref{7.2} with the mass density 1.

As is known, the Newtonian potential ${N_{\Omega '}}\left( {\bfx}' \right)$ induced by $\Omega'$ with the mass density 1,
irrespective of its shape, is the solution of the equation
\begin{align}\label{1.21}
\begin{split}
{\boldsymbol{\Delta}_{x'}}{N_{\Omega '}}\left( {\bfx}' \right) = \chi_{\Omega} \left( \bfx' \right) \;\;\text{{in}} \;\;\mathbb{R}^3.
\end{split}
\end{align}

Then there comes a theorem which sets the stage for both the previous proofs \citep{Kang2008,Ammari2010} concerning the
isotropic material and the current proof concerning the cubic material.
\begin{theorem}\label{TheoremNP}
 Let $\Omega'$ be a bounded domain with a Lipschitz boundary. The relation
 \beqs\label{eq:TheoremNP}
 \frac{1}{4\pi}\int \limits_{\Omega'} \frac{1}{{|\bfz' -\textbf{y}'|}}d \textbf{y}' =quadratic,\quad \bfz'\in\Omega'
 \eeqs
holds if and only if $\Omega'$ is an ellipsoid~\citep{Kang2008}.
\end{theorem}
{  We note that the left-hand side of \eqref{eq:TheoremNP} is the negative of the Newtonian potential ${N_{\Omega' }}\left( \bfz' \right) $ induced by $\Omega'$.
The proof of Theorem~\ref{TheoremNP} was first obtained by \cite{Dive1931} and \cite{Nikliborc1932} for $C^1$ domains, and then given by \cite{Kang2008} for Lipschitz domains. It is noted that \cite{Liu2007,Liu2021} defined the inclusions $\Omega'$ that satisfy (\ref{eq:TheoremNP})
as ``E-inclusions". In this sense, single ellipsoidal inclusions are E-inclusions; however, \cite{Liu2007,Liu2021} proved that for multiple inclusions, there exist non-ellipsoidal E-inclusions, and importantly, they proved that these non-ellipsoidal E-inclusions have the
Eshelby uniformity property for the isotropic medium.
}

Inserting \eqref{121.9} into \eqref{1.21}  gives rise to
\[\frac{{{d^2}}}{{d{x_1}'{^2}}}{\psi}\left( x_1' \right) = {\text{constant}},\]
which indicates that ${\psi}\left( x_1'\right)$ can only be a constant, linear or quadratic function of $x_1'$.

Due to {Theorem \ref{TheoremNP}}, the validation of the quadratic form of the right-hand side of \eqref{121.9} within $\Omega'$ leads to the substantiation that $\Omega'$  can only be of ellipsoidal shape.
The verification that $\Omega'$ can only be of ellipsoidal shape results in the conclusion that $\Omega$ can only be of ellipsoidal shape, for $\Omega$ is constructed by stretching $\Omega'$ along axis $x_{3}$ by proportion $\zeta=1/t$. Therefore, \eqref{newstrategy} is verified so that the substantiation of  Theorem \ref{cubictheorem} is fulfilled for the cubic material.

\subsection*{{(2) Transversely isotropic material}}

For a transversely isotropic material, let a Cartesian coordinate system $\bfx= \left( {{x_1},{x_2},{x_3}} \right)$ be set in $\mathbb{R}^{3}$ with axis $x_{3}$ normal to the plane of isotropy of the infinite homogeneous transversely isotropic medium and the origin placed within the inclusion $\Omega$.
Hence, the five independent elastic parameters of the medium are ${C_{11}},\;{C_{12}},\;{C_{13}},\;{C_{33}}\;\text{and}\;{C_{44}}$.
The positive definiteness of the strain energy implies
\begin{align}\label{1.8}
\begin{split}
&{C_{44}} > 0,\;{C_{11}} - {C_{12}} > 0,\;{C_{11}} + {C_{12}} + {C_{33}} > 0,\;\mathrm{and}\;\left( {\;{C_{11}} + {C_{12}}} \right){C_{33}} > 2{C_{13}}^2.
\end{split}
\end{align}

{In this  part, to prove \eqref{newstrategy} for transversely isotropic materials, we will prove
\begin{align}\label{1.4}
\begin{split}
\exists\; \boldsymbol{\overline{\epsilon}^*}\in \{\boldsymbol{\varepsilon^*}\}\quad s.t.\;\;\forall {\bfC}\in\{\bfC^{\;\text{trans}}\},\;\;\;\{F(\boldsymbol{\overline{\epsilon}^*}, {\bfC})\}=\left\{ E \right\}\;,
\end{split}
\end{align}
where $\{\bfC^{\;\text{trans}}\}$ denotes the set of elastic tensors of transversely isotropic materials.}

{That \eqref{1.4} is true means that \eqref{newstrategy} is true, by choosing $\boldsymbol{\overline{\epsilon}^*}$ that leads to \eqref{1.4} as $\boldsymbol{\overline{\epsilon}^{(1)}}$ , and choosing another arbitrary uniform eigenstrain which is linearly independent of $\boldsymbol{\overline{\epsilon}^*}$ as $\boldsymbol{\overline{\epsilon}^{(2)}}$.
Therefore,  for transversely isotropic materials, Theorem \ref{cubictheorem} can be seen as a corollary of the following theorem embodied by \eqref{1.4}:}
\begin{theorem}\label{transtheorem}
{Let $\Omega\subset \mathbb{R}^{3}$ be a { one-component connected bounded open domain} with a Lipschitz boundary. There exists a uniform eigenstrain $\boldsymbol{\overline{\varepsilon}^*}$, such that $\forall {\bfC}\in\{\bfC^{\;\text{trans}}\}$,
 {$\bfu(\bfx)$ in \eqref{Fourier} is linear inside $\Omega$} for
$(\overline{\boldsymbol{\varepsilon}}^*,{\bfC})$,  if and only if $\Omega$ is of ellipsoidal shape.}
\end{theorem}

Now, we turn to prove Theorem \ref{transtheorem}.
Firstly, we need to specify the eigenstrain to get the displacement field from \eqref{Fourier}.
We choose a particular kind of uniform eigenstrain $\boldsymbol{\overline{\varepsilon}^*}$ that belongs to
the transversely isotropic category, i.e.,
\beas
\begin{split}
 \boldsymbol{\overline{\varepsilon}^*}\in\{\boldsymbol{\overline{\varepsilon}^*}|
\boldsymbol{\overline{\varepsilon}^*}=\overline{\varepsilon}^*_{11}\boldsymbol{\tilde\alpha}+\overline{\varepsilon}^*_{33}\boldsymbol{\tilde\beta},\;\;\;
\;\;\overline{\varepsilon}^*_{11},\overline{\varepsilon}^*_{33}\in R \},
\end{split}
\eeas
where the two tensors
\beqs\label{tildealphabeta}
\boldsymbol{\tilde\alpha}{=} \boldsymbol{{\tilde I}}-\boldsymbol{\tilde\beta}\quad
\mathrm{and}\quad \boldsymbol{\tilde\beta}{=} \boldsymbol{n}\otimes\boldsymbol{n}
 \eeqs
facilitate description of transverse isotropy~\citep{Walpole1981}, with
$\boldsymbol{n}$ denoting the unit vector along the axis of symmetry of the transverse isotropic material.
 Such a uniform eigenstrain can be realized by the uniform eigenstress $\boldsymbol{\overline{\sigma}^*}$ based on the constitutive relation:
\begin{align}\label{1.11}
\begin{split}
\overline{\varepsilon}^*_{11} &=\overline{\varepsilon}^*_{22}= \frac{C_{33}\overline{\sigma}^*_{11}-{{{{C_{13}} }}\overline{\sigma}^*_{33} }}{\left( {\;{C_{11}} + {C_{12}}} \right){C_{33}} - 2{C_{13}}^2},\quad\overline{\varepsilon}^*_{33} = \frac{{( {{C_{11}} + {C_{12}}} )\overline{\sigma}^*_{33} }-2C_{33}\overline{\sigma}^*_{11}}{\left( {\;{C_{11}} + {C_{12}}} \right){C_{33}} - 2{C_{13}}^2},\quad\overline{\varepsilon}^*_{ij} = 0\;\left( {i \ne j} \right).
\end{split}
\end{align}
In this proof, we require that  $\overline{\sigma}^*_{11}\neq 0$ or $\overline{\sigma}^*_{33}\neq 0$ to ensure that at least one of
$\overline{\varepsilon}^*_{11}\ (\overline{\varepsilon}^*_{22})$ and $\overline{\varepsilon}^*_{33}$ is nonzero.

Because the structure of $L_{mn}(\boldsymbol{\xi})$ introduced in \eqref{L} relies on the elastic parameters, we consider the following two cases separately:
\begin{subequations}
\begin{align}
&C_{13}+C_{44}= 0\label{transconstrain1}\\
\quad &C_{13}+C_{44}\neq 0.\label{transconstrain12}
\end{align}
\end{subequations}

For the case \eqref{transconstrain1}, we require $\overline{\sigma}^*_{11}\neq 0$, and then substitution of \eqref{1.11} into \eqref{Fourier} yields
\begin{align}\label{transversedegenerate}
\begin{split}
&u_1(\bfx)=\frac{-\mathrm{i}\;\overline{\sigma}^*_{11}}{(2\pi)^3}\int_{\mathbb{R}^3}\frac{\xi_1}{C_{11}(\xi_1^2+\xi_2^2)+C_{44}\xi_3^2}\int_{\Omega}e^{-\mathrm{i}\boldsymbol{\xi}\cdot( \bfx - \bfy)}d\bfy d\boldsymbol{\xi},\\
&u_2(\bfx)=\frac{-\mathrm{i}\;\overline{\sigma}^*_{11}}{(2\pi)^3}\int_{\mathbb{R}^3}\frac{\xi_2}{C_{11}(\xi_1^2+\xi_2^2)+C_{44}\xi_3^2}\int_{\Omega}e^{-\mathrm{i}\boldsymbol{\xi}\cdot( \bfx - \bfy)}d\bfy d\boldsymbol{\xi},\\
&u_3(\bfx)=\frac{-\mathrm{i}\;\overline{\sigma}^*_{33}}{(2\pi)^3}\int_{\mathbb{R}^3}\frac{\xi_3}{C_{44}(\xi_1^2+\xi_2^2)+C_{33}\xi_3^2}\int_{\Omega}e^{-\mathrm{i}\boldsymbol{\xi}\cdot( \bfx - \bfy)}d\bfy d\boldsymbol{\xi}.\\
\end{split}
\end{align}

By assuming that $\Omega$  { possesses the {  Eshelby uniformity property}} defined in \eqref{4.4}, it can be derived from $\eqref{transversedegenerate}_1$ and $\eqref{transversedegenerate}_2$ that
\begin{align}\label{transverselinear}
\begin{split}
&\frac{\partial}{\partial x_1} U(\bfx) =\mathcal{L}_1(\bfx);\quad\frac{\partial}{\partial x_2} U(\bfx)=\mathcal{L}_2(\bfx),\;\bfx\in\Omega
\end{split}
\end{align}
where $\mathcal{L}_i(\bfx)\;(i=1,2)$ still denote linear scalar functions of $\bfx$; $U(x)$ has the same form as that in \eqref{NewtonianpotentialFourier}; and the boundary conditions concerning $U(x)$ at infinity has the same structure as those in \eqref{laplacianboundary}.

Then by following the same procedure from \eqref{cubiclinear} to \eqref{7.2}, we can derive from \eqref{transverselinear} that
\begin{align}\label{degeneratetransnewtonian}
\begin{split}
{N_{\Omega '}}\left( {\bfx}' \right) =&-\int_{\Omega'}\frac{1}{{4\pi | \bfx'-\bfy' |}}d\bfy'={{t}\left[ {q}\left( {{x_1}',{x_2}',\frac{{{x_3}'}}{{{t}}}} \right) + {\psi}\left( \frac{{{x_3}'}}{{{t}}}\right) \right]},\;\bfx'\in\Omega',
\end{split}
\end{align}
where $t=\sqrt{\frac{C_{44}}{C_{11}}}$ still holds; $q$ also represents a quadratic function; $\psi$ also represents an unknown function; and  ${N_{\Omega'}}\left( \bfx' \right)$ represents the Newtonian potential induced by the inclusion $\Omega'$ that is transformed from the original inclusion $\Omega$ via \eqref{7.2}.
Substituting \eqref{degeneratetransnewtonian} into \eqref{1.21} shows that ${\psi}\left( \frac{{{x_3}'}}{{{t}}}\right)$ can only be a constant, linear or quadratic function of $x_3'$. Accordingly, we conclude that $\Omega $  can only be of ellipsoidal shape, which implies \eqref{1.4} and thus the verification of  Theorem \ref{transtheorem} for the case \eqref{transconstrain1}.

For the case \eqref{transconstrain12}, by requiring  $\overline{\sigma}^*_{11}\neq 0$ and assuming ${\overline{\sigma}^*_{33}}=\gamma\overline{\sigma}^*_{11}$, we can get
\begin{align}\label{transverse0}
\begin{split}
&\begin{bmatrix}L_{1n}(\boldsymbol{\xi})\overline{\sigma}^*_{ns}\xi_s\\L_{2n}(\boldsymbol{\xi})\overline{\sigma} ^*_{ns}\xi_s\\L_{3n}(\boldsymbol{\xi})\overline{\sigma}^*_{ns}\xi_s \end{bmatrix}=\frac{1}{\eta(\boldsymbol{\xi})}\cdot\begin{bmatrix}\xi_1 (\xi_3^2 (C_{33} - C_{13} \gamma) + C_{44} (\xi_1^2 + \xi_2^2 - \xi_3^2 \gamma))\\ \xi_2 (\xi_3^2 (C_{33} - C_{13} \gamma) + C_{44} (\xi_1^2 + \xi_2^2 - \xi_3^2 \gamma))\\ \xi_3 (-C_{13} (\xi_1^2 + \xi_2^2) + C_{11} (\xi_1^2 + \xi_2^2) \gamma - C_{44} (\xi_1^2 + \xi_2^2 - \xi_3^2 \gamma))\end{bmatrix},
\end{split}
\end{align}
with
\beas
\begin{split}
\eta(\boldsymbol{\xi})=&C_{11} (\xi_1^2 + \xi_2^2) (C_{44} (\xi_1^2 + \xi_2^2) + C_{33} \xi_3^2)+\xi_1^2 (-C_{13}^2 (\xi_1^2 + \xi_2^2) - 2 C_{13} C_{44} (\xi_1^2 + \xi_2^2) + C_{33} C_{44} \xi_3^2).
\end{split}
\eeas
We note that if
\begin{align}\label{gamma}
\begin{split}
\gamma=\frac{\overline{\sigma}^*_{33}}{\overline{\sigma}^*_{11}}=\frac{C_{11}C_{33}-C_{33}C_{44}v^2}{(C_{13}+C_{44})C_{11}},
\end{split}
\end{align}
then substitution of \eqref{gamma} into \eqref{transverse0} yields
 \begin{align}\label{transverse}
\begin{bmatrix}L_{1n}(\boldsymbol{\xi})\overline{\sigma}^*_{ns}\xi_s\\L_{2n}(\boldsymbol{\xi})\overline{\sigma}^*_{ns}\xi_s\\L_{3n}(\boldsymbol{\xi})\overline{\sigma}^*_{ns}\xi_s \end{bmatrix}=\frac{\overline{\sigma}^*_{11}}{(\xi_1^2+\xi_2^2+\frac{1}{v^2}\xi_3^2)}\begin{bmatrix}\frac{1}{C_{11}}\xi_1\\ \frac{1}{C_{11}}\xi_2\\ \frac{(C_{11}-C_{44}v^2)}{v^2(C_{13}+C_{44})C_{11}}\xi_3\end{bmatrix},
\end{align}
where $v$ is a constant solved by
 \begin{align}\label{root}
 C_{33}C_{44}v^4-(C_{11}C_{33}+C_{44}^2-(C_{13}+C_{44})^2)v^2+C_{11}C_{44}=0.
 \end{align}
Substitution of \eqref{transverse} back into \eqref{Fourier} yields
\begin{align}\label{1.12}
\begin{split}
u_m(\bfx)= K_{mi}\frac{\partial U(\bfx)}{\partial x_i},
\end{split}
\end{align}
where $U(\bfx)$ is shown in \eqref{NewtonianpotentialFourier} with $t$ replaced by $v$, and
$\bfK$ is a second-order tensor satisfying
\begin{align}\label{1.12b}
\begin{split}
&K_{11}=K_{22} =\frac{\overline{\sigma}^*_{11}}{C_{11}};\quad K_{33}=\frac{(C_{11}-C_{44}v^2)\overline{\sigma}^*_{11}}{v^2(C_{13}+C_{44})C_{11}};\quad K_{mn}=0,\;\;\;\text{for}\;\;m\neq n.
\end{split}
\end{align}

Since $U(\bfx)$ satisfies the boundary conditions of the same structure as those in \eqref{laplacianboundary} along all three directions at infinity, by using the same method as that for addressing Lemma \ref{lemma1}, $U(\bfx)$ must be expressed as
\beqs\label{Ufortrans}
U(\bfx)=-\int_\Omega \frac{v}{4\pi{{R_v}\left(\bfx-\bfy\right)}}d\bfy,
\eeqs
where $R_v(\bfx-\bfy){:=}\sqrt{(x_1-y_1)^2+(x_2-y_2)^2+v^2(x_3-y_3)^2}$. Then
 substituting \eqref{Ufortrans} back into \eqref{1.12} leads to
\begin{align}\label{finalsolution1}
\begin{split}
\bfu\left( \bfx \right)=&-\bfK \cdot {\boldsymbol{\nabla}_\bfx} \int_\Omega \frac{v}{4\pi{{R_v}\left(\bfx-\bfy\right)}}d\bfy.
\end{split}
\end{align}

Since $\Omega$ is supposed to {possess the {  Eshelby uniformity property}} defined in \eqref{4.4}, then via the transformation in \eqref{cotrans} with $t$ replaced by $v$, it follows from \eqref{finalsolution1}  that
\begin{align}\label{77777.5c}
\begin{split}
{N_{\Omega'}}\left( \bfx' \right)=-\int_{\Omega'} \frac{1}{{4\pi | \bfx'-\bfy' |}}d\bfy'=\text{quadratic},\;\bfx'\in\Omega',
\end{split}
\end{align}
with
\begin{align}\label{7.2new}
\begin{split}
\Omega' {:=} \left\{ {\bfy'\left| {\left( {\begin{array}{*{20}{c}}
  1&0&0 \\
  0&1&0 \\
  0&0&{\frac{1}{{{v}}}}
\end{array}} \right)\cdot\bfy'} \right. \in \Omega } \right\}.
\end{split}
\end{align}

Comparison of \eqref{77777.5c} with  Theorem \ref{TheoremNP} means that $\Omega'$  can only be of ellipsoidal shape. Based on \eqref{7.2new}, we conclude that $\Omega $  can only be ellipsoidal; thus \eqref{1.4} is verified so that tthe substantiation of  Theorem \ref{transtheorem} for the case \eqref{transconstrain12} is achieved, which together with the proof for the case \eqref{transconstrain1} constitutes the complete proof of Theorem \ref{transtheorem} and thus leads to the proof Theorem \ref{cubictheorem} for transversely isotropic materials.

Note that for the case \eqref{transconstrain12}, there is an alternative way to show \eqref{finalsolution1} and thus \eqref{77777.5c}
by using the explicit Green function { in the real space}, which is derived by 
\cite{Pan1976} for a transversely isotropic material and will be given in Appendix A. However, the method in {Appendix A} is not capable of dealing with a generally anisotropic material whose explicit expression of the Green function is not available, while the above method is applicable to any anisotropic material.
In addition, as a comparison, a theorem which is also subjected to stronger constraints than the generalized weak version of the Eshelby conjecture but concerned with the material parameters of the transversely isotropic material is presented and proved in {Appendix B} to reveal the impact of the material symmetry.

 It is worth mentioning that~\cite{Liu2008} indicated that through linear transformations of the coordinates and the displacements, the solutions to the Eshelby conjectures for the isotropic elastic tensor can be extended to cases of the elastic tensors that are transformed from the isotropic elastic tensor. Here, we find that the coordinate transformations \eqref{cotrans} and the inverse of $\bfK$ satisfying \eqref{1.12b}
  can serve as the linear transformations mentioned by \cite{Liu2008}, which can transform the transversely isotropic tensor satisfying the condition \eqref{transconstrain12} into
 the special structure given by~\cite{Liu2008} (equation (4.2) in Section 4 of ~\cite{Liu2008}). Thus, in this sense,
 the method of~\cite{Liu2008} can be utilized to prove the generalized weak version of the Eshelby conjecture for transverse isotropy.


\subsection*{{(3) Orthotropic material}}

For an orthotropic material, let the Cartesian coordinate system $\bfx= \left( {{x_1},{x_2},{x_3}} \right)$ be set in $\mathbb{R}^{3}$ with the $x_2-$axis and $x_3-$axis coinciding with the 2-fold axes of rotational symmetry of the infinite homogeneous orthotropic medium. The origin is placed within the inclusion $\Omega$. Given this, the nine independent elastic parameters of the medium are ${C_{11}}$, ${C_{22}}$, ${C_{33}}$, ${C_{44}}$, ${C_{55}}$, ${C_{66}}$, ${C_{12}}$, ${C_{13}}$ and ${C_{23}}$.
The positive definiteness of the strain energy requires
\begin{align}\label{122.1}
\begin{split}
&{C_{11}},{C_{22}},{C_{33}},{C_{44}},{C_{55}},{C_{66}}> 0,\;{C_{11}}{C_{22}}>{C_{12}^2},\;C_{11}C_{22}C_{33}+2C_{12}C_{23}C_{13}>C_{11}C_{23}^2+C_{22}C_{13}^2+C_{33}C_{12}^2.
\end{split}
\end{align}

We still choose the particular eigenstrains $\boldsymbol{\overline{\varepsilon}^{(1)}},\boldsymbol{\overline{\varepsilon}^{(2)}}$ introduced in  \eqref{cubiceigenstrain}, and thus the corresponding eigenstresses $\boldsymbol{\overline{\sigma}^{(1)}},\boldsymbol{\overline{\sigma}^{(2)}}$ are given as
\beqs\label{othoeigenstress}
\begin{split}
\boldsymbol{\overline{\sigma}^{(1)}}=\overline{\varepsilon}^{(1)}_{33}\begin{bmatrix} C_{12}&0&0\\0&C_{23}&0\\0&0&C_{33}\end{bmatrix},\;\boldsymbol{\overline{\sigma}^{(2)}}=2\overline{\varepsilon}^{(2)}_{23}\begin{bmatrix} 0&0&0\\0&0&C_{44}\\0&C_{44}&0\end{bmatrix}.
\end{split}
\eeqs

Then we consider the stiffness ${\bfC}$ satisfying the conditions
\begin{align}\label{orthoconstrain}
\begin{split}
&C_{33}\neq C_{44} \neq C_{55},\;C_{12}+C_{66}=0,\;C_{13}+C_{55}=0,\;C_{23}+C_{44}=0.
\end{split}
\end{align}
The inequality in  \eqref{orthoconstrain} guarantees that such an orthotropic material does not degenerate into a cubic material.

By substituting \eqref{orthoconstrain} and the expressions of $\boldsymbol{\overline{\sigma}^{(1)}}$ and $\boldsymbol{\overline{\sigma}^{(2)}}$ defined in \eqref{othoeigenstress} into \eqref{Fourier}, we obtain
\begin{align}\label{122.2}
\begin{split}
&u_1^{(1)}(\bfx)=\frac{-\mathrm{i}\;\overline{\varepsilon}^{(1)}_{33}}{(2\pi)^3}\int_{\mathbb{R}^3}\frac{C_{13}\xi_1}{C_{11} \xi_1^2 + C_{66} \xi_2^2 +C_{55}  \xi_3^2}\int_{\Omega}e^{-\mathrm{i}\boldsymbol{\xi}\cdot( \bfx - \bfy)}d\bfy d\boldsymbol{\xi},\\
&u_2^{(1)}(\bfx)=\frac{-\mathrm{i}\;\overline{\varepsilon}^{(1)}_{33}}{(2\pi)^3}\int_{\mathbb{R}^3}\frac{C_{23}\xi_2}{C_{66} \xi_1^2 + C_{22} \xi_2^2 +C_{44}  \xi_3^2}\int_{\Omega}e^{-\mathrm{i}\boldsymbol{\xi}\cdot( \bfx - \bfy)}d\bfy d\boldsymbol{\xi},\\
&u_3^{(1)}(\bfx)=\frac{-\mathrm{i}\;\overline{\varepsilon}^{(1)}_{33}}{(2\pi)^3}\int_{\mathbb{R}^3}\frac{C_{33}\xi_3}{C_{55} \xi_1^2 + C_{44} \xi_2^2 +C_{33}  \xi_3^2}\int_{\Omega}e^{-\mathrm{i}\boldsymbol{\xi}\cdot( \bfx - \bfy)}d\bfy d\boldsymbol{\xi}.\\
\end{split}
\end{align}
and
\begin{align}\label{122.3}
\begin{split}
&u_1^{(2)}(\bfx)=0,\\
&u_2^{(2)}(\bfx)=\frac{-\mathrm{i}\;\overline{\varepsilon}^{(2)}_{23}}{(2\pi)^3}\int_{\mathbb{R}^3}\frac{2C_{44}\xi_3}{C_{66} \xi_1^2 + C_{22} \xi_2^2 +C_{44}  \xi_3^2}\int_{\Omega}e^{-\mathrm{i}\boldsymbol{\xi}\cdot( \bfx - \bfy)}d\bfy d\boldsymbol{\xi},\\
&u_3^{(2)}(\bfx)=\frac{-\mathrm{i}\;\overline{\varepsilon}^{(2)}_{23}}{(2\pi)^3}\int_{\mathbb{R}^3}\frac{2C_{44}\xi_2}{C_{55} \xi_1^2 + C_{44} \xi_2^2 +C_{33}  \xi_3^2}\int_{\Omega}e^{-\mathrm{i}\boldsymbol{\xi}\cdot( \bfx - \bfy)}d\bfy d\boldsymbol{\xi}.\\
\end{split}
\end{align}
Likewise, by requiring the linear form of the right-hand side of $\eqref{122.2}_3$ and $\eqref{122.3}_3$ with respect to $\bfx$ inside $\Omega$, we note that

\begin{align}\label{122.4}
\begin{split}
&\frac{\partial}{\partial x_2} V(\bfx)=\mathcal{L}_1(\bfx);\;\frac{\partial}{\partial x_3}  V(\bfx)=\mathcal{L}_2(\bfx),\;\bfx\in\Omega,
\end{split}
\end{align}
where $\mathcal{L}_i(\bfx)\;(i=1,2)$ denote linear scalar functions of $\bfx$, and
\begin{align}\label{122.5}
\begin{split}
V(\bfx){:=}\frac{1}{(2\pi)^3}\int_{\mathbb{R}^3}\frac{1}{ \frac{C_{55}}{C_{33}}\xi_1^2 + \frac{C_{44}}{C_{33}} \xi_2^2 + \xi_3^2}\int_{\Omega}e^{-\mathrm{i}\boldsymbol{\xi}\cdot( \bfx - \bfy)}d\bfy d\boldsymbol{\xi}
\end{split}
\end{align}
is the solution to
\begin{align}\label{122.6}
\begin{split}
\Delta^s V(\bfx)=\chi_\Omega(\bfx)\;\;\text{on}\;\;\mathbb{R}^3,
\end{split}
\end{align}
with
\beas
\begin{split}
\Delta^s{:=}\frac{\partial^2}{s_1^2\partial x_1^2}+\frac{\partial^2}{s_2^2\partial x_2^2}+\frac{\partial^2}{\partial x_3^2}.
\end{split}
\eeas
Here $s_1{:=}\sqrt{\frac{C_{33}}{C_{55}}}, s_2{:=}\sqrt{\frac{C_{33}}{C_{44}}}$, and $V(\bfx)$ must satisfy the boundary conditions
\begin{align}\label{boundaryconditionortho}
\begin{split}
&\left.\frac{\partial}{\partial x_2}V(\bfx)\right|_{|\bfx|\rightarrow +\infty}=\left.\frac{C_{33}}{C_{44}\overline{\varepsilon}_{23}^{(2)}}u_3^{(2)}(\bfx)\right|_{|\bfx|\rightarrow +\infty},\quad\left.\frac{\partial}{\partial x_3}V(\bfx)\right|_{|\bfx|\rightarrow +\infty}=\left.\frac{1}{\overline{\varepsilon}_{33}^{(1)}}u_3^{(1)}(\bfx)\right|_{|\bfx|\rightarrow +\infty}.
\end{split}
\end{align}

{By using the same mathod as that for addressing Lemma \ref{lemma1}, we can prove that the solution of \eqref{122.6} under the condition \eqref{boundaryconditionortho} must be expressed as}
\begin{align}\label{122.8}
\begin{split}
V(\bfx)=-s_1s_2\int_\Omega \frac{1}{4\pi{{R_s}\left(\bfx-\bfy\right)}}d\bfy+\hat{\mathcal{L}}(x_1),
\end{split}
\end{align}
where $R_s(\bfx-\bfy){:=}\sqrt{s_1^2(x_1-y_1)^2+s_2^2(x_2-y_2)^2+(x_3-y_3)^2}$, and $\hat{\mathcal{L}}(x_1)$ denotes a linear scalar function of $x_1$.

Then by substituting \eqref{122.8} into \eqref{122.4}, we obtain
\begin{align}
\begin{split}
&\frac{\partial}{\partial x_2} \int_\Omega \frac{1}{{{R_s}\left(\bfx-\bfy\right)}}d\bfy=\tilde{\mathcal{L}}_1(\bfx);\quad \frac{\partial}{\partial x_3} \int_\Omega\frac{1}{{{R_s}\left(\bfx-\bfy\right)}}d\bfy=\tilde{\mathcal{L}}_2(\bfx),\;\bfx\in\Omega
\end{split}
\end{align}
where $\tilde{\mathcal{L}}_i(\bfx)\;(i=1,2)$ denote linear scalar functions of $\bfx$,
which lead to
\begin{align}\label{orthosolution}
\begin{split}
-\int_\Omega \frac{1}{4\pi{{R_s}\left(\bfx-\bfy\right)}}d\bfy=q(\bfx)+\psi(x_1),\;\bfx\in\Omega,
\end{split}
\end{align}
where $q(\bfx)$ also denotes a quadratic function of $\bfx$, and $\psi(x_1)$ denotes an unknown function.

By introducing new transformations
\begin{align}\label{newtransformation}
\bfx' {:=} \left( {\begin{array}{*{20}{c}}
  s_1&0&0 \\
  0&s_2&0 \\
  0&0&1
\end{array}} \right) \cdot \bfx,\ \ \ \  \
\bfy' {:=} \left( {\begin{array}{*{20}{c}}
  s_1&0&0 \\
  0&s_2&0 \\
  0&0&1
\end{array}} \right) \cdot \bfy,
\end{align}
and then substituting \eqref{newtransformation} into \eqref{orthosolution}, we obtain
\begin{align}\label{Newtonianforortho}
\begin{split}
{N_{\Omega'}}\left( \bfx' \right)&=-\int_{\Omega '} \frac{1}{{4\pi | \bfx'-\bfy' |}}d\bfy'=s_1s_2\left[q\left(\frac{x_1'}{s_1},\frac{x_2'}{s_2},x_3'\right)+\psi\left(\frac{x_1'}{s_1}\right)\right],\;\bfx'\in\Omega',
\end{split}
\end{align}
with
\begin{align}\label{newtansomega}
\begin{split}
\Omega'{:=} \left\{ {\bfy'\left| {\left( {\begin{array}{*{20}{c}}
  \frac{1}{s_1}&0&0 \\
  0&\frac{1}{s_2}&0 \\
  0&0&1
\end{array}} \right)\cdot\bfy'}\right. \in \Omega } \right\}.
\end{split}
\end{align}
  ${N_{\Omega'}}\left( \bfx' \right)$ denotes the Newtonian potential induced by the inclusion $\Omega'$ that is transformed from the original inclusion $\Omega$ via \eqref{newtansomega}.

  Then substitution of \eqref{Newtonianforortho} into \eqref{1.21} yields the constant, quadratic or linear form of $\psi\left(\frac{x_1'}{s_1}\right)$, which implies that ${N_{\Omega'}}\left( \bfx' \right)$  in \eqref{Newtonianforortho} is quadratic within $\Omega'$. Then, owing to  Theorem \ref{TheoremNP},  $\Omega'$ can only be of ellipsoidal shape, and so is $\Omega$. Thus we have proved \eqref{newstrategy}, which leads to the proof of  Theorem \ref{cubictheorem} for orthotropic materials.

\subsection*{{(4) Monoclinic material}}

Under the condition that a monoclinic material possesses one 2-fold axis of rotational symmetry, let the Cartesian coordinate system $\bfx= \left( {{x_1},{x_2},{x_3}} \right)$ be set in $\mathbb{R}^{3}$ with the $x_3$-axis coinciding with the 2-fold axis of rotational symmetry of the infinite homogeneous monoclinic material; the origin is placed within the inclusion $\Omega$.
Then the thirteen independent elastic parameters of the medium are ${C_{11}}$, ${C_{22}}$, ${C_{33}}$, ${C_{44}}$, ${C_{55}}$, ${C_{66}}$, ${C_{12}}$, ${C_{13}}$, ${C_{23}}$, ${C_{16}}$,${C_{26}}$, ${C_{36}}$ and ${C_{45}}$.
The positive definiteness of the strain energy requires
\begin{align}\label{122.9}
\begin{split}
&{C_{11}},{C_{22}},{C_{33}},{C_{44}},{C_{55}},{C_{66}}> 0,\;{C_{11}}{C_{22}}>{C_{12}^2},\\
\; \;&C_{44} C_{55}> C_{45}^2,\;C_{11}C_{22}C_{33}+2C_{12}C_{23}C_{13}>C_{11}C_{23}^2+C_{22}C_{13}^2+C_{33}C_{12}^2.
\end{split}
\end{align}

In this case, we choose $\boldsymbol{\overline{\varepsilon}^{(1)}}$ and $\boldsymbol{\overline{\varepsilon}^{(2)}}$ which results in
\begin{align}\label{monoclinicstress}
\begin{split}
\boldsymbol{\overline{\sigma}^{(1)}}=\begin{bmatrix} 0&0&0\\0&0&0\\0&0&\overline{\sigma}^{(1)}_{33}\end{bmatrix},\;\boldsymbol{\overline{\sigma}^{(2)}}=\begin{bmatrix} 0&0&0\\0&0&\overline{\sigma}^{(2)}_{23}\\0&\overline{\sigma}^{(2)}_{23}&0\end{bmatrix}.
\end{split}
\end{align}

Then we impose some constraints on ${\bfC}$
\begin{align}\label{monoclinicconstrain}
\begin{split}
&C_{16}\neq 0,\;C_{36}=0,\;C_{45}=0,\;C_{13}+C_{55}=0,\;C_{23}+C_{44}=0.
\end{split}
\end{align}

The  conditions  in \eqref{monoclinicconstrain} guarantees that such a monoclinic material will not degenerate into an orthotropic material.
By substituting \eqref{monoclinicconstrain} and $\boldsymbol{\overline{\sigma}^{(1)}}$ and $\boldsymbol{\overline{\sigma}^{(2)}}$ in \eqref{monoclinicstress} into \eqref{Fourier}, we obtain
\begin{align}\label{122.10}
\begin{split}
&u_1^{(1)}(\bfx)=u_2^{(1)}(\bfx)=0,\quad u_3^{(1)}(\bfx)=\frac{-\mathrm{i}\overline{\sigma}^{(1)}_{33}}{(2\pi)^3}\int_{\mathbb{R}^3}\frac{\xi_3}{C_{55} \xi_1^2 + C_{44} \xi_2^2 +C_{33}  \xi_3^2}\int_{\Omega}e^{-\mathrm{i}\boldsymbol{\xi}\cdot( \bfx - \bfy)}d\bfy d\boldsymbol{\xi}.\\
\end{split}
\end{align}
and
\begin{align}\label{122.11}
\begin{split}
&u_1^{(2)}(\bfx)=\frac{-\mathrm{i}\overline{\sigma}^{(2)}_{23}}{(2\pi)^3}\int_{\mathbb{R}^3}\frac{M_1(\boldsymbol{\xi})}{M_3(\boldsymbol{\xi})}\int_{\Omega}e^{-\mathrm{i}\boldsymbol{\xi}\cdot( \bfx - \bfy)}d\bfy d\boldsymbol{\xi},\\
&u_2^{(2)}(\bfx)=\frac{-\mathrm{i}\overline{\sigma}^{(2)}_{23}}{(2\pi)^3}\int_{\mathbb{R}^3}\frac{M_2(\boldsymbol{\xi})}{M_3(\boldsymbol{\xi})}\int_{\Omega}e^{-\mathrm{i}\boldsymbol{\xi}\cdot( \bfx - \bfy)}d\bfy d\boldsymbol{\xi},\\
&u_3^{(2)}(\bfx)=\frac{-\mathrm{i}\overline{\sigma}^{(2)}_{23}}{(2\pi)^3}\int_{\mathbb{R}^3}\frac{\xi_2}{C_{55} \xi_1^2 + C_{44} \xi_2^2 +C_{33}  \xi_3^2}\int_{\Omega}e^{-\mathrm{i}\boldsymbol{\xi}\cdot( \bfx - \bfy)}d\bfy d\boldsymbol{\xi},\\
\end{split}
\end{align}
where
\beas
\begin{split}
M_1(\boldsymbol{\xi}){:=}&(C_{16} \xi_1^2 + (C_{12} + C_{66}) \xi_1 \xi_2 + C_{26} \xi_2^2) \xi_3,\\
M_2(\boldsymbol{\xi}){:=}& (C_{11} \xi_1^2 + 2 C_{16} \xi_1 \xi_2 + C_{66} \xi_2^2 + C_{55} \xi_3^2)\xi_3\\
M_3(\boldsymbol{\xi}){:=} &-(C_{16} \xi_1^2 + (C_{12} + C_{66}) \xi_1 \xi_2 + C_{26} \xi_2^2)^2 \\
&+ (C_{66} \xi_1^2 +
    2 C_{26} \xi_1 \xi_2 + C_{22} \xi_2^2 + C_{44} \xi_3^2)  \cdot(C_{11} \xi_1^2 + 2 C_{16} \xi_1 \xi_2 +
    C_{66} \xi_2^2 + C_{55} \xi_3^2).
\end{split}
\eeas

Then through the same procedure from \eqref{122.2} to \eqref{Newtonianforortho}, it can be derived from $\eqref{122.10}_3$ and $\eqref{122.11}_3$ that \eqref{Newtonianforortho} still holds, whose right-hand side can be proved to be constant, linear or quadratic  for $\bfx'\in\Omega'$ by substituting \eqref{Newtonianforortho} into \eqref{1.21}. Owing to  Theorem \ref{TheoremNP}, we conclude that $\Omega'$ can only be of ellipsoidal shape, and thus $\Omega$ must be ellipsoidal, which leads to the substantiation of \eqref{newstrategy}, and thus the proof of  Theorem \ref{cubictheorem} for monoclinic materials.

Up to now, we have
proved that in these three-dimensional
anisotropic media that possess cubic, transversely isotropic, orthotropic, and monoclinic symmetries, only an ellipsoid  can transform {\it all} uniform eigenstrains into uniform elastic strains. However, this conclusion does not exclude the possibility that there may exist some non-ellipsoidal inclusions that can transform {\it some} uniform
eigenstrains (not {\it all} uniform eigenstrains) into uniform elastic strains in these anisotropic media; more generally,
whether there exist some non-ellipsoidal inclusions that can transform {\it some} polynomial
eigenstrains (not {\it all} polynomial eigenstrains) into polynomial elastic strains of the same degree in these anisotropic media is not known.
We shall explore these issues in the next sections.

\section{Counter-examples to the generalized strong version of the Eshelby conjecture for quadratic eigenstrains}

It is straightforward to derive from \eqref{Fourier} that
\beqs\label{Fourierstrain}
\begin{split}
\varepsilon_{ij}(\bfx)&=\frac{1}{2}(\frac{\partial u_j}{\partial x_i}+\frac{\partial u_i}{\partial x_j})=\frac{-1}{(2\pi)^3}\int_{\mathbb{R}^3}\frac{1}{2}\left(L_{in}(\boldsymbol{\xi})\xi_j+L_{jn}(\boldsymbol{\xi})\xi_i\right)C_{nspq}\varepsilon^*_{pq}(\bfy)\xi_s\int_{\Omega}e^{-\mathrm{i}\boldsymbol{\xi}\cdot( \bfx-\bfy)}d\bfy d\boldsymbol{\xi}.
\end{split}
\eeqs

Note that the Newtonian potential $N_\Omega[\rho]$ induced by $\Omega$ with mass density $\rho$ is the solution to
\beqs\label{Newtonianroot}
\Delta N_\Omega[\rho]=\chi_\Omega\rho,
\eeqs
 whose Hessian matrix $H_{ij}(N_\Omega[\rho])$, by the Fourier analysis, can be expressed as
\beqs\label{Netownianfourier}
H_{ij}(N_\Omega[\rho])=\frac{\partial^2 N_\Omega[\rho]}{\partial x_i\partial x_j} =\frac{1}{(2\pi)^3}\int_{\mathbb{R}^3}\frac{\xi_i\xi_j}{\xi_1^2+\xi_2^2+\xi_3^2}\int_{\Omega}\rho(\bfy)e^{-\mathrm{i}\boldsymbol{\xi}\cdot( \bfx-\bfy)}d\bfy d\boldsymbol{\xi}.
\eeqs

Previously, 
\cite{Liu2008} pointed out that there is a correlation between $H_{ij}$ in \eqref{Netownianfourier} and $\varepsilon_{ij}$ in \eqref{Fourierstrain}  when $C_{nspq}$ is isotropic, and when $\varepsilon^*_{pq}(\bfy)$, which is appropriately chosen, and $\rho(\bfy)$ are both constant functions of $\bfy$.
In this work, we find that there still exists a relationship between  $H_{ij}$ in \eqref{Netownianfourier} and $\varepsilon_{ij}$ in \eqref{Fourierstrain} even when $C_{nspq}$ is anisotropic, and $\varepsilon^*_{pq}(\bfy)$ and $\rho(\bfy)$ are in some particular polynomial forms.

{Then we present the following theorem, which provides counter-examples to the generalized strong version of the high-order Eshelby conjecture when the eigenstrains are in  quadratic forms in both anisotropic media 
and the isotropic medium. The cases when the eigenstrains are in polynomial forms of even degrees will be studied in the next section.

\begin{theorem}\label{cubicstrong}
 There exists a non-ellipsoidal and { one-component connected bounded open domain} $\Omega\subset \mathbb{R}^{3}$ with a Lipschitz boundary that   possesses Eshelby's polynomial conservation property for a combination $(\boldsymbol{\varepsilon}^*,\bfC)$, where  $\boldsymbol{\varepsilon}^*$ is a quadratic eigenstrain, and
$\bfC$ is the elastic tensor of an isotropic material, or an anisotropic material with one of the cubic, transversely isotropic, orthotropic, and monoclinic symmetries.
 \end{theorem}
 }

\subsection{{Proof of Theorem \ref{cubicstrong}}}


\subsection*{{(1) Cubic material}}
We consider a quadratic eigenstrain such that the corresponding eigenstress $\sigma^*_{ij}=C_{ijmn}\varepsilon^*_{mn}$ takes the form
\beqs\label{eigenstresscubic}
\sigma^*_{ij}(\bfx)=\rho(\bfx)P_{ij},
\eeqs
where
\beqs\label{rhoquadratic}
\rho(\bfx):=-\sum_{k=1}^3 c_kx_k^2,
\eeqs
with $c_k\;(k=1,2,3)$
being real constants,
 and $P_{ij}$ denotes the uniaxial stress state. Here, we just consider the case where
\beqs\label{Pcubic}
\bfP{:=}\begin{bmatrix}0& 0& 0\\0& 0& 0\\0&0&1
\end{bmatrix},
\eeqs
and the other two cases can be analysed in the same way.

 Under the condition \eqref{cubiconstrain}, by substituting \eqref{eigenstresscubic} along with \eqref{rhoquadratic} and \eqref{Pcubic}  into \eqref{Fourierstrain}, we can obtain
\beqs\label{cubicstrain}
\begin{split}
&\varepsilon_{11}=0,\;\;\varepsilon_{22}=0,\;\;\varepsilon_{12}=0,\\
&\varepsilon_{13}(\bfx)=-\frac{1}{2}\frac{1}{(2\pi)^3}\int_{\mathbb{R}^3}\frac{\xi_1\xi_3}{C_{11} \xi_3^2 + C_{44} (\xi_1^2 + \xi_2^2)}\int_{\Omega}\rho(\bfy)e^{-\mathrm{i}\boldsymbol{\xi}\cdot( \bfx-\bfy)}d\bfy d\boldsymbol{\xi},\\
&\varepsilon_{23}(\bfx)=-\frac{1}{2}\frac{1}{(2\pi)^3}\int_{\mathbb{R}^3}\frac{\xi_2\xi_3}{C_{11} \xi_3^2 + C_{44} (\xi_1^2 + \xi_2^2)}\int_{\Omega}\rho(\bfy)e^{-\mathrm{i}\boldsymbol{\xi}\cdot( \bfx-\bfy)}d\bfy d\boldsymbol{\xi},\\
&\varepsilon_{33}(\bfx)=-\frac{1}{(2\pi)^3}\int_{\mathbb{R}^3}\frac{\xi^2_3}{C_{11} \xi_3^2 + C_{44} (\xi_1^2 + \xi_2^2)}\int_{\Omega}\rho(\bfy)e^{-\mathrm{i}\boldsymbol{\xi}\cdot(\bfx-\bfy)}d\bfy d\boldsymbol{\xi}.
\end{split}
\eeqs

 By  transformations of coordinates
 \begin{align}\label{cotransp2}
\bfx' {:=}\tilde{\bfQ}\cdot\bfx,\ \ \ \  \
\bfy' {:=} \tilde{\bfQ}\cdot\bfy,\ \ \ \  \
\boldsymbol{\xi}' {:=}{\tilde{\bfQ}}^{-1} \cdot\boldsymbol{\xi}
\end{align}
with
\beqs\label{Qtrans}
\tilde{\bfQ}{:=}\begin{bmatrix}
  1&0&0 \\
  0&1&0 \\
  0&0&s
\end{bmatrix}
\eeqs
and then substitution of \eqref{cotransp2} into \eqref{cubicstrain} with $s{:=}\sqrt{\frac{C_{44}}{C_{11}}}$, we obtain
\beqs\label{cubicstraintrans}
\begin{split}
&\varepsilon_{11}=0,\;\;\varepsilon_{22}=0,\;\;\varepsilon_{12}=0,\\
&\varepsilon_{13}(\bfx')=-\frac{1}{2\sqrt{C_{11}C_{44}}}\frac{1}{(2\pi)^3}\int_{\mathbb{R}^3}\frac{\xi'_1\xi'_3}{  {\xi_1'}^2 + {\xi_2'}^2+{\xi_3'}^2 }\int_{\Omega'}\rho(\bfy')e^{-\mathrm{i}\boldsymbol{\xi}'\cdot( \bfx'-\bfy')}d\bfy' d\boldsymbol{\xi}',\\
&\varepsilon_{23}(\bfx')=-\frac{1}{2\sqrt{C_{11}C_{44}}}\frac{1}{(2\pi)^3}\int_{\mathbb{R}^3}\frac{\xi'_2\xi'_3}{{\xi_1'}^2 + {\xi_2'}^2+{\xi_3'}^2 }\int_{\Omega'}\rho(\bfy')e^{-\mathrm{i}\boldsymbol{\xi}'\cdot( \bfx'-\bfy')}d\bfy' d\boldsymbol{\xi}',\\
&\varepsilon_{33}(\bfx')=-\frac{1}{\sqrt{C_{11}}}\frac{1}{(2\pi)^3}\int_{\mathbb{R}^3}\frac{{\xi'_3}^2}{{\xi_1'}^2 + {\xi_2'}^2+{\xi_3'}^2 }\int_{\Omega'}\rho(\bfy')e^{-\mathrm{i}\boldsymbol{\xi}'\cdot(\bfx'-\bfy')}d\bfy' d\boldsymbol{\xi}',
\end{split}
\eeqs
with
\begin{align}\label{Omega}
\begin{split}
\Omega' {:=} \left\{\bfy'\left| {{\tilde{\bfQ}}^{-1}\cdot\bfy'} \right. \in \Omega \right\}.
\end{split}
\end{align}

Through comparison of \eqref{Netownianfourier} with \eqref{cubicstraintrans}, we see that
\beqs\label{keyequation}
{\varepsilon}_{ij}(\bfx')=\frac{1}{2C_{44}}\left(\tilde{Q}_{il}P_{jm}\tilde{Q}_{mq}\frac{\partial^2 N_{\Omega'}[\rho](\bfx')}{\partial x'_l \partial x'_q}+\tilde{Q}_{jm}P_{il}\tilde{Q}_{ls}\frac{\partial^2 N_{\Omega'}[\rho](\bfx')}{\partial x'_m \partial x'_s}\right),
\eeqs
where

\beqs
N_{\Omega'}[\rho](\bfx')=-\frac{1}{(2\pi)^3}\int_{\mathbb{R}^3}\frac{1}{{\xi_1'}^2 + {\xi_2'}^2+{\xi_3'}^2 }\int_{\Omega'}\rho(\bfy')e^{-\mathrm{i}\boldsymbol{\xi}'\cdot(\bfx'-\bfy')}d\bfy' d\boldsymbol{\xi}'
\eeqs
is the Newtonian potential induced by $\Omega'$ with the mass density $\rho$.

 We note that if we can find a non-ellipsoidal $\Omega'$ that leads to $N_{\Omega'}[\rho](\bfx')$ being a quartic polynomial, then due to \eqref{keyequation}, the strain field $\boldsymbol{\varepsilon}(\bfx')$ induced by $\Omega$ will be a quadratic function of $\bfx'$ and thus $\bfx$ via the transformation \eqref{cotransp2}.
 Hence $\Omega$  must be the non-ellipsoidal inclusion that possesses Eshelby's polynomial conservation property. To this end, we prove the following lemma.

\begin{lemma}\label{keylemma}
For  $\rho(\bfx')=-\sum_{k=1}^3 {x'_k}^2$, there exists a non-ellipsoidal and { one-component connected bounded open domain} $\Omega'\subset \mathbb{R}^{3}$ with a Lipschitz boundary that leads to
\beqs\label{construct1}
N_{\Omega'}[\rho](\bfx'){:=}-\int_{\Omega'} \frac{\rho(\bfy')}{4\pi|\bfx'-\bfy'|} d\bfy'=\varphi(\bfx'), \;\;\bfx'\in \Omega',
\eeqs
where
\beqs\label{phi001}
\varphi(\bfx'){:=}C-\frac{1}{12}\left({x'_1}^4+{x'_2}^4+{x'_3}^4\right)
\eeqs
 with $C$ a positive real constant.
\end{lemma}

It is straightforward to verify that the right-hand side of \eqref{construct1} satisfies the definition \eqref{Newtonianroot} of the Newtonian potential.
The proof of Lemma \ref{keylemma} via a variational method is given in Appendix C.
 Then, via substitution of \eqref{construct1} into \eqref{keyequation}, it is proved that a non-ellipsoidal $\Omega'$ that yields \eqref{construct1} can lead to the quadratic strain field induced by $\Omega$.
 The shape of a counter-example non-ellipsoidal inclusion in this case is shown as $\Omega^{(1)}$ in Figure~\ref{counterexample1} in Appendix D.

We note that based on \eqref{keyequation}, the quartic form of $N_{\Omega'}[\rho]$ is not a necessity for the existence of a non-ellipsoidal $\Omega'$ and thus a non-ellipsoidal $\Omega$ that leads to quadratic strain fields. For example, we assume that there exists a { one-component connected bounded open domain} $\Omega'$ and consider the case where
 $N_{\Omega'}[\rho]$ consists of quartic terms and a non-polynomial term as follows:
\beqs\label{NewformnNewtonian}
N_{\Omega'}[\rho](\bfx')=\varphi(\bfx')+\omega(x_1',x_2'),\;\bfx'\in\Omega'
\eeqs
where
\beqs\label{omegaform}
\omega(x_1',x_2'){:=}-\beta\log \frac{(x_1'-12\sqrt{C})^2+(x_2'-12\sqrt{C})^2}{36C}
\eeqs
 with $\beta$ a positive real constant. It is easily seen that $\Delta_{x'}\omega=0$ so that $\omega$ is harmonic,  which guarantees that the right-hand side of \eqref{NewformnNewtonian} satisfies the definition \eqref{Newtonianroot} of the Newtonian potential.
Then substitution of \eqref{NewformnNewtonian} along with \eqref{omegaform} into \eqref{keyequation} yields the quadratic strain field.  In this case, it is straightforward that $\Omega'$ cannot be ellipsoidal due to the purely quartic polynomial form of the Newtonian potential induced by ellipsoids. The existence of such a non-ellipsoidal $\Omega'$
that yields \eqref{NewformnNewtonian} is proved in Appendix E, via the same method as that in the proof of Lemma \ref{keylemma}.  Moreover, the existence of an $\Omega'$ that yields \eqref{NewformnNewtonian} also inspires us to construct more counter-examples to deal with the high-order Eshelby conjecture for polynomial eigenstrains of any even degree in Section 5.

\subsection*{{ (2) Transversely isotropic material}}


 For the case \eqref{transconstrain1},  we still choose the eigentress $\sigma^*_{ij}$ in \eqref{eigenstresscubic}, substitution of which along with \eqref{rhoquadratic} and \eqref{Pcubic}  into \eqref{Fourierstrain}  still generates \eqref{cubicstrain}, only with $C_{11}$ replaced by $C_{33}$.   Then \eqref{cubicstraintrans} can be derived from \eqref{cubicstrain} via  the same transformation as \eqref{cotransp2} with the  magnitude of $s$ replaced by $s{:=}\sqrt{\frac{C_{44}}{C_{33}}}$.
By comparison of \eqref{cubicstraintrans} with \eqref{Netownianfourier}, we can obtain the same results in \eqref{keyequation}. Therefore, by following the same discussion from \eqref{keyequation} to \eqref{omegaform}, we verify the existence of a non-ellipsoidal inclusion that possesses Eshelby's polynomial conservation property.

For the case  \eqref{transconstrain12}, we specify the particular eigentrain that makes the corresponding eigenstress belong to the transversely isotropic category, i.e.,
\begin{align}\label{sigmatransdegenerate}
\begin{split}
 \boldsymbol{{\sigma}^*}(\bfx)\in\{\boldsymbol{{\sigma}^*}(\bfx)|
\boldsymbol{{\sigma}^*}(\bfx)=\rho(\bfx)(\overline{\sigma}^*_{11}\boldsymbol{\tilde\alpha}+\overline{\sigma}^*_{33}\boldsymbol{\tilde\beta}),\;\;\;
\;\;\overline{\sigma}^*_{11},\overline{\sigma}^*_{33}\in R,\;\bfx\in\mathbb{R}^3 \}
\end{split}
\end{align}
where $\boldsymbol{\tilde\alpha}$ and $\boldsymbol{\tilde\beta}$ are defined in \eqref{tildealphabeta}, and $\rho$ is defined in \eqref{rhoquadratic}.
And we additionally require $\overline{\sigma}^*_{33}=\gamma{\overline{\sigma}^*_{11}}$ where $\gamma$ is defined in \eqref{gamma}

By substituting \eqref{sigmatransdegenerate} along with \eqref{tildealphabeta}, \eqref{gamma} and \eqref{rhoquadratic} into \eqref{Fourierstrain},
 we can get a concise expression of the strain field, i.e.,
\beqs\label{keyequationdetrans}
\boldsymbol{\varepsilon}(\bfx)=\frac{1}{2}\left[\nabla \otimes(\bfK^*\cdot\nabla u^*)+(\bfK^*\cdot\nabla u^*)\otimes\nabla\right],
\eeqs
with $\bfK^*$ possessing the same components as those in \eqref{1.12b} and
\beas
u^*(\bfx){:=}\frac{1}{(2\pi)^3}\int_{\mathbb{R}^3}\frac{1}{ \xi_1^2 + \xi_2^2+\frac{1}{v^2} \xi_3^2}\int_{\Omega}\rho(\bfy)e^{-\mathrm{i}\boldsymbol{\xi}\cdot( \bfx-\bfy)}d\bfy d\boldsymbol{\xi}
\eeas
where $v$ is the root of \eqref{root}.

Then comparing \eqref{keyequationdetrans} with \eqref{Netownianfourier}  yields
\beqs\label{keyequationtrans}
{\varepsilon}_{ij}(\bfx')=\frac{1}{2}\left(\tilde{Q}_{ip}K^*_{jl}\tilde{Q}_{lm}\frac{\partial^2 N_{\Omega'}[\rho](\bfx')}{\partial x'_p \partial x'_m} +\tilde{Q}_{jl}K^*_{ip}\tilde{Q}_{pq}\frac{\partial^2 N_{\Omega'}[\rho](\bfx')}{\partial x'_l \partial x'_q} \right).
\eeqs
Via comparison of \eqref{keyequationtrans} with Lemma \ref{keylemma},  we verify the existence of a non-ellipsoidal $\Omega'$ that leads to the quadratic strain field induced by $\Omega$, which proves Theorem \ref{cubicstrong} for transversely isotropic materials.

 \subsection*{{ (3) Orthotropic material}}

We still consider the eigentress $\sigma^*_{ij}$ in \eqref{eigenstresscubic}. Under the condition \eqref{orthoconstrain}, substituting \eqref{eigenstresscubic} along with \eqref{rhoquadratic} and \eqref{Pcubic}  into \eqref{Fourierstrain} gives
\beqs\label{orthostrain}
\begin{split}
&\varepsilon_{11}=0,\;\;\varepsilon_{22}=0,\;\;\varepsilon_{12}=0,\\
&\varepsilon_{13}(\bfx)=-\frac{1}{2}\frac{1}{(2\pi)^3}\int_{\mathbb{R}^3}\frac{\xi_1\xi_3}{{C_{55}}\xi_1^2 +{C_{44}} \xi_2^2 + {C_{33}}\xi_3^2}\int_{\Omega}\rho(\bfy)e^{-\mathrm{i}\boldsymbol{\xi}\cdot( \bfx-\bfy)}d\bfy d\boldsymbol{\xi},\\
&\varepsilon_{23}(\bfx)=-\frac{1}{2}\frac{1}{(2\pi)^3}\int_{\mathbb{R}^3}\frac{\xi_2\xi_3}{{C_{55}}\xi_1^2 +{C_{44}} \xi_2^2 + {C_{33}}\xi_3^2}\int_{\Omega}\rho(\bfy)e^{-\mathrm{i}\boldsymbol{\xi}\cdot( \bfx-\bfy)}d\bfy d\boldsymbol{\xi},\\
&\varepsilon_{33}(\bfx)=-\frac{1}{(2\pi)^3}\int_{\mathbb{R}^3}\frac{\xi^2_3}{{C_{55}}\xi_1^2 +{C_{44}} \xi_2^2 + {C_{33}}\xi_3^2}\int_{\Omega}\rho(\bfy)e^{-\mathrm{i}\boldsymbol{\xi}\cdot(\bfx-\bfy)}d\bfy d\boldsymbol{\xi}.
\end{split}
\eeqs

 By  transformations of coordinates
 \begin{align}\label{newtransortho}
\bfx'' {:=}\hat{\bfQ}\cdot\bfx,\ \ \ \  \
\bfy'' {:=} \hat{\bfQ}\cdot\bfy,\ \ \ \  \
\boldsymbol{\xi}'' {:=}{\hat{\bfQ}}^{-1} \cdot\boldsymbol{\xi}
\end{align}
with
\beqs\label{Qtransortho}
\hat{\bfQ}{:=}\begin{bmatrix}
  s_1&0&0 \\
  0&s_2&0 \\
  0&0&1
\end{bmatrix}
\eeqs
and then substitution of \eqref{Qtransortho} into \eqref{orthostrain} with $s_1{:=}\sqrt{\frac{C_{33}}{C_{55}}}, s_2{:=}\sqrt{\frac{C_{33}}{C_{44}}}$, it follows that
\beqs\label{orthostraintrans}
\begin{split}
&\varepsilon_{11}=0,\;\;\varepsilon_{22}=0,\;\;\varepsilon_{12}=0,\\
&\varepsilon_{13}(\bfx'')=-\frac{1}{2\sqrt{C_{33}C_{55}}}\frac{1}{(2\pi)^3}\int_{\mathbb{R}^3}\frac{\xi''_1\xi''_3}{  {\xi_1''}^2 + {\xi_2''}^2+{\xi_3''}^2 }\int_{\Omega''}\rho(\bfy'')e^{-\mathrm{i}\boldsymbol{\xi}'\cdot( \bfx''-\bfy'')}d\bfy'' d\boldsymbol{\xi}'',\\
&\varepsilon_{23}(\bfx'')=-\frac{1}{2\sqrt{C_{33}C_{44}}}\frac{1}{(2\pi)^3}\int_{\mathbb{R}^3}\frac{\xi''_2\xi''_3}{{\xi_1''}^2 + {\xi_2''}^2+{\xi_3''}^2 }\int_{\Omega''}\rho(\bfy'')e^{-\mathrm{i}\boldsymbol{\xi}''\cdot( \bfx''-\bfy'')}d\bfy'' d\boldsymbol{\xi}'',\\
&\varepsilon_{33}(\bfx'')=-\frac{1}{C_{33}}\frac{1}{(2\pi)^3}\int_{\mathbb{R}^3}\frac{{\xi''_3}^2}{{\xi_1''}^2 + {\xi_2''}^2+{\xi_3''}^2 }\int_{\Omega''}\rho(\bfy'')e^{-\mathrm{i}\boldsymbol{\xi}''\cdot(\bfx''-\bfy'')}d\bfy'' d\boldsymbol{\xi}'',
\end{split}
\eeqs
with
\begin{align}\label{Omegaortho}
\begin{split}
\Omega'' {:=} \left\{\bfy''\left| {{\hat{\bfQ}}^{-1}\cdot\bfy''} \right. \in \Omega \right\}.
\end{split}
\end{align}

By comparing \eqref{orthostraintrans} with \eqref{Netownianfourier}, we find
\beqs\label{keyequationortho}
{\varepsilon}_{ij}(\bfx'')=\frac{1}{2C_{33}}\left(\hat{Q}_{il}P_{jm}\hat{Q}_{mq}\frac{\partial^2 N_{\Omega'}[\rho](\bfx'')}{\partial x''_l \partial x''_q}+\hat{Q}_{jm}P_{il}\hat{Q}_{ls}\frac{\partial^2 N_{\Omega'}[\rho](\bfx'')}{\partial x''_m \partial x''_s}\right).
\eeqs

Resorting to Lemma \ref{keylemma}, there exists a non-ellipsoid $\Omega''$ leading to the quadratic polynomial form of the right-hand side of \eqref{keyequationortho}, which implies 
the existence of a non-ellipsoidal $\Omega$ possessing Eshelby's polynomial conservation property --  $\Omega$ is constructed by stretching $\Omega''$ along the $x_1''$-axis and $x_2''$-axis by proportions $\frac{1}{s_1}$ and $\frac{1}{s_2}$, respectively, in terms of \eqref{Omegaortho} and \eqref{Qtransortho}.

\subsection*{{(4) Monoclinic material}}


We still consider the eigentress $\sigma^*_{ij}$ in \eqref{eigenstresscubic}. Under the condition \eqref{monoclinicconstrain}, substitution of \eqref{eigenstresscubic} along with \eqref{rhoquadratic} and \eqref{Pcubic}  into \eqref{Fourierstrain} still yields \eqref{orthostrain}.
  Likewise, through the same analysis as before, \eqref{keyequationortho} is derived from \eqref{orthostrain}, comparison of which with Lemma \ref{keylemma} implies the existence of a non-ellipsoidal $\Omega''$ that leads to the quadratic strain field, and thus the existence of the corresponding non-ellipsoidal $\Omega$ which is constructed by the inverse transformation of \eqref{Omegaortho}.

\subsection*{{(5) Isotropic material}}
For an isotropic material whose elastic tensor takes the form
 \beqs\label{isoelastictensor}
C_{ijkl}=\lambda\delta_{ij}\delta_{kl}+\mu(\delta_{ik}\delta_{jl}+\delta_{il}\delta_{jk})\;\;(i,j,k,l=1,2,3)
\eeqs
with $\lambda$ and $\mu$  being Lam\'{e} constants, we take a particular uniform eigenstrain $\varepsilon^*_{ij}=\rho(\bfx)\delta_{ij}$.  {By substituting $\varepsilon^*_{ij}=\rho(\bfx)\delta_{ij}$ and \eqref{isoelastictensor} into \eqref{Fourierstrain} and then comparing \eqref{Fourierstrain} with \eqref{Netownianfourier}, we obtain} a linear relationship between the Hessian matrix of the Newtonian potential $N_{\Omega}[\rho](\bfx)$ induced by the domain $\Omega$ of the inclusion with mass density $\rho(\bfx)$ and the strain field $\varepsilon_{ij}(\bfx)$ inside, i.e.,
  \beqs\label{strainiso}
\varepsilon_{ij}(\bfx)= \frac{1}{2}\left(\frac{\partial u_i}{\partial x_j}+\frac{\partial u_j}{\partial x_i}\right)=\frac{3\lambda+2\mu}{\lambda+2\mu}\frac{\partial^2 N_{\Omega}[\rho](\bfx)}{\partial x_i\partial x_j},\;\;\bfx\in \Omega.
\eeqs
 Combining \eqref{strainiso} with Lemma \ref{keylemma}  { leads to the conclusion that there exists a non-ellipsoidal $\Omega$ which possesses Eshelby's polynomial conservation property in the isotropic medium for the quadratic eigenstrain.}


\section{Counter-examples to the generalized strong version of the high-order Eshelby conjecture for polynomial eigenstrains of even degrees}

 {Here, we will show that the conclusion of Theorem \ref{cubicstrong} can be extended to polynomial eigenstrains of any even degree in the anisotropic media.
 The particular eigenstress $\boldsymbol{\sigma^*}$, which is assumed to be linear with respect to $\rho(\bfx)$, will be taken as \eqref{eigenstresscubic}.
But} instead of the quadratic $\rho(\bfx)$ in (\ref{rhoquadratic}), we now consider a polynomial $\rho(\bfx)$ of even degrees as follows:
\beqs\label{rhohighorder}
\rho(\bfx):=-\sum_{i=1}^3 d_ix_i^n
\eeqs
 with $d_i\;(i=1,2,3)$ denoting real constants, and $n\in \{n\;|\;n{:=}2k,\;k\geq0,\;k\in \mathrm{Z}\}$, while noting that $n=2$ have been studied in Section~4.

Then by substituting \eqref{eigenstresscubic} along with \eqref{rhohighorder} into \eqref{Fourierstrain}, we can obtain the strain fields $\boldsymbol{\varepsilon}[\rho](\bfx)$
\beqs\label{inducedstrainhighorder}
\begin{split}
&\varepsilon_{11}=0,\;\;\varepsilon_{22}=0,\;\;\varepsilon_{12}=0,\\
&\varepsilon_{13}(\bfx)\propto-\frac{1}{(2\pi)^3}\int_{\mathbb{R}^3}\frac{\xi_1\xi_3}{a\xi_1^2 +b \xi_2^2 + c\xi_3^2}\int_{\Omega}\rho(\bfy)e^{-\mathrm{i}\boldsymbol{\xi}\cdot( \bfx-\bfy)}d\bfy d\boldsymbol{\xi},\\
&\varepsilon_{23}(\bfx)\propto-\frac{1}{(2\pi)^3}\int_{\mathbb{R}^3}\frac{\xi_2\xi_3}{a\xi_1^2 +b \xi_2^2 + c\xi_3^2}\int_{\Omega}\rho(\bfy)e^{-\mathrm{i}\boldsymbol{\xi}\cdot( \bfx-\bfy)}d\bfy d\boldsymbol{\xi},\\
&\varepsilon_{33}(\bfx)\propto-\frac{1}{(2\pi)^3}\int_{\mathbb{R}^3}\frac{\xi^2_3}{a\xi_1^2 +b\xi_2^2 + c\xi_3^2}\int_{\Omega}\rho(\bfy)e^{-\mathrm{i}\boldsymbol{\xi}\cdot(\bfx-\bfy)}d\bfy d\boldsymbol{\xi},
\end{split}
\eeqs
where $a,b,c$ are three positive real constants related to the elastic constants of the specific anisotropic material as follows:
\beqs\label{abc}
\begin{split}
&a=b=C_{44},\quad c=C_{11} \quad \mathrm{\it for \;cubic \;materials};\\
&a=b=C_{44},\quad c=C_{33}  \quad \mathrm{\it for\; transversely\; isotropic \;materials};\\
&a=C_{55},\quad b=C_{44}, \quad c=C_{33} \quad \mathrm{\it for \;orthotropic\; and \;monoclinic \;materials}.\\
\end{split}
\eeqs

If we define new coordinate transformations
\beqs\label{highordertransformation}
\bfx'{:=}\bar{\bfQ}\cdot\bfx,\ \ \ \  \
\bfy' {:=} \bar{\bfQ}\cdot\bfy,\ \ \ \  \
\boldsymbol{\xi}'{:=}{\bar{\bfQ}}^{-1} \cdot\boldsymbol{\xi}
\eeqs
with
\beqs\label{Qtranshighoer}
\bar{\bfQ}{:=}\begin{bmatrix}
  \sqrt{\frac{c}{a}}&0&0 \\
  0&\sqrt{\frac{c}{b}}&0 \\
  0&0&1
\end{bmatrix},
\eeqs
then substitution of \eqref{highordertransformation}  into \eqref{inducedstrainhighorder}  leads to
\beqs\label{highoerstraintrans}
\begin{split}
&\varepsilon_{11}=0,\;\;\varepsilon_{22}=0,\;\;\varepsilon_{12}=0,\\
&\varepsilon_{13}(\bfx')\propto-\frac{1}{(2\pi)^3}\int_{\mathbb{R}^3}\frac{\xi'_1\xi'_3}{  {\xi_1'}^2 + {\xi_2'}^2+{\xi_3'}^2 }\int_{\bar{\Omega}}\rho(\bfy')e^{-\mathrm{i}\boldsymbol{\xi}'\cdot( \bfx'-\bfy')}d\bfy' d\boldsymbol{\xi}',\\
&\varepsilon_{23}(\bfx')\propto-\frac{1}{(2\pi)^3}\int_{\mathbb{R}^3}\frac{\xi'_2\xi'_3}{{\xi_1'}^2 + {\xi_2'}^2+{\xi_3'}^2 }\int_{\bar{\Omega}}\rho(\bfy')e^{-\mathrm{i}\boldsymbol{\xi}'\cdot( \bfx'-\bfy')}d\bfy' d\boldsymbol{\xi}',\\
&\varepsilon_{33}(\bfx')\propto-\frac{1}{(2\pi)^3}\int_{\mathbb{R}^3}\frac{{\xi'_3}^2}{{\xi_1'}^2 + {\xi_2'}^2+{\xi_3'}^2 }\int_{\bar{\Omega}}\rho(\bfy')e^{-\mathrm{i}\boldsymbol{\xi}'\cdot(\bfx'-\bfy')}d\bfy' d\boldsymbol{\xi}',
\end{split}
\eeqs
comparison of which with \eqref{cubicstraintrans} yields
\beqs\label{keyequationforanydegree}
{\varepsilon}_{ij}[\rho](\bfx')\propto \left(\bar{Q}_{il}P_{jm}\bar{Q}_{mq}\frac{\partial^2 N_{\bar{\Omega}}[\rho](\bfx')}{\partial x'_l \partial x'_q}+\bar{Q}_{jm}P_{il}\bar{Q}_{ls}\frac{\partial^2 N_{\bar{\Omega}}[\rho](\bfx')}{\partial x'_m \partial x'_s}\right),
\eeqs
where
\begin{align}\label{Omegahighorder}
\begin{split}
\bar{\Omega} {:=} \left\{\bfy'\left| {{\bar{\bfQ}}^{-1}\cdot\bfy'} \right. \in \Omega \right\}.
\end{split}
\end{align}

Specifically, we choose  $\rho(\bfx')=-\sum_{i=1}^3 {x'_i}^n$, which corresponds to  $\rho(\bfx)=\big[ \left({\frac{c}{a}}\right)^{\frac{n}{2}}{x_1}^n+\left({\frac{c}{b}}\right)^{\frac{n}{2}}{x_2}^n+{x_3}^n\big]$ in \eqref{rhohighorder}. Then we assume that there exists a { one-component connected bounded open domain} $\bar{\Omega}$ and consider the case where
 $N_{\bar{\Omega}}[\rho](\bfx')$ consists of polynomial terms and a non-polynomial term as follows:
\beqs\label{highdegreeN}
N_{\bar{\Omega}}[\rho](\bfx')=-\int_{\bar{\Omega}} \frac{\rho(\bfy')}{4\pi|\bfx'-\bfy'|} d\bfy'=\hat{\varphi}(\bfx',n)+\hat{\omega}(x_1',x_2'), \;\;\bfx'\in \bar{\Omega},
\eeqs
with $\hat{\varphi}(\bfx',n)$ denoting a polynomial function of $\bfx'$ with degree $n+2$ that satisfies $\Delta_{x'} \hat{\varphi}(\bfx',n)=\rho(\bfx')$, and $\hat{\omega}(x_1',x_2')$ denoting a harmonic function that is non-polynomial and Lipschitz continuous, which guarantees that the right-hand side of \eqref{highdegreeN} satisfies the definition \eqref{Newtonianroot} of the Newtonian potential. In this case,
because of the existence of  the non-polynomial term $\hat{\omega}$ in \eqref{highdegreeN},
the shape of $\bar{\Omega}$ cannot be ellipsoidal due to the pure polynomial forms of the Newtonian potential induced by ellipsoids.
Then $\forall\;n\in \{n\;|\;n{:=}2k,\;k\geq0,\;k\in \mathrm{Z}\}$, by  substituting \eqref{highdegreeN}  into \eqref{keyequationforanydegree}, we can see that the non-ellipsoidal $\bar{\Omega}$ that generates \eqref{highdegreeN} will lead to the polynomial strain field of degree $n$, which verifies Eshelby's polynomial conservation property of $\Omega$,
and thus constitutes count-examples to the generalized strong version of the high-order Eshelby conjecture for polynomial eigenstrains of any even degree.

 However, it is left to prove the existence of an $\bar{\Omega}$ that yields \eqref{highdegreeN} for any $n\in \{n\;|\;n{:=}2k,\;k\geq0,\;k\in \mathrm{Z}\}$.
To this end, we firstly consider the case when $n=0$. In this case, $\hat{\varphi}(\bfx',n)$ in (\ref{highdegreeN}) becomes a quadratic function, and $\rho(\bfy')$ becomes a constant, which means that (\ref{highdegreeN}) can be expressed as
\beqs\label{Liuinvalid}
N_{\bar{\Omega}}(\bfx')=-\int_{\bar{\Omega}} \frac{1}{4\pi|\bfx'-\bfy'|} d\bfy'=q'(\bfx')+\omega'(x_1',x_2'), \;\;\bfx'\in \bar{\Omega},
\eeqs
where $q'(\bfx')$ denotes a quadratic function, and $\omega'(x_1',x_2')$ still dentes a non-polynomial function.
Then, in terms of the work of \cite{Liu2008} (Section 3 of the work of \cite{Liu2008}), there exists a non-ellipsoidal $\bar{\Omega}$ that makes \eqref{Liuinvalid} hold.
The non-ellipsoidal inclusion $\bar{\Omega}$ that makes \eqref{Liuinvalid} hold in the work of \cite{Liu2008} is initially constructed as a counter-example for the isotropic medium. Here we have provided a counter-example to the generalized strong version of the Eshelby conjecture for uniform eigenstrains in the anisotropic media.

 With the case $n=2$ has already been proved in Section 4, we turn to prove the case when $n>2$.
 We take
 the non-polynomial function $\hat{\omega}(x_1',x_2')$ in \eqref{highdegreeN} as $\omega^*(x_1',x_2')$ defined in \eqref{Omegastar}, and we take the polynomial function $\hat{\varphi}(\bfx',n)$ in \eqref{highdegreeN} as
\beqs\label{newphiansio}
\tilde{\varphi}(\bfx'):=\hat{C}-\frac{1}{(n+2)(n+1)}\left({x'_1}^{n+2}+{x'_2}^{n+2}+{x'_3}^{n+2}\right), \;\;\;\;\bfx\in\mathbb{R}^3,
\eeqs
 where $\hat{C}$ is a positive real constant. Then \eqref{highdegreeN} can be expressed as
 \beqs\label{specialcaseN}
N_{\bar{\Omega}}[\rho](\bfx')=-\int_{\bar{\Omega}} \frac{-({y'_1}^n+{y'_2}^n+{y'_3}^n)}{4\pi|\bfx'-\bfy'|} d\bfy'=\tilde{\varphi}(\bfx')+\omega^*(x_1',x_2'), \;\;\bfx'\in \bar{\Omega}.
\eeqs
It can be verified that the right-hand side of \eqref{specialcaseN} satisfies the definition \eqref{Newtonianroot} of the Newtonian potential by
substituting \eqref{newphiansio} and  \eqref{Omegastar} into \eqref{specialcaseN}.

 Then, our aim is to prove the existence of a non-ellipsoidal $\bar{\Omega}$ that makes \eqref{specialcaseN} hold.
We introduce
 \beqs\label{newphihighdegree}
\begin{split}
\hat{\phi}^*(\bfx'){:=}\left\{ {\begin{array}{*{20}{c}}
  {\tilde{\varphi}(\bfx')+\omega^*(x_1',x_2')\;\;\;\;\;\;\;\;\;\;\;\;\;\;\bfx' \in \hat{U}, } \\
  {\;\;-\hat{C}+\omega^*(x_1',x_2')\;\;\;\;\;\;\;\;\;\;\bfx' \in \mathbb{R}^3\setminus \hat{U},}
\end{array}\;\;\;\;\;\;\;\;\;\;\;\;\;\;\;\;\;\;\;\;\;\;\;\;\;\;\;\;\;\;}  \right.,
\end{split}
\eeqs
where
\beqs\label{Uhat}
\hat{U}{:=}\{\;\bfx'\;|\;\;{x_1'}^{n+2}+{x_2'}^{n+2}+{x_3'}^{n+2}\leq 2(n+1)(n+2)\hat{C},\;\bfx'\in\mathbb{R}^3\}.
 \eeqs
 Here $\hat{U}$ is bounded due to $\hat{U}$ being contained in another bounded domain $\{\;\bfx'\;|\;\;|{x_1'}|\leq (2(n+1)(n+2)\hat{C})^{\frac{1}{n+2}},\;\bfx'\in\mathbb{R}^3\}$.
Then we will prove ${\phi}^*$ is an obstacle function that is defined in Appendix C for the proof of Lemma \ref{keylemma}.
According to the definition in Appendix C, an obstacle function needs to satisfy four conditions. We will show that  $\hat{\phi}^*$ in (\ref{newphihighdegree}) satisfies all of these conditions.

  Firstly, since $\omega^*\in C^{0}(\mathbb{R}^{3})$ due to \eqref{Omegastar}, and it can be derived from \eqref{newphihighdegree} that
 \beas
\hat{\phi}^*|_{\partial \hat{U}^-}=\hat{\phi}^*|_{\partial \hat{U}^+}=-\hat{C},
\eeas
 we see $\hat{\phi}^*\in C^{0}(\mathbb{R}^{3})$.
Further, substituting \eqref{newphihighdegree} into \eqref{normdefinition} yields
\beqs\label{normphistar}
||\hat{\phi}^*||_{0,1}\leq \sup_{\bfx'\in \hat{U}}|\tilde{\varphi}(\bfx')|+\sup_{\bfx'\in \hat{U}}|\nabla_{x'}\tilde{\varphi}(\bfx')|+\sup_{\bfx'\in U^\omega}|\omega(x_1',x_2')|+\sup_{\bfx'\in  U^\omega}|\nabla_{x'}\omega(x_1',x_2')|,
\eeqs
where $\omega$ is defined in \eqref{omegaform}, and $ U^\omega\in \mathbb{R}^3$ is a bounded domain defined in \eqref{Omegastar}.  $|\omega(x_1',x_2')|$ and $|\nabla_{x'}\omega(x_1',x_2')|$ are bounded in $U^\omega$ owing to \eqref{omegaform}, and $|\tilde{\varphi}(\bfx')|$ and $|\nabla_{x'}\tilde{\varphi}(\bfx')|$ are bounded in $\hat{U}$ owing to \eqref{newphiansio}, which means that
 the norm $||\hat{\phi}^*||_{0,1}$ of $\hat{\phi}^*$ is bounded, and thus $\hat{\phi}^*\in C^{0,1}(\mathbb{R}^3)$. Therefore, $\hat{\phi}^*$ satisfies the first condition of an obstacle function.

Secondly, we let $\hat{r}=(2(n+1)(n+2)\hat{C})^{\frac{1}{n+2}}$.
 Since \eqref{Omegastar} and \eqref{omegaform} imply $\omega^*(x'_1, x_2')\leq 0$ for any $\bfx'$,  then due to \eqref{newphiansio},  we conclude that $\forall \;|\bfx'|\geq \hat{r}, \;\hat{\phi}^*(\bfx')=-\hat{C}+\omega^*(x_1',x_2')\leq 0$, which proves that $\hat{\phi}^*$ satisfies the second condition of an obstacle function.

Thirdly, we take a sphere $B_{\hat{r}}=\{\bfx'||\bfx'|\leq \hat{r},\;\bfx'\in\mathbb{R}^3\}$, and thus $\hat{U}\subset B_{\hat{r}}$. Then for $\bfx'\in B_{\hat{r}}\setminus U^*$, with $U^*\subset B_{r_0}$ being defined in Appendix C as the set of the singular points where the norm $|\nabla_{x'}\otimes\nabla_{x'}\hat{\phi}^*|$ of $\nabla_{x'}\otimes\nabla_{x'}\hat{\phi}^*$ is unbounded, it can be derived from \eqref{newphihighdegree} along with \eqref{newphiansio} and \eqref{Omegastar} that
\beas
\begin{split}
|\Delta_{x'}\hat{\phi}^*(\bfx')|=\left\{ {\begin{array}{*{20}{l}}
  {0\;\;\;\;\;\;\;\;\;\;\;\;\;\;\quad\quad\quad\quad\quad\;\;\;\bfx' \in B_{\hat{r}}\setminus \hat{U}, } \\
  {|{x'_1}^n+{x'_2}^n+{x'_3}^n|\;\;\;\;\;\;\;\;\;\;\;\;\bfx' \in \hat{U}\setminus U^*,}
\end{array}\;\;\;\;\;\;\;\;\;\;\;\;\;\;\;\;\;\;\;\;\;\;\;\;\;\;\;\;\;\;}  \right.
\end{split}
\eeas
which indicates that $|\Delta_{x'}\hat{\phi}^*(\bfx')|\leq 3(2(n+1)(n+2)\hat{C})^{\frac{n}{n+2}}$ for $\bfx'\in B_{\hat{r}}\setminus U^*$.
 Thus we conclude that $|\Delta_{x'}\hat{\phi}^*(\bfx')|$ is bounded in $B_{\hat{r}}\setminus U^*$, which proves that $\hat{\phi}^*$ satisfies the third condition of an obstacle function.

Fourthly, it can be derived from \eqref{newphihighdegree}  that $\forall \boldsymbol{\zeta}\in \bfR^3$ with $|\boldsymbol{\zeta}|=1$,
\beqs\label{judgeChat1}
\begin{split}
&\int_{U^{\vartheta}}\frac{\partial^2 \vartheta }{\partial \boldsymbol{\zeta}^2}\left(\hat{\phi}^*+\frac{1}{2}\hat{C}|\bfx'|^2\right)d\bfx'
\\
&\geq \int_{U^{\vartheta}} \vartheta \left(\hat{C}-\sup_{\bfx'\in \hat{U}}\left|\frac{\partial \tilde{\varphi}}{\partial \boldsymbol{\zeta}}\right|-\sup_{\bfx' \in \hat{U}}\left|\frac{\partial^2 \tilde{\varphi}}{\partial \boldsymbol{\zeta}^2}\right|-\sup_{\bfx'\in U^\omega}\left|\frac{\partial \omega}{\partial \boldsymbol{\zeta}}\right|-\sup_{\bfx'\in U^\omega}\left|\frac{\partial^2 \omega}{\partial \boldsymbol{\zeta}^2}\right|\right)d\bfx',
\end{split}
\eeqs
where $\vartheta$ is defined in \eqref{smooth} with a compact support $U^{\vartheta}$, on which $\vartheta\geq 0$.
Then it follows from \eqref{judgeChat1} that $\exists \hat{C}\in\bfR$ satisfying
\beas\label{judgeChat3}
\hat{C}\geq\sup_{\bfx'\in \hat{U}}\left|\frac{\partial \tilde{\varphi}}{\partial \boldsymbol{\zeta}}\right|+\sup_{\bfx' \in \hat{U}}\left|\frac{\partial^2 \tilde{\varphi}}{\partial \boldsymbol{\zeta}^2}\right|+\sup_{\bfx'\in U^\omega}\left|\frac{\partial \omega}{\partial \boldsymbol{\zeta}}\right|+\sup_{\bfx'\in U^\omega}\left|\frac{\partial^2 \omega}{\partial \boldsymbol{\zeta}^2}\right|,
\eeas
such that  $\forall \boldsymbol{\zeta}\in \bfR^3$ with $|\boldsymbol{\zeta}|=1$,
\beas
\int_{U^{\vartheta}}\frac{\partial^2 \vartheta }{\partial \boldsymbol{\zeta}^2}\left(\hat{\phi}^*+\frac{1}{2}\hat{C}|\bfx'|^2\right)d\bfx' \geq 0,
\eeas
which proves that $\hat{\phi}^*$ satisfies the fourth condition of an obstacle function.

Therefore, according to Appendix C,  $\hat{\phi}^*$ that satisfies  all the conditions of an obstacle function will result in the existence of a coincident set $\bar{\Omega}\subseteq B_{\hat{r}}$, where $N_{\bar{\Omega}}[\rho](\bfx')=\hat{\phi}^*(\bfx')$ for $\bfx' \in \bar{\Omega}$.  $\hat{\phi}^*$ in \eqref{newphihighdegree} is a piece-wise function, and thus $\hat{\phi}^*$ has two possible expressions, i.e., $-\hat{C}+\omega^*(x_1',x_2')$ and $\tilde{\varphi}(\bfx')+\omega^*(x_1',x_2')$.
However,
since the case $N_{\bar{\Omega}}[\rho](\bfx')=-\hat{C}+\omega^*(x_1',x_2')$ for $\bfx' \in \bar{\Omega}$ contradicts the definition \eqref{Newtonianroot} of the Newtonian potential $N_{\bar{\Omega}}[\rho]$ induced by $\bar{\Omega}$ with mass density $\rho$, we conclude that  $N_{\bar{\Omega}}[\rho](\bfx')=\tilde{\varphi}(\bfx')+\omega^*(x_1',x_2')$ for $\bfx' \in \bar{\Omega}$ and also $\bar{\Omega}\subseteq\hat{U}$. Furthermore, because $\tilde{\varphi}(\bfx')+\omega^*(x_1',x_2')$ is not the Newtonian potential caused by an ellipsoid, $\bar{\Omega}$ must be non-ellipsoidal. Hence we have proved the existence of a non-ellipsoidal $\bar{\Omega}$ that makes \eqref{specialcaseN} hold.
The shape of a counter-example non-ellipsoidal inclusion for a quartic eigenstrain is shown as $\Omega^{(2)}$ in Figure~\ref{counterexample2} in Appendix D.

In retrospect, we see that \eqref{highdegreeN} is actually a sufficient condition for the existence of a non-ellipsoidal $\bar{\Omega}$ and thus the existence of a corresponding counter-example $\Omega$, which can be simply verified by substitution of \eqref{highdegreeN} into \eqref{keyequationforanydegree}. Consequently, by appropriately choosing $\rho(\bfy')$, $\hat{\varphi}(\bfx',n)$ and $\hat{\omega}(x_1',x_2')$ in \eqref{highdegreeN} to make \eqref{highdegreeN} hold for some $\bar{\Omega}$ with the utilization of the variational method used in the proof of Lemma \ref{keylemma}, we can construct more counter-example $\Omega$, each of which corresponds to a specific $\bar{\Omega}$, to the generalized strong version of the high-order Eshelby conjecture for polynomial eigenstrains of any even degree in the anisotropic media.

\section{Conclusions and discussion}

Firstly,
we have presented proofs of the generalized weak version of the Eshelby conjecture for an inclusion in three-dimensional anisotropic media that possess cubic, transversely isotropic, orthotropic, and monoclinic symmetries, which means that only the ellipsoidal shape can transform {\it all} uniform eigenstrains into uniform elastic strain fields in a solitary inclusion in infinite media possessing these symmetries.
Secondly, we prove that in these anisotropic media, there indeed exist non-ellipsoidal inclusions that can transform  particular polynomial eigenstrains of even degrees into polynomial elastic strain fields of the same even degrees in them, and also in the isotropic medium, there exist non-ellipsoidal inclusions that can transform  particular quadratic eigenstrains into quadratic elastic strain fields in them, which constitutes counter-examples to the generalized strong version of the high-order Eshelby conjecture for polynomial eigenstrain of even degrees in these anisotropic media and also in the isotropic medium (quadratic eigenstrain only). On one hand, the findings in this work reveal that in anisotropic media, a striking rich class of inclusions beyond ellipsoids can exhibit the uniformity between the eigenstrains and the induced elastic strains.
On the other hand, the results can be regarded as generalization of Eshelby's
polynomial conservation theorem to non-ellipsoidal inclusions in these anisotropic media which are a much wider regime than the isotropic medium.

Several interesting issues related to the present work may deserve further investigations. First, it is worth noting that for the particular eigenstrains we have studied in this work, an infinite number of non-ellipsoidal inclusions in the anisotropic media have the generalized  Eshelby property, that is, they can transform the uniform eigenstains into uniform elastic strains, and transform the polynomial eigenstrains of even degrees into elastic strains of the same degrees. However, whether there are other forms of particular uniform eigenstrains and polynomial eigenstrains  that can be transformed into
 the corresponding uniform elastic strains and polynomial elastic strains by non-ellipsoidal inclusions is not known.
 {  Second, exploring the solutions to the generalized Eshelby conjectures through linear transformations of the coordinates and the displacements to transform the anisotropic elastic tensors into the isotropic elastic tensor in the work of \cite{Liu2008} is a very interesting topic (though we have found the corresponding transformations for the transversely isotropic material).}
Third, in contrast to the generalized strong version, the problem concerning the generalized weak version of the high-order Eshelby conjecture is hard and remains to be solved. Finally,
 when the eigenstrain is in the expression of the polynomial of odd degrees, the variational method utilized in the proof of Lemma \ref{keylemma} for polynomial eigenstrains of even degrees is inapplicable. Thus new approaches need to be devised to deal with this case.

\begin{flushleft}
\textbf{\emph{Declaration of Competing Interest}}
\end{flushleft}

The authors declare no competing interests.

\begin{flushleft}
\textbf{\emph{Acknowledgements}}
\end{flushleft}

The authors thank the National Natural Science Foundation of China (Grant Nos. 11521202, 11890681) for support of this work. Part of the work was completed when Yuan and Wang were visiting the Department of Mechanics and Aerospace Engineering of Southern University of Science and Technology, the support of which is acknowledged. We also thank Professor Bhushan L. Karihaloo of Cardiff University, Professor Shaoqiang Tang of Peking University, and Mr Shuo Zhang of University of Minnesota for helpful discussions. {  The authors thank an anonymous reviewer whose insightful comments and suggestions improved the technical content and presentational quality of this work.}

\appendix
\setcounter{equation}{0}
\renewcommand\theequation{A.\arabic{equation}}

\section*{Appendix}
\section*{A. \quad An alternative method to prove Theorem \ref{transtheorem}}

The explicit Green function for a transversely isotropic material has  been derived by \cite{Pan1976} under the condition \eqref{transconstrain12}. Based on the explicit expression of the Green function, we prove Theorem \ref{transtheorem} via an alternative method, which not only validates the correctness of the result given in the main text but also helps us prove another theorem in relation to the material parameters of the transversely isotropic material.

We will follow all of the notation in the main text.
We consider two cases concerning the possible degeneracy of the elastic parameters of the transversely isotropic material, where $\sqrt {{C_{11}}{C_{33}}}  - {C_{13}} - 2{C_{44}}>0$ (non-degenerate) and $\sqrt {{C_{11}}{C_{33}}}  - {C_{13}} - 2{C_{44}}=0$ (degenerate).

\subsection*{(1)\quad Non-degenerate case}

The non-degenerate case refers to transverse isotropy that satisfies the following condition:
\begin{align}\label{8.3ak}
\begin{split}
\sqrt {{C_{11}}{C_{33}}}  - {C_{13}} - 2{C_{44}} > 0.
\end{split}
\end{align}
The Green function in \eqref{1.2} can be expressed as \citep{Withers1989}
\begin{align}\label{1.5k}
\begin{split}
{G_{11}}\left( \bfx \right) &= \mathop \sum \limits_{i = 1}^2 2{v_i}H_i\frac{{x_2^{2}R_i^{2} - v_i^2x_1^2x_3^2}}{{{\rho ^4}\left( \bfx \right){R_i}\left( \bfx \right)}} + \frac{1}{{4\pi {C_{44}}{v_3}}}\frac{{x_1^2R_3^2 - v_3^2x_2^2x_3^2}}{{{\rho ^4}\left( \bfx \right){R_3}\left( \bfx \right)}},\\
{G_{12}}\left( \bfx \right) &= \mathop \sum \limits_{i = 1}^2  - 2{v_i}H_i\frac{{{x_1}{x_2}\left( {{\rho ^2} + 2v_i^2x_3^2} \right)}}{{{\rho ^4}\left( \bfx \right){R_i}\left( \bfx \right)}} + \frac{1}{{4\pi {C_{44}}{v_3}}}\frac{{{x_1}{x_2}\left( {{\rho ^2} + 2v_3^2x_3^2} \right)}}{{{\rho ^4}\left( \bfx \right){R_3}\left( \bfx \right)}},\\
{G_{22}}\left( \bfx \right) &= \mathop \sum \limits_{i = 1}^2 2{v_i}H_i\frac{{x_1^2R_i^2 - v_i^2x_2^2x_3^2}}{{{\rho ^4}\left( \bfx \right){R_i}\left( \bfx \right)}} + \frac{1}{{4\pi {C_{44}}{v_3}}}\frac{{x_2^2R_3^2 - v_3^2x_1^2x_3^2}}{{{\rho ^4}\left( \bfx \right){R_3}\left( \bfx \right)}},\\
{G_{13}}\left( \bfx \right) &= -\mathop \sum \limits_{i = 1}^2 {v_i}{A_i}\frac{{{v_i}{x_1}{x_3}}}{{{\rho ^2}\left( \bfx \right){R_i}\left( \bfx \right)}},\;\;\;{G_{23}}\left( \bfx \right) = -\mathop \sum \limits_{i = 1}^2 {v_i}{A_i}\frac{{{v_i}{x_2}{x_3}}}{{{\rho ^2}\left( \bfx \right){R_i}\left( \bfx \right)}},\\
{G_{33}}\left( \bfx \right)& = \mathop \sum \limits_{i = 1}^2 v_i^2\frac{{{k_i}{A_i}}}{{{R_i}\left( \bfx \right)}},
\;\;\;\;\;\;\;\;\;\;\;\;\bfx\ne \textbf{0},
\end{split}
\end{align}
where $\rho \left( \bfx \right) = \sqrt {x_1^2 + x_2^2}$; $ R_i (\bfx)=\sqrt{\rho^2 \left( \bfx \right)+v_i^2 x_3^2 }$; and $ v_i, k_i, A_i$ and $ H_i\ (i=1,2,3)$ are all constants determined by the elastic parameters, i.e.,
\begin{align}\label{1.7k}
\begin{split}
{v_1} \!&=\! \sqrt {\frac{{\left( {\sqrt {{C_{11}}{C_{33}}}  \!- \!{C_{13}}} \right)\left( {\sqrt {{C_{11}}{C_{33}}} \! +\! {C_{13}} \!+\! 2{C_{44}}} \right)}}{{4{C_{33}}{C_{44}}}}}\!+\! \sqrt {\frac{{\left( {\sqrt {{C_{11}}{C_{33}}} \! + \!{C_{13}}} \right)\left( {\sqrt {{C_{11}}{C_{33}}} \! - \!{C_{13}} \!- \!2{C_{44}}} \right)}}{{4{C_{33}}{C_{44}}}}} ,\\
{v_2}\! &= \!\sqrt {\frac{{\left( {\sqrt {{C_{11}}{C_{33}}} \! -\! {C_{13}}} \right)\left( {\sqrt {{C_{11}}{C_{33}}} \! +\! {C_{13}}\! + \!2{C_{44}}} \right)}}{{4{C_{33}}{C_{44}}}}}  \!- \!\sqrt {\frac{{\left( {\sqrt {{C_{11}}{C_{33}}}\!  +\! {C_{13}}} \right)\left( {\sqrt {{C_{11}}{C_{33}}} \! - \! {C_{13}} \!- \!2{C_{44}}} \right)}}{{4{C_{33}}{C_{44}}}}} ,\\
{v_3}\! &=\! \sqrt {\frac{{{C_{11}}\! -\! {C_{12}}}}{{2{C_{44}}}}} ,\;{A_i} \!=\! \frac{{{{\left( { \!-\! 1} \right)}^{i\! +\! 1}}\left( {{C_{13}} \!+\! {C_{44}}} \right)}}{{4\pi {C_{33}}{C_{44}}\left( {v_2^2 \!-\! v_1^2} \right){v_i}}},\;{k_i}\! =\! \frac{{{C_{11}}/v_i^2\! -\! {C_{44}}}}{{{C_{13}} \!+ \! {C_{44}}}},\;{H_i}{\text{}} \!=\! \frac{{{{\left( { - 1} \right)}^i}( {{C_{44}} \!-\! {C_{33}}v_i^2} )}}{{8\pi {C_{33}}{C_{44}}\left( {v_2^2\! -\! v_1^2} \right){v_i^2}}}.
\end{split}
\end{align}
Based on \eqref{1.7k}, it is straightforward to verify that $v_1$ and $v_2$ are the roots of \eqref{root}.


By substituting the Green function in \eqref{1.5k} along with the eigenstrain \eqref{1.11} into \eqref{4.4}, we obtain
\begin{align}\label{1.12k}
\begin{split}
& {\left[ u_{1}\left( \bfx \right), u_{2}\left( \bfx \right),u_{3}\left( \bfx \right)\right]^{T}}\\
= &\mathop \int \limits_\Omega ^{} {\left( {\begin{array}{*{20}{c}}
{ \!\!\! \mathop \sum \limits_{i = 1}^2 \frac{\left(2{H_i}v_i\overline{\sigma}^*_{11}-A_iv_i^2\overline{\sigma}^*_{33}\right){\left( {{x_1} - {y_1}} \right)}}{{R_i^3\left( \bfx - \bfy \right)}}},
\!\!\!\!\!&{ \mathop \sum \limits_{i = 1}^2 \frac{\left(2{H_i}v_i\overline{\sigma}^*_{11}-A_iv_i^2\overline{\sigma}^*_{33}\right){\left( {{x_2} - {y_2}} \right)}}{{R_i^3\left( \bfx - \bfy \right)}}},
\!\!\!\!\!&{{ - \mathop \sum \limits_{i = 1}^2 \frac{{\left({{A_iv^2_i}}\overline{\sigma}^*_{11}+{k_i}{A_i}v_i^4\overline{\sigma}^*_{33}\right)\left( {{x_3} - {y_3}} \right)}}{{R_i^3\left( \bfx - \bfy \right)}}}}
\end{array}} \right)^T}d \bfy \\
 =  &\sum \limits_{i = 1}^2 \mathop \int \limits_\Omega ^{} {\left( {\begin{array}{*{20}{c}}
{ \!\!\! \frac{v_i\left(2{H_i}\overline{\sigma}^*_{11}-A_iv_i\overline{\sigma}^*_{33}\right){\left( {{x_1} - {y_1}} \right)}}{{R_i^3\left( \bfx - \bfy\right)}}},
\!\!\!\!\!&{   \frac{v_i\left(2{H_i}\overline{\sigma}^*_{11}-A_iv_i\overline{\sigma}^*_{33}\right){\left( {{x_2} - {y_2}} \right)}}{{R_i^3\left( \bfx - \bfy\right)}}},
\!\!\!\!\!&{{ -  \frac{{A_i}{\left(\overline{\sigma}^*_{11}+k_iv_i^2\overline{\sigma}^*_{33}\right)\left( {{x_3} - {y_3}} \right)}}{{R_i^3\left( \bfx - \bfy \right)}}}}
\end{array}} \right)^T}d \bfy\\
 =  &- \sum \limits_{i = 1}^2 \bfK^i \cdot {\boldsymbol{\nabla}_\bfx}\mathop \int \limits_\Omega  \frac{1}{{{R_i}\left(\bfx - \bfy \right)}}dV\bfy=\mathcal{L}(\bfx),\;\;\;\bfx\in \Omega,
\end{split}
\end{align}
with $\mathcal{L}(\bfx)$ denoting a linear vector function of $\bfx$ and
\begin{align}\label{1.12bk}
\begin{split}
\bfK^i {:=} \left( {\begin{array}{*{20}{c}}
  v_i\left(2{H_i}\overline{\sigma}^*_{11}-A_iv_i\overline{\sigma}^*_{33}\right)&0&0 \\
  0&v_i\left(2{H_i}\overline{\sigma}^*_{11}-A_iv_i\overline{\sigma}^*_{33}\right)&0 \\
  0&0&{-{A_i}\left(\overline{\sigma}^*_{11}+k_iv_i^2\overline{\sigma}^*_{33}\right)}
\end{array}} \right).
\end{split}
\end{align}

 It follows from \eqref{1.12k} along with \eqref{1.12bk} that
\begin{align}\label{1.14k}
\begin{split}
&\frac{\partial }{{\partial {x_1}}}\left( {\mathop \int \limits_\Omega \frac{{\alpha _1}}{{{R_1}\left(\bfx - \bfy \right)}}d \bfy  + \mathop \int \limits_\Omega  \frac{{\alpha _2}}{{{R_2}\left(\bfx - \bfy \right)}}d \bfy } \!\!\right) = {\mathcal{L}_1}(\bfx),\\
&\frac{\partial }{{\partial {x_2}}}\left( {\mathop \int \limits_\Omega\frac{{\alpha _1}}{{{R_1}\left( \bfx - \bfy \right)}}d \bfy  + \mathop \int \limits_\Omega\frac{{\alpha _2}}{{{R_2}\left(\bfx - \bfy \right)}}d\bfy} \!\!\right) = {\mathcal{L}_2}(\bfx),\\
&\frac{\partial }{{\partial {x_3}}}\left( {\mathop \int \limits_\Omega\frac{{\alpha _1}'}{{{R_1}\left( \bfx - \bfy\right)}}d \bfy + \mathop \int \limits_\Omega \frac{{\alpha _2}'}{{{R_2}\left( \bfx - \bfy \right)}}d\bfy } \!\!\right) = {\mathcal{L}_3}(\bfx),
\end{split}
\end{align}
with ${\alpha _i} {:=} K^i_{11}$ and ${\alpha _i}'{:=}K^i_{33}\;\;(i=1,2)$.

To continue the analysis, we define
\begin{align}\label{1.15k}
\begin{split}
&{\omega _1}( {{x_1},{x_2},{x_3}} ) {:=}  - \mathop \int \limits_{\Omega} \frac{1}{{4\pi {R_1}\left(\bfx - \bfy\right)}}d \bfy ,\;{\omega _2}\left( {{x_1},{x_2},{x_3}} \right) {:=}  - \mathop \int \limits_{\Omega} \frac{1}{{4\pi {R_2}\left(\bfx - \bfy \right)}}d \bfy .
\end{split}
\end{align}
Then by substituting \eqref{1.15k} into \eqref{1.14k}, we obtain
\begin{align}\label{1.16k}
\begin{split}
& - 4\pi \left( {{\alpha _1}{\omega _1}\left( {{x_1},{x_2},{x_3}} \right) + {\alpha _2}{\omega _2}\left( {{x_1},{x_2},{x_3}} \right)} \right) = {q_1}\left( {{x_1},{x_2},{x_3}} \right) + {\varphi _1}\left( {{x_3}} \right),\\
& - 4\pi \left( {{\alpha _1}'{\omega _1}\left( {{x_1},{x_2},{x_3}} \right) + {\alpha _2}'{\omega _2}\left( {{x_1},{x_2},{x_3}} \right)} \right)= {q_2}\left( {{x_1},{x_2},{x_3}} \right) + {\varphi _2}\left( {{x_1},{x_2}} \right).
\end{split}
\end{align}
 where $\varphi_1(x_3 )$ and $\varphi_2(x_1,x_2)$ represent two unknown functions, and $q_1(x_1,x_2,x_3 )$ and $q_2(x_1,x_2,x_3 )$ represent two quadratic functions which are related to $\mathcal{L}_i(x_1,x_2,x_3 )$ through the following conditions:
\begin{align}\label{1.18k}
\begin{split}
&\frac{\partial }{{\partial {x_1}}}{q_1}\left( {{x_1},{x_2},{x_3}} \right) = {\mathcal{L}_1}\left( {{x_1},{x_2},{x_3}} \right),\\
\quad &\frac{\partial }{{\partial {x_2}}}{q_1}\left( {{x_1},{x_2},{x_3}} \right) = {\mathcal{L}_2}\left( {{x_1},{x_2},{x_3}} \right),\\
\quad &\frac{\partial }{{\partial {x_3}}}{q_2}\left( {{x_1},{x_2},{x_3}} \right) = {\mathcal{L}_3}\left( {{x_1},{x_2},{x_3}} \right).
\end{split}
\end{align}

It is always possible to find $\overline{\sigma}^*_{11}$ and $\overline{\sigma}^*_{33}$ that give rise to ${\alpha _1}\ne0,\;\alpha_2=0$, that is,

\begin{align}\label{3333.1k}
\begin{split}
\left\{{\begin{array}{*{20}{c}}
 {2{H_1}v_1\overline{\sigma}^*_{11}-A_1v_1^2\overline{\sigma}^*_{33}={\alpha _1}}\ne0,\\
 {2{H_2}v_2\overline{\sigma}^*_{11}-A_2v_2^2\overline{\sigma}^*_{33}={\alpha _2}=0}.
\end{array}} \right.
\end{split}
\end{align}
The existence of a set of $\overline{\sigma}^*_{11}$ and $\overline{\sigma}^*_{33}$  satisfying \eqref{3333.1k} requires
\begin{align}\label{3333.2k}
\begin{split}
\left|{\begin{array}{*{20}{c}}
  {2{H_1}v_1}&{-A_1v_1^2}\\
  {2{H_2}v_2}&{-A_2v_2^2}
\end{array}} \right|={-\frac{(C_{13}+C_{44})}{16\pi^2C^2_{33}C_{44}(v_2^2-v_1^2)v_1^2v_2^2}\ne 0,}
\end{split}
\end{align}
which is satisfied owing to \eqref{1.8}, \eqref{transconstrain12} and \eqref{1.7k}.

Then by substituting \eqref{3333.1k} into $\eqref{1.16k}_1$, we obtain
\beqs\label{newomegav1}
{\omega _1}\left( x_1,x_2,x_3\right)= -\frac{ {q_1}\left(x_1,x_2,x_3 \right) + {\varphi _1}\left(x_3 \right) }{4\pi \alpha_1}.
\eeqs
In terms of \eqref{newomegav1}, by the transformations introduced in \eqref{cotrans} with $t$ replaced by $v_1$,
the Newtonian potential induced by the inclusion $\Omega'$ that is transformed from the original inclusion $\Omega$ via \eqref{7.2} is

\begin{align}\label{1.20k}
\begin{split}
{N_{\Omega '}}\left( {\bfx}' \right) =&- \int \limits_{\Omega '} \frac{1}{{4\pi | \bfx'-\bfy'  |}}d\bfy'={v_1}{\omega _1}\left( {{x_1}',{x_2}',\frac{{{x_3}'}}{{{v_1}}}} \right)= \frac{{{v_1}\left[ {q_1}\left( {{x_1}',{x_2}',\frac{{{x_3}'}}{{{v_1}}}} \right) + {\varphi _1}\left( {\frac{{{x_3}'}}{{{v_1}}}} \right) \right]}}{4\pi {\alpha_1}}.
\end{split}
\end{align}
Substituting \eqref{1.20k} into \eqref{1.21}  demonstrates that $\varphi_1(x_3 )$ is a constant, linear or quadratic function; thus the Newtonian potential induced by $\Omega'$ is quadratic due to \eqref{1.20k}. Based on  Theorem \ref{TheoremNP}  , we claim that $\Omega'$ can only be of ellipsoidal shape, and so is $\Omega$ due to \eqref{7.2}, which accomplishes the proof of Theorem \ref{transtheorem} for the non-degenerate transversely isotropic medium.

Moreover, we see that for the case ${\alpha _1}\ne0,\;\alpha_2=0$ which results in \eqref{1.20k},
\begin{align}\label{122k.1}
\begin{split}
\alpha_2=v_2\left(2{H_2}\overline{\sigma}^*_{11}-A_2v_2\overline{\sigma}^*_{33}\right)=0\;\;\;\Rightarrow\;\;\;{\overline{\sigma}^*_{33}}=\frac{2H_2}{A_2v_2}\overline{\sigma}^*_{11}.
\end{split}
\end{align}
Substitution of \eqref{1.7k} into \eqref{122k.1} leads to
\begin{align}\label{122k.2}
\begin{split}
\frac{\overline{\sigma}^*_{33}}{\overline{\sigma}^*_{11}}=\frac{2H_2}{A_2v_2}=\frac{C_{11}C_{33}-C_{33}C_{44}v_1^2}{(C_{13}+C_{44})C_{11}},
\end{split}
\end{align}
which is consistent with the ratio \eqref{gamma} of $\overline{\sigma}^*_{33}$ to $\overline{\sigma}^*_{11}$ we require in the main text.

 Further, substituting \eqref{122k.2} back into \eqref{1.12k} leads to
\begin{align}\label{finalsolution1k}
\begin{split}
\bfu\left( \bfx \right)=&-\bfK^1 \cdot {\boldsymbol{\nabla}_\bfx} \int_\Omega \frac{1}{{{R_1}\left(\bfx-\bfy\right)}}d\bfy,
\end{split}
\end{align}
where $\bfK^1$ is diagonal with
\begin{align}\label{122.3k}
\begin{split}
&K^1_{11}=K^1_{22}=\frac{v_1\overline{\sigma}^*_{11}}{4\pi C_{11}},\quad K^1_{33}=\frac{(C_{11}C_{33}-C_{33}C_{44}v_1^2)\overline{\sigma}^*_{11}}{ C_{33}v_1(C_{13}+C_{44})C_{11}}.
\end{split}
\end{align}

Based on  \eqref{122.3k}, it is straightforward to verify that \eqref{finalsolution1k} is exactly the result we obtain in \eqref{finalsolution1} with $v$ replaced by $v_1$, which validates the correctness of \eqref{finalsolution1} derived by using Fourier forms of the Eshelby formalism in the main text.

By following the same procedure from \eqref{122k.1} to \eqref{122.3k}, we can also verify that the displacement $\bfu\left( \bfx \right)$ derived under the condition ${\alpha _1}=0,\;\alpha_2\neq 0$ is the same as that shown in \eqref{finalsolution1}.

\subsection*{(2)\quad Degenerate case}

The degenerate case refers to transverse isotropy that satisfies the following condition:
\begin{align}\label{8.3k}
\begin{split}
\sqrt {{C_{11}}{C_{33}}}  - {C_{13}} - 2{C_{44}} =0,
\end{split}
\end{align}
under which \eqref{root} will admit a unique solution, and the Green function used for the non-degenerate case becomes invalid,  so we must use other explicit expressions of the Green function derived by \cite{Ding2006}, i.e.,
\begin{align}\label{3.2k}
\begin{split}
&{G_{11}}\left( \bfx \right) = \frac{{{{\left( {{R_3}\left( \bfx \right) + {v_3}\left| {{x_3}} \right|} \right){R_3}\left( \bfx \right)}} - x_2^2}}{{4\pi {C_{44}}{v_3}{{\left( {{R_3}\left( \bfx \right) + {v_3}\left| {{x_3}} \right|} \right)}^2}{R_3}\left( \bfx \right)}} + \frac{1}{{8\pi v{R_0}\left( \bfx \right)}}\left( {\frac{1}{{{C_{33}}{v^2}}} + \frac{{{\rho ^2}\left( \bfx \right)}}{{{C_{44}}{{\left( {{R_0}\left( \bfx \right) + v\left| {{x_3}} \right|} \right)}^2}}}} \right) + Ux_1^2,\\
&{G_{12}}\left( \bfx \right) = \frac{{{x_1}{x_2}}}{{4\pi {C_{44}}{v_3}{{\left( {{R_3}\left( \textbf{\emph{x }}\right) + {v_3}\left| {{x_3}} \right|} \right)}^2}{R_3}\left( \bfx \right)}} + U{x_1}{x_2},\\
&{G_{22}}\left( \bfx \right) = \frac{{{{\left( {{R_3}\left( \bfx \right) + {v_3}\left| {{x_3}} \right|} \right){R_3}\left( \bfx \right)}} - x_1^2}}{{4\pi {C_{44}}{v_3}{{\left( {{R_3}\left( \bfx \right) + {v_3}\left| {{x_3}} \right|} \right)}^2}{R_3}\left( \bfx \right)}} + \frac{1}{{8\pi v{R_0}\left(\bfx\right)}}\left( {\frac{1}{{{C_{33}}{v^2}}} + \frac{{{\rho ^2}\left( \bfx \right)}}{{{C_{44}}{{\left( {{R_0}\left( \bfx \right) + v\left| {{x_3}} \right|} \right)}^2}}}} \right) \!\!+ Ux_2^2,\\
&{G_{13}}\left( \bfx \right) =\frac{{\left( {{C_{13}} + {C_{44}}} \right){x_1}{x_3}}}{{8\pi {C_{33}}{C_{44}}vR_0^3\left( \bfx \right)}},\;\;\;{G_{23}}\left( \bfx \right) = \frac{{\left( {{C_{13}} + {C_{44}}} \right){x_2}{x_3}}}{{8\pi {C_{33}}{C_{44}}vR_0^3\left( \bfx \right)}},\\
&{G_{33}}\left( \bfx \right) = \frac{\left( v^2C_{33}+ {C_{44}} \right){\rho ^2}\left( \bfx \right) + 2{C_{33}}v^4x_3^2}{{8\pi {C_{33}}{C_{44}}vR_0^3\left( \bfx \right)}},\;\;\;\;\;\;\;\;\;\bfx\ne \textbf{0},
\end{split}
\end{align}
with
\begin{align}\label{3.1k}
\begin{split}
&v =\sqrt {\frac{{\left( {\sqrt {{C_{11}}{C_{33}}}  - {C_{13}}} \right)\left( {\sqrt {{C_{11}}{C_{33}}}  + {C_{13}} + 2{C_{44}}} \right)}}{{4{C_{33}}{C_{44}}}}{\text{}}}  =\left(\frac{C_{11}}{C_{33}}\right)^{\frac{1}{4}},\;\;v_3=\sqrt{\frac{C_{11}-C_{12}}{2C_{44}}},\\
&{R_0}\left( \bfx \right) = \sqrt {{\rho ^2}(\bfx) + {v^2}x_3^2},\ \ \ U = \frac{{ - 1}}{{8\pi {C_{33}}{v^3}R_0^3\left( \bfx \right)}} + \frac{{2{v^2}x_3^2{{\left( {{R_0}\left( x \right) + v\left| {{x_3}} \right|} \right)}^2} - {\rho ^4}\left( \bfx \right)}}{{8\pi {C_{44}}{v}R_0^3\left( \bfx \right){{\left( {{R_0}\left( \bfx \right) + v\left| {{x_3}} \right|} \right)}^4}}}.
\end{split}
\end{align}

 We stress that based on \eqref{3.1k}, it can be verified that $v$ is the unique root of \eqref{root} under the condition \eqref{8.3k}.

By substituting the Green function in \eqref{3.2k} along with the eigenstrain \eqref{1.11} into \eqref{4.4}, we obtain
\begin{align}\label{3.3kk}
\begin{split}
\left[ {\begin{array}{*{20}{c}}
 u_{1}\left( \bfx \right)\\
 u_{2}\left( \bfx \right)\\
 u_{3}\left( \bfx \right)\end{array}}\right]=&-\int \limits_\Omega\left[ {\begin{array}{*{20}{c}}
 \frac{\left((C_{33}v^2+C_{44})\overline{\sigma}^*_{11}-(C_{13}+C_{44})v^2\overline{\sigma}^*_{33}\right)\left( {{x_1} - {y_1}} \right)}{8\pi C_{33}C_{44}v^3{R_0^3\left( \bfx-\bfy\right)}}\\
 \frac{\left((C_{33}v^2+C_{44})\overline{\sigma}^*_{11}-(C_{13}+C_{44})v^2\overline{\sigma}^*_{33}\right)\left( {{x_2} - {y_2}} \right)}{8\pi C_{33}C_{44}v^3{R_0^3\left( \bfx-\bfy\right)}}\\
\frac{\left((C_{13}+C_{44})\overline{\sigma}^*_{11}-(C_{33}v^2-3C_{44})v^2\overline{\sigma}^*_{33}\right)\left( {{x_3} - {y_3}} \right)}{8\pi C_{33}C_{44}v{R_0^3\left( \bfx-\bfy\right)}}
\end{array}}\right]d \bfy\\
+&\int \limits_\Omega\left[ {\begin{array}{*{20}{c}}
 \frac{3\left((C_{33}v^2-C_{44})\overline{\sigma}^*_{11}-(C_{13}+C_{44})v^2\overline{\sigma}^*_{33}\right)\left( {{x_1} - {y_1}} \right)\left(x_3-y_3\right)^2}{8\pi C_{33}C_{44}v{R_0^5\left( \bfx-\bfy \right)}}\\
 \frac{3\left((C_{33}v^2-C_{44})\overline{\sigma}^*_{11}-(C_{13}+C_{44})v^2\overline{\sigma}^*_{33}\right)\left( {{x_2} - {y_2}} \right)\left(x_3-y_3\right)^2}{8\pi C_{33}C_{44}v{R_0^5\left( \bfx-\bfy\right)}}\\
 \frac{3\left(\left(C_{13}+C_{44}\right)v^2\overline{\sigma}^*_{11}-\left(C_{33}v^2-C_{44}\right)v^4\overline{\sigma}^*_{33}\right)\left(x_3-y_3\right)^3}{8\pi C_{33}C_{44}v{R_0^5\left( \bfx-\bfy\right)}}
\end{array}}\right]d\bfy \\
=&\bfK^1 \cdot {\boldsymbol{\nabla}_\bfx}\mathop \int \limits_\Omega\frac{1}{{{R_0}\left( \bfx-\bfy\right)}}d\bfy+\bfK^2 \cdot {\boldsymbol{\nabla}_\bfx}\mathop \int \limits_\Omega\frac{(x_3-y_3)^2}{{{R^3_0}\left(\bfx-\bfy\right)}}d\bfy =\mathcal{L}(\bfx),\;\;\;\bfx\in \Omega,
\end{split}
\end{align}
with $\mathcal{L}(\bfx)$ denoting a linear vector function of $\bfx$, and
\begin{align}\label{3.3ak}
\begin{split}
\bfK^1 {:=}\frac{(C_{33}v^2+C_{44})\overline{\sigma}^*_{11}-(C_{13}+C_{44})v^2\overline{\sigma}^*_{33}}{8\pi C_{33}C_{44}v^3}\cdot
\left( {\begin{array}{*{20}{c}}
   1&0&0 \\
  0&1&0 \\
  0&0&{\frac{-\left(C_{13}+C_{44}\right)\overline{\sigma}^*_{11}+(C_{33}v^2+C_{44})v^2\overline{\sigma}^*_{33}}{(C_{33}v^2+C_{44})\overline{\sigma}^*_{11}-(C_{13}+C_{44})v^2\overline{\sigma}^*_{33}}}
\end{array}} \right)
\end{split}
\end{align}
and
\begin{align}\label{3.3bk}
\begin{split}
\bfK^2 {:=}-\frac{(C_{33}v^2-C_{44})\overline{\sigma}^*_{11}-(C_{13}+C_{44})v^2\overline{\sigma}^*_{33}}{8\pi C_{33}C_{44}v}\cdot
\left( {\begin{array}{*{20}{c}}
   1&0&0 \\
  0&1&0 \\
  0&0&\frac{\left(C_{13}+C_{44}\right)\overline{\sigma}^*_{11}-\left(C_{33}v^2-C_{44}\right)v^2\overline{\sigma}^*_{33}}{(C_{33}v^2-C_{44})\overline{\sigma}^*_{11}-(C_{13}+C_{44})v^2\overline{\sigma}^*_{33}}
\end{array}} \right).
\end{split}
\end{align}

It follows from \eqref{3.3kk} that
\begin{align}\label{3.4k}
\begin{split}
&\frac{\partial }{{\partial {x_1}}}\left( \beta_1\int \limits_\Omega\frac{1}{{{R_0}\left( \bfx-\bfy \right)}}d\bfy + \beta_2 \int \limits_\Omega \frac{{{{\left( {{x_3} - {y_3}} \right)}^2}}}{{R_0^3\left( \bfx-\bfy\right)}}d\bfy \right) = {\mathcal{L}_1}\left( {{x_1},{x_2},{x_3}} \right), \\
&\frac{\partial }{{\partial {x_2}}}\left( \beta_1\int \limits_\Omega\frac{1}{{{R_0}\left( \bfx-\bfy \right)}}d\bfy + \beta_2 \int \limits_\Omega \frac{{{{\left( {{x_3} - {y_3}} \right)}^2}}}{{R_0^3\left( {\textbf{\emph{x - y}}} \right)}}d\bfy \right) = {\mathcal{L}_2}\left( {{x_1},{x_2},{x_3}} \right), \\
&\frac{\partial }{{\partial {x_3}}}\left( \beta_1'\int \limits_\Omega\frac{1}{{{R_0}\left( \bfx-\bfy\right)}}d\bfy + \beta_2' \int \limits_\Omega \frac{{{{\left( {{x_3} - {y_3}} \right)}^2}}}{{R_0^3\left( \bfx-\bfy \right)}}d\bfy \right) = {\mathcal{L}_3}\left( {{x_1},{x_2},{x_3}} \right),
\end{split}
\end{align}
with
\begin{align}\label{4444.0k}
\begin{split}
&{\beta _1} {:=} \frac{(C_{33}v^2+C_{44})\overline{\sigma}^*_{11}-(C_{13}+C_{44})v^2\overline{\sigma}^*_{33}}{8\pi C_{33}C_{44}v^3}, \quad{\beta _2}{:=}-\frac{(C_{33}v^2-C_{44})\overline{\sigma}^*_{11}-(C_{13}+C_{44})v^2\overline{\sigma}^*_{33}}{8\pi C_{33}C_{44}v},\\
&{\beta _1'}{:=}\frac{-\left(C_{13}+C_{44}\right)\overline{\sigma}^*_{11}+(C_{33}v^2+C_{44})v^2\overline{\sigma}^*_{33}}{8\pi C_{33}C_{44}v^3},\quad{\beta _2'}{:=}-\frac{\left(C_{13}+C_{44}\right)\overline{\sigma}^*_{11}-\left(C_{33}v^2-C_{44}\right)v^2\overline{\sigma}^*_{33}}{8\pi C_{33}C_{44}v}.
\end{split}
\end{align}

Similarly, we define
\begin{align}\label{4444.1k}
\begin{split}
{\omega _0}( {{x_1},{x_2},{x_3}} ) {:=}  - \int \limits_{\Omega} \frac{1}{{4\pi {R_0}\left(\bfx - \bfy \right)}}d\bfy,\;\;\tilde{{\omega }}_0( {{x_1},{x_2},{x_3}} ) {:=}  - \int \limits_{\Omega} \frac{(x_3-y_3)^2}{{4\pi {R^3_0}\left(\bfx - \bfy\right)}}d\bfy.
\end{split}
\end{align}
Then by substituting \eqref{4444.1k} into \eqref{3.4k}, we obtain
\begin{align}\label{3.5k}
\begin{split}
-4\pi(\beta_1{\omega _0}( {{x_1},{x_2},{x_3}} )+\beta_2\tilde{\omega _0}( {{x_1},{x_2},{x_3}} ))&=q_1(x_1,x_2,x_3 )+\varphi_1(x_3 ),\\
-4\pi(\beta_1'{\omega _0}( {{x_1},{x_2},{x_3}} )+\beta_2'\tilde{\omega _0}( {{x_1},{x_2},{x_3}} ))&=q_2(x_1,x_2,x_3 )+\varphi_2(x_1,x_2),
\end{split}
\end{align}
where $\varphi_1(x_3 )$ and $\varphi_2(x_1,x_2)$ still represent two unknown functions, and $q_1(x_1,x_2,x_3 )$ and $q_2(x_1,x_2,x_3 )$ still represent two quadratic functions restricted by \eqref{1.18k}.

We can always arrive at ${\beta _1}\ne0,\;\beta_2=0$ by choosing
$\overline{\sigma}^*_{11}$ and $\overline{\sigma}^*_{33}$, that is,
\begin{align}\label{5555.1k}
\begin{split}
\left\{{\begin{array}{*{20}{c}}
{(C_{33}v^2+C_{44})\overline{\sigma}^*_{11}-(C_{13}+C_{44})v^2\overline{\sigma}^*_{33}}={\beta _1}\ne0,\\
{(C_{33}v^2-C_{44})\overline{\sigma}^*_{11}-(C_{13}+C_{44})v^2\overline{\sigma}^*_{33}}={\beta _2}=0.
\end{array}} \right.
\end{split}
\end{align}
The existence of the solution to  \eqref{5555.1k} requires that
\begin{align}\label{5555.2k}
\begin{split}
\left|{\begin{array}{*{20}{c}}
  {C_{33}v^2+C_{44}}&{-(C_{13}+C_{44})v^2}\\
  {C_{33}v^2-C_{44}}&{-(C_{13}+C_{44})v^2}
\end{array}} \right|=-2C_{44}(C_{13}+C_{44})v^4\ne 0,
\end{split}
\end{align}
which is satisfied owing to  \eqref{1.8}, \eqref{transconstrain12} and \eqref{3.1k}.

Then, through substitution of \eqref{5555.1k} into $\eqref{3.5k}_1$, we obtain

\begin{align}\label{3.5bk}
\begin{split}
{\omega _0}\left( {{x_1},{x_2},{x_3}} \right)=-\frac{{q_1}\left( {{x_1},{x_2},{x_3}} \right) + {\varphi _1}\left( {{x_3}} \right)}{4\pi\beta_1}.
\end{split}
\end{align}
According to \eqref{3.5bk}, by the transformations introduced in \eqref{cotrans} with $t$ replaced by $v$,
the Newtonian potential induced by the inclusion $\Omega'$ that is transformed from the original inclusion $\Omega$ by \eqref{7.2} is
\begin{align}\label{3.6k}
\begin{split}
{N_{\Omega '}}\left( {\bfx}' \right)=- \frac{1}{{4\pi }}\mathop \int \limits_{\Omega '}\frac{1}{{|\bfx' - \bfy'|}}d {\bfy}' ={v}{\omega _0}\left( {{x_1}',{x_2}',\frac{{{x_3}'}}{{{v}}}} \right) = \frac{{{v}\left[ {q_1}\left( {{x_1}',{x_2}',\frac{{{x_3}'}}{{{v}}}} \right) + {\varphi _1}\left( {\frac{{{x_3}'}}{{{v}}}} \right) \right]}}{4\pi {\beta_1}}.
\end{split}
\end{align}

Likewise, substitution of \eqref{3.6k} into \eqref{1.21} verifies that $\varphi_1(x_3 )$ is a constant, linear or quadratic function, which indicates that the right-hand side of \eqref{3.6k} is quadratic. Therefore, based on Theorem \ref{TheoremNP}  , we claim that $\Omega'$ can only be of ellipsoidal shape, and thus $\Omega$ must be ellipsoidal due to \eqref{7.2}, which fulfills the proof of Theorem \ref{transtheorem}  for the degenerate transversely isotropic medium.

Moreover, we see that for the case  ${\beta _1}\ne0,\;\beta_2=0$ which results in \eqref{3.6k},
\begin{align}\label{122.4k}
\begin{split}
\beta_2=-\frac{(C_{33}v^2-C_{44})\overline{\sigma}^*_{11}-(C_{13}+C_{44})v^2\overline{\sigma}^*_{33}}{8\pi C_{33}C_{44}v}=0\;\;\;\Rightarrow\;\;\;{\overline{\sigma}^*_{33}}=\frac{(C_{33}v^2-C_{44})}{(C_{13}+C_{44})v^2}\overline{\sigma}^*_{11}.
\end{split}
\end{align}

Then substitution of \eqref{3.1k} into \eqref{122.4k} generates
\begin{align}\label{122.5k}
\begin{split}
\frac{\overline{\sigma}^*_{33}}{\overline{\sigma}^*_{11}}=\frac{(C_{33}v^2-C_{44})}{(C_{13}+C_{44})v^2}=\frac{C_{11}C_{33}-C_{33}C_{44}v^2}{(C_{13}+C_{44})C_{11}},
\end{split}
\end{align}
which is consistent with the ratio \eqref{gamma} of $\overline{\sigma}^*_{33}$ to $\overline{\sigma}^*_{11}$ we require in the main text.

Further, substituting \eqref{122.5k} back into \eqref{3.3kk} leads to \eqref{finalsolution1k} with $v_1$ replaced by $v$, which has been proved to coincide with \eqref{finalsolution1} and thus demonstrates the correctness of \eqref{finalsolution1} derived by using Fourier forms of the Eshelby formalism in the main text.

\setcounter{equation}{0}
\renewcommand\theequation{B.\arabic{equation}}
\renewcommand\thetheorem{B.\arabic{theorem}}

\section*{B. \quad Another theorem that proves the generalized weak version for the transversely isotropic material based on material symmetry}

The material symmetry brings a new dimension regarding the investigation into the Eshelby conjecture. Here we present another theorem that proves the generalized weak version for the transversely isotropic material, i.e., 
\begin{theorem}\label{transnewtheorem}
{Let $\Omega\subset \mathbb{R}^{3}$ be a { one-component connected bounded open domain} with a Lipschitz boundary.}
 There exist combinations $( \boldsymbol{\overline{\varepsilon}^*},\bfC^{(1)})$ and $( \boldsymbol{\overline{\varepsilon}^*},\bfC^{(2)})$,  where $\boldsymbol{\overline{\varepsilon}^*}$ is a nonzero uniform eigenstrain defined by \eqref{1.11}, and $\bfC^{(1)}$ and $\bfC^{(2)}$
are {two  linearly independent elastic tensors with the transversely isotropic symmetry}, such that  \eqref{4.4} holds for
$( \boldsymbol{\overline{\varepsilon}^*},\bfC^{(1)})$ and $( \boldsymbol{\overline{\varepsilon}^*},\bfC^{(2)})$ simultaneously,
if and only if $\Omega$ is of ellipsoidal shape.
\end{theorem}

Theorem \ref{transnewtheorem} means
\begin{equation}\label{77777.2k}
\begin{split}
\exists\;  \boldsymbol{\overline{\varepsilon} }^*\in \{\boldsymbol{{\varepsilon }}^*\}\;, \text{and} \;\; \bfC^{(1)},\bfC^{(2)}\in \{\bfC^{\;\text{trans}}\}, \;s.t.\;\;\;\{F(\boldsymbol{\overline{\varepsilon }}^*, \bfC^{(1)})\}\cap \{F(\boldsymbol{\overline{\varepsilon }}^*, \bfC^{(2)})\}=\{ E\},
\end{split}
\end{equation}

{It is straightforward to see
\begin{align}\label{222.1k}
\begin{split}
\cap \{\{F(\boldsymbol{\varepsilon ^*}, \bfC)\}\}\subseteq\{F(\boldsymbol{\overline{\varepsilon }}^*, \bfC^{(1)})\}\cap \{F(\boldsymbol{\overline{\varepsilon }}^*, \bfC^{(2)})\}.\\
\end{split}
\end{align}
Substitution of \eqref{222.1k} into \eqref{77777.2k} yields \eqref{assist4}. Then combination of \eqref{assist4} with \eqref{assist3} leads to \eqref{weak}. }

We also consider two cases concerning the possible degeneracy of the elastic parameters of the transversely isotropic material.

\subsection*{(1)\quad Non-degenerate case}

For the case \eqref{8.3ak},
we can start the analysis from \eqref{1.16k}. { To prove \eqref{77777.2k},
 we then choose two different transversely isotropic materials whose elastic tensors are linearly independent by the
procedure below.}

1. Firstly, we choose material 1 and fix its elastic tensor, so the five independent elastic parameters of material 1 are regarded as constants in the subsequent derivations. In particular, the material parameters $v_{1}$ and $v_{2}$  defined in \eqref{1.7k} are denoted by  $v_1^{(1)}, v_{2}^{(1)}$, with the superscript (1) representing material 1.

2. Secondly,  we choose material 2, whose elastic parameters satisfy, with the superscript (2) representing material 2,
\beqs\label{02.1kkk}
\begin{split}
v_i^{(2)}=v_i^{(1)}\quad(i=1,2).
\end{split}
\eeqs

 Likewise, all of the parameters in the sequel will be distinguished by the superscripts (1) and (2) that correspond to material 1 and 2, respectively, except that for brevity, we will re-express $v_1^{(1)}$ and $v_1^{(2)}$  as $v_1$ and re-express $v_2^{(1)}$ and $v_2^{(2)}$ as $v_2$ owing to \eqref{02.1kkk},  which means
  \beqs\label{02.1k}
  v_1\equiv v_1^{(1)}=v_1^{(2)},\quad v_2\equiv v_2^{(1)}=v_2^{(2)}.
  \eeqs

Under the condition \eqref{02.1k}, we can at most make three of the five independent elastic parameters of material 2 different from those of material 1, which { helps us to ensure the linear independence of the elastic tensors of material 1 and material 2}.

Then based on \eqref{02.1k}, $\eqref{1.16k}_1$ for material 1 and material 2 are
\begin{align}\label{77777.3k}
\begin{split}
 - 4\pi \left( {{\alpha^{(1)} _1}{\omega _1}\left( {{x_1},{x_2},{x_3}} \right) + {\alpha^{(1)} _2}{\omega _2}\left( {{x_1},{x_2},{x_3}} \right)} \right) = {q_1}^{(1)}\left( {{x_1},{x_2},{x_3}} \right) + {\varphi _1}^{(1)}\left( {{x_3}} \right),\\
 - 4\pi \left( {{\alpha^{(2)} _1}{\omega _1}\left( {{x_1},{x_2},{x_3}} \right) + {\alpha^{(2)} _2}{\omega _2}\left( {{x_1},{x_2},{x_3}} \right)} \right) = {q_1}^{(2)}\left( {{x_1},{x_2},{x_3}} \right) + {\varphi _1}^{(2)}\left( {{x_3}} \right),
\end{split}
\end{align}
where ${\omega _1}\left( {{x_1},{x_2},{x_3}} \right)$ and ${\omega _2}\left( {{x_1},{x_2},{x_3}} \right)$ defined in \eqref{1.15k} only depend on $\Omega$ now since $v_1$ and $v_2$ are fixed constants here.

It can be derived from \eqref{77777.3k} that

\begin{align}\label{77777.4k}
\begin{split}
{\omega _1}\left( {{x_1},{x_2},{x_3}} \right)=-\frac{q_3\left( {{x_1},{x_2},{x_3}} \right) + \varphi_3\left( {{x_3}} \right) }{4\pi\left(\alpha^{(1)}_1\alpha^{(2)}_2-\alpha^{(1)}_2\alpha^{(2)}_1\right)},
\end{split}
\end{align}
where $q_3\left( {{x_1},{x_2},{x_3}} \right){:=}\alpha^{(2)}_2{q_1}^{(1)}\left( {{x_1},{x_2},{x_3}} \right)-\alpha^{(1)}_2{q_1}^{(2)}\left( {{x_1},{x_2},{x_3}} \right)$ and $\varphi_3\left( {{x_3}} \right){:=}\alpha^{(2)}_2$ ${\varphi^{(1)} _1}$ $( {{x_3}} )-\alpha^{(1)}_2$ ${\varphi^{(2)} _1}\left( {{x_3}} \right)$.
It is noted that the validity of \eqref{77777.4k} requires that $\alpha^{(1)}_1\alpha^{(2)}_2-\alpha^{(1)}_2\alpha^{(2)}_1\neq 0$, which will be shown to be satisfied by the selection of material 2.

According to \eqref{77777.4k}, by the transformations introduced in \eqref{cotrans} with $t$ replaced by $v_1$,
the Newtonian potential induced by the inclusion $\Omega'$ that is transformed from the original inclusion $\Omega$ by \eqref{7.2} is
\begin{align}\label{01.2k}
\begin{split}
{N_{\Omega '}}\left( {\bfx}' \right) &= - \frac{1}{{4\pi }}\mathop \int \limits_{\Omega '}\frac{1}{{|\bfx' - \bfy'|}}dV\left( {\bfy}' \right)={v_1}{\omega _1}\left( {{x_1}',{x_2}',\frac{{{x_3}'}}{{{v_1}}}} \right) = \frac{{{v_1}\left[ {q_3}\left( {{x_1}',{x_2}',\frac{{{x_3}'}}{{{v_1}}}} \right) + {\varphi _3}\left( {\frac{{{x_3}'}}{{{v_1}}}} \right) \right]}}{4\pi\left(\alpha^{(1)}_1\alpha^{(2)}_2-\alpha^{(1)}_2\alpha^{(2)}_1\right)}.
\end{split}
\end{align}

Then substituting \eqref{01.2k} into \eqref{1.21} generates the constant, linear or quadratic form of $\varphi_1\left( {{x_3}} \right)$. Similarly, the quadratic form of the Newtonian potential induced by $\Omega'$ is obtained  owing to \eqref{01.2k}. In terms of Theorem \ref{TheoremNP}, we draw the conclusion that $\Omega'$ must be an ellipsoid, and so is $\Omega$ due to \eqref{7.2}, which will prove Theorem \ref{transnewtheorem} for the non-degenerate transversely isotropic material.

However, it is left to prove that $\alpha^{(1)}_1\alpha^{(2)}_2-\alpha^{(1)}_2\alpha^{(2)}_1\neq 0$ can be realized, that is,

\begin{align}\label{77777.6k}
\begin{split}
\exists\;\; \alpha^{(i)}_j=2H^{(i)}_jv^{(i)}_j\overline{\sigma}^*_{11}-A^{(i)}_j{v^{(i)}_j}^2\overline{\sigma}^*_{33},\;\;s.t.\;\;\alpha^{(1)}_1\alpha^{(2)}_2-\alpha^{(1)}_2\alpha^{(2)}_1\ne 0\;\;(i,j=1,2).
\end{split}
\end{align}

The inequality  in \eqref{77777.6k} can be reformulated in the expression of the elastic parameters and the eigenstresses
\begin{align}\label{77777.9k}
\begin{split}
{2\left(C^{(2)}_{33}C^{(1)}_{44}-C^{(1)}_{33}C^{(2)}_{44}\right){\overline{\sigma}^*_{11}}^2+
\left(C^{(2)}_{44}\left(C^{(1)}_{13}+C^{(1)}_{44}\right)-C^{(1)}_{44}\left(C^{(2)}_{13}+
C^{(2)}_{44}\right)\right){\overline{\sigma}^*_{11}}{\overline{\sigma}^*_{33}}\neq0.}
\end{split}
\end{align}
In terms of \eqref{02.1k}, our goal is to verify that when material 1 is chosen and fixed, we can always choose the elastic parameters of material 2 yielding \eqref{77777.9k}.

It can be derived form \eqref{02.1k} and \eqref{1.7k} that
\begin{align}\label{03.1k}
\begin{split}
\frac{C^{(2)}_{11}}{C^{(2)}_{33}}&=v_1^2v_2^2,\;\;\;\;\;\;\frac{C^{(2)}_{11}C^{(2)}_{33}+{C^{(2)}_{44}}^2-(C^{(2)}_{13}+C^{(2)}_{44})^2}{C^{(2)}_{33}C^{(2)}_{44}}=v_1^2+v_2^2.
\end{split}
\end{align}
Since $v_1$ and $v_2$ are fixed constants, $C^{(2)}_{11}$ and $C^{(2)}_{13}$ can be determined via \eqref{03.1k} once $C^{(2)}_{12},C^{(2)}_{33}$ and $C^{(2)}_{44}$ are chosen.
Thus we take $C^{(2)}_{12},C^{(2)}_{33}$ and $C^{(2)}_{44}$ as three independent elastic parameters of material 2.
Given this, we let {  $\frac{C^{(2)}_{12}}{C^{(2)}_{44}}\neq \frac{C^{(1)}_{12}}{C^{(1)}_{44}}$, which guarantees the linear independence of the elastic tensors of material 1 and material 2}, and fix $C^{(2)}_{12}$ and $ C^{(2)}_{44}$, which means only $C^{(2)}_{33}$ remains to be chosen.

By substituting \eqref{03.1k} into \eqref{77777.9k}, we obtain

\begin{align}\label{77777.10k}
\begin{split}
&2\left(C^{(2)}_{33}C^{(1)}_{44}-C^{(1)}_{33}C^{(2)}_{44}\right){\overline{\sigma}^*_{11}}\\
&+\left(C^{(2)}_{44}\left(C^{(1)}_{13}+C^{(1)}_{44}\right)-C^{(1)}_{44}\sqrt{v_1^2v_2^2{C^{(2)}_{33}}^2+{C^{(2)}_{44}}^2-(v_1^2+v_2^2)C^{(2)}_{33}C^{(2)}_{44}}\right){\overline{\sigma}^*_{33}}\neq0,
\end{split}
\end{align}
which is a nonlinear inequality only with respect to  $C^{(2)}_{33}$, where $C^{(1)}_{13},C^{(1)}_{33},C^{(1)}_{44},C^{(2)}_{44},v_1,v_2$ are all treated as constants.

If we can find some $C^{(2)}_{33}$ which makes (\ref{77777.10k}) valid, we will complete the proof of  \eqref{77777.2k}.
If we assume that \eqref{77777.10k} is invalid for $C^{(2)}_{33}$, we can get

\begin{align}\label{77777.12k}
\begin{split}
\forall C^{(2)}_{33}>0,\;\;\; &{C^{(1)}_{44}}^2\left(4{\overline{\sigma}^*_{11}}^2-v_1^2v_2^2{\overline{\sigma}^*_{33}}^2\right){C^{(2)}_{33}}^2-{C^{(1)}_{44}}\left(4\left({C^{(2)}_{44}}\left({C^{(1)}_{13}}+{C^{(1)}_{44}}\right){\overline{\sigma}^*_{33}}-2{C^{(1)}_{33}}{C^{(2)}_{44}}{\overline{\sigma}^*_{11}}\right){\overline{\sigma}^*_{11}}\right.\\
&\left.-(v_1^2+v_2^2){C^{(1)}_{44}}{C^{(2)}_{44}}{\overline{\sigma}^*_{33}}^2\right){C^{(2)}_{33}}+\left({C^{(2)}_{44}}\left({C^{(1)}_{13}}+{C^{(1)}_{44}}\right){\overline{\sigma}^*_{33}}-2{C^{(1)}_{33}}{C^{(2)}_{44}}{\overline{\sigma}^*_{11}}\right)^2\\
&-{C^{(1)}_{44}}^2{C^{(2)}_{44}}^2{\overline{\sigma}^*_{33}}^2=0,
\end{split}
\end{align}
Note that $\overline{\sigma}^*_{11}$ and $\overline{\sigma}^*_{33}$ in \eqref{1.11} are required not to be equal to zero simultaneously. Hence, according to \eqref{02.1k} and \eqref{1.7k},  we can guarantee \beqs\label{fonditionforc33}
{C^{(1)}_{44}}^2\left(4{\overline{\sigma}^*_{11}}^2-v_1^2v_2^2{\overline{\sigma}^*_{33}}^2\right)\neq 0
\eeqs
by the initial selection of material 1.
Given \eqref{fonditionforc33}, the quadratic equation in \eqref{77777.12k} has a limited number of roots, and it is possible to choose $C^{(2)}_{33}>0$ other than these roots to
make the inequality \eqref{77777.10k} hold.

Thus, the proof of Theorem \ref{transnewtheorem} for the non-degenerate transversely isotropic medium is completed.

\subsection*{(2)\quad Degenerate case}

For the case \eqref{8.3k}, we can start our derivations from $\eqref{3.5k}$. Then the steps of the proof are the same as those for the non-degenerate case. We choose material 1 and fix its four independent elastic parameters, and then choose material 2 { whose elastic tensor is linearly independent of that of material 1} according to the needs for achieving the proof of \eqref{77777.2k}.
To this end, the material parameter $v$  defined in \eqref{3.1k} is denoted by  $v^{(1)}$ and $v^{(2)}$ for the two materials, respectively,
and we let
\beqs\label{06.1k}
\begin{split}
v^{(1)}=v^{(2)}\equiv v.
\end{split}
\eeqs

Under the condition \eqref{06.1k}, we can at most make three of the four independent elastic parameters of material 2 different from those of  material 1. Based on \eqref{06.1k}, we can derive from $\eqref{3.5k}_1$ that
\begin{align}\label{88888.1k}
\begin{split}
-4\pi(\beta^{(1)}_1{\omega _0}( {{x_1},{x_2},{x_3}} )+\beta^{(1)}_2\tilde{\omega _0}( {{x_1},{x_2},{x_3}} ))&=q^{(1)}_1(x_1,x_2,x_3 )+\varphi^{(1)}_1(x_3 ),\\
-4\pi(\beta^{(2)}_1{\omega _0}( {{x_1},{x_2},{x_3}} )+\beta^{(2)}_2\tilde{\omega _0}( {{x_1},{x_2},{x_3}} ))&=q^{(2)}_1(x_1,x_2,x_3 )+\varphi^{(2)}_1(x_3 ),
\end{split}
\end{align}
where $\tilde{\omega _0}( {{x_1},{x_2},{x_3}} ))$ defined in \eqref{4444.1k} only depends on $\Omega$ since $v$ is now a constant that is fixed.

It can be derived from \eqref{88888.1k} that

\begin{align}\label{88888.2k}
\begin{split}
{\omega _0}\left( {{x_1},{x_2},{x_3}} \right)=-\frac{q_3\left( {{x_1},{x_2},{x_3}} \right) + \varphi_3\left( {{x_3}} \right) }{4\pi\left(\beta^{(1)}_1\beta^{(2)}_2-\beta^{(1)}_2\beta^{(2)}_1\right)},
\end{split}
\end{align}
where $q_3\left( {{x_1},{x_2},{x_3}} \right){:=}\beta^{(2)}_2{q_1}^{(1)}\left( {{x_1},{x_2},{x_3}} \right)-\beta^{(1)}_2{q_1}^{(2)}\left( {{x_1},{x_2},{x_3}} \right)$ and $\varphi_3$ $\left( {{x_3}} \right)$ $=\beta^{(2)}_2$ ${\varphi^{(1)} _1}$ $\left( {{x_3}} \right)-\beta^{(1)}_2$ ${\varphi^{(2)} _1}\left( {{x_3}} \right)$.
{ Here we also see that $\beta^{(1)}_1\beta^{(2)}_2-\beta^{(1)}_2\beta^{(2)}_1$ cannot be zero, which will be shown in the sequel.}

According to \eqref{88888.2k}, by the transformations introduced in \eqref{cotrans} with $t$ replaced by $v$,
the Newtonian potential induced by the inclusion $\Omega'$ that is transformed from the original inclusion $\Omega$ by \eqref{7.2} is
\begin{align}\label{88888.3k}
\begin{split}
{N_{\Omega '}}\left( {\bfx}' \right)=- \frac{1}{{4\pi }}\mathop \int \limits_{\Omega '}\frac{1}{{|\bfx' - \bfy'|}}d {\bfy}' = {v}{\omega _0}\left( {{x_1}',{x_2}',\frac{{{x_3}'}}{{{v}}}} \right) = \frac{{{v}\left[ {q_3}\left( {{x_1}',{x_2}',\frac{{{x_3}'}}{{{v}}}} \right) + {\varphi _3}\left( {\frac{{{x_3}'}}{{{v}}}} \right) \right]}}{4\pi\left(\beta^{(1)}_1\beta^{(2)}_2-\beta^{(1)}_2\beta^{(2)}_1\right)}.
\end{split}
\end{align}

Similarly, via substitution of \eqref{88888.3k} into \eqref{1.21}, we see that $\varphi_3\left( {{x_3}} \right)$ is a constant, linear or quadratic function of $x_3$. Given this, comparison of \eqref{88888.3k} with Theorem \ref{TheoremNP} yields that $\Omega'$ can only be of ellipsoidal shape, which means that $\Omega $ can only be of ellipsoidal shape due to \eqref{7.2}. Thus the proof of Theorem \ref{transnewtheorem}  is achieved for the degenerate transversely isotropic material.

However, the proof of Theorem \ref{transnewtheorem} has not been completed yet unless

\begin{align}\label{88888.4k}
\begin{split}
&\exists\;\; \beta^{(i)}_j= \frac{(C_{44}-(-1)^jC_{33}v^2)v^{2j-2}\overline{\sigma}^*_{11}+(-1)^j(C_{13}+C_{44})v^{2j}\overline{\sigma}^*_{33}}{8\pi C_{33}C_{44}v^{3}},\;\;\\
&s.t.\;\;\beta^{(1)}_1\beta^{(2)}_2-\beta^{(1)}_2\beta^{(2)}_1\ne 0\;\;(i,j=1,2).
\end{split}
\end{align}

The inequality in \eqref{88888.4k} can be expressed in the form
\begin{align}\label{88888.6k}
\begin{split}
\left(C^{(2)}_{33}C^{(1)}_{44}-C^{(1)}_{33}C^{(2)}_{44}\right){\overline{\sigma}^*_{11}}^2-\left(C^{(2)}_{13}C^{(1)}_{44}-C^{(1)}_{13}C^{(2)}_{44}\right)\overline{\sigma}^*_{11}\overline{\sigma}^*_{33}\neq0,
\end{split}
\end{align}

Based on \eqref{06.1k},  our goal is to verify that when material 1 is chosen and fixed, we can always choose the elastic parameters of material 2 yielding \eqref{88888.6k}.

It follows from \eqref{06.1k} and \eqref{3.1k} that
\begin{align}\label{88888.7k}
\begin{split}
\frac{C^{(2)}_{11}}{C^{(2)}_{33}}=v^4,
\end{split}
\end{align}
which implies $C^{(2)}_{11}$ can be determined via \eqref{88888.7k} once $C^{(2)}_{33}$ is chosen, since $v$ is a fixed constant.

Thus we take $C^{(2)}_{12},C^{(2)}_{33}$ and $C^{(2)}_{44}$ as three independent elastic parameters of material 2. And we let {  $\frac{C^{(2)}_{12}}{C^{(2)}_{44}}\neq \frac{C^{(1)}_{12}}{C^{(1)}_{44}}$ to ensure the linear independence of the elastic tensors of material 1 and material 2}, and fix $C^{(2)}_{12}$ and $ C^{(2)}_{44}$, which signifies that only $C^{(2)}_{33}$ remains to be chosen.

Then substitution of \eqref{88888.7k} and \eqref{8.3k} into \eqref{88888.6k} yields
\begin{align}\label{88888.8k}
\begin{split}
\left(C^{(2)}_{33}C^{(1)}_{44}-C^{(1)}_{33}C^{(2)}_{44}\right){\overline{\sigma}^*_{11}}-\left(\left(v^2C^{(2)}_{33}-2C^{(2)}_{44}\right)C^{(1)}_{44}-C^{(1)}_{13}C^{(2)}_{44}\right)\overline{\sigma}^*_{33}\neq0,
\end{split}
\end{align}
which is a linear inequality only with respect to $C^{(2)}_{33}$, where $C^{(1)}_{13},C^{(1)}_{33},C^{(1)}_{44},C^{(2)}_{44},v$ are all treated as constants.

If we can find some $C^{(2)}_{33}$ which makes \eqref{88888.8k} valid, we will complete the proof of  \eqref{77777.2k}.
If we assume that \eqref{88888.8k} is invalid for $C^{(2)}_{33}$, we can get
\begin{align}\label{88888.8kk1}
\begin{split}
\forall C^{(2)}_{33}>0,\;\;\; C^{(1)}_{44}({\overline{\sigma}^*_{11}}-v^2\overline{\sigma}^*_{33})C^{(2)}_{33}-\left(C^{(1)}_{33}C^{(2)}_{44}{\overline{\sigma}^*_{11}}-\left(2C^{(2)}_{44}C^{(1)}_{44}+C^{(1)}_{13}C^{(2)}_{44}\right)\overline{\sigma}^*_{33}\right)=0.
\end{split}
\end{align}
Note that $\overline{\sigma}^*_{11}$ and $\overline{\sigma}^*_{33}$ in \eqref{1.11} are required not to be equal to zero simultaneously. Hence, according to \eqref{06.1k} and \eqref{3.1k}, we can guarantee
\beqs\label{fonditionforc332}
C^{(1)}_{44}({\overline{\sigma}^*_{11}}-v^2\overline{\sigma}^*_{33})\neq 0,
\eeqs
by the initial selection of material 1.
Given \eqref{fonditionforc332}, the linear equation in \eqref{88888.8kk1} only admits a unique root, and it is possible to choose $C^{(2)}_{33}>0$ other than such root to make the inequality \eqref{88888.8k} hold.

Ultimately, the proofs of Theorem \ref{transnewtheorem} for the non-degenerate and the degenerate transversely isotropic media are completed.

\setcounter{equation}{0}
\renewcommand\theequation{C.\arabic{equation}}
\renewcommand\thetheorem{C.\arabic{theorem}}
\renewcommand\thefigure{C.\arabic{figure}}

\section*{C.\quad Proof of Lemma \ref{keylemma}}

The proof of Lemma \ref{keylemma} is divided into two parts.
Firstly, we will verify the existence of an $\Omega'$ that yields \eqref{construct1}. Secondly, we will verify an $\Omega'$ leading to \eqref{construct1} can not be ellipsoidal, which is equivalent to proving
\beqs\label{construct2}
\forall E\subset \mathbb{R}^{3},\;\;\;N_E[\rho](\bfx'){:=}-\int_E \frac{\rho(\bfy')}{4\pi|\bfx'-\bfy'|} d\bfy'\neq \varphi(\bfx'),\;\;\bfx'\in E,
\eeqs
where $\varphi(\bfx')$ is given in \eqref{phi001} as the expression of the Newtonian potential $N_{\Omega'}[\rho](\bfx')$ induced by $\Omega'$ with the mass density $\rho$.

\subsection*{(1)\quad Part 1: the verification of the existence of an $\Omega'$ that yields \eqref{construct1}}
 \quad\\
In this part, searching for $\Omega'$ that generates \eqref{construct1} is mathematically a free boundary problem when the boundary of $\Omega'$ is undetermined.
To handle the free boundary problem, 
\cite{Friedman1982} has set up a variational inequality to analyze a series of potential problems. Further, the variational method has been extended by 
\cite{Liu2008},
achieving the construction of non-ellipsoidal extremal structures that possess the Eshelby uniformity property {in a medium} with a fourth-order isotropic elastic tensor of three elastic constants by solving a particular over-determined problem concerning the Newtonian potential with a constant mass density.

We note that the variational scheme proposed by 
\cite{Liu2008} can also be applied to proving the existence of non-ellipsoidal inclusions that possess Eshelby's polynomial conservation property {in anisotropic media} by solving a corresponding Newtonian potential problem but with a quadratic mass density, as is shown in \eqref{construct1}.

First of all, let us recall the variational method utilized by \cite{Liu2008}.
According to the work of \cite{Liu2008}, we know that for an obstacle function $\phi$ satisfying:
\begin{enumerate}
\item $\phi\in C^{0,1}(\mathbb{R}^{3})$, where $C^{0,1}(\mathbb{R}^{3})$ denotes the set of Lipschitz continuous functions defined on $\mathbb{R}^{3}$ with the norm
\beqs\label{normdefinition}
||\phi||_{0,1}=\sup_{\bfx'\in\mathbb{R}^{3}}|\phi(\bfx')|+\sup_{\bfx',
\bfy'\in\mathbb{R}^{3}}\frac{|\phi(\bfx')-\phi(\bfy')|}{|\bfx'-\bfy'|};
\eeqs
\item there exists $r_0>0$, such that $\forall |\bfx'|\geq r_0$, $\phi(\bfx')\leq0$;
\item $|\Delta\phi|$ is bounded in $B_{r_0}\setminus U^*$, with  $B_{r_0}=\{\bfx'||\bfx'|\leq r_0,\bfx'\in\mathbb{R}^3\}$ and $U^*\subset B_{r_0}$ the set of the singular points where {$|\nabla\otimes\nabla\phi|$, which denotes the norm of the second-order tensor $\nabla\otimes\nabla\phi$,} is unbounded, in the sense of distribution;
\item $\exists C^\phi \in \bfR$, such that $\forall \boldsymbol{\zeta}\in \bfR^3$ with $|\boldsymbol{\zeta}|=1$,
\beqs\label{constrainphi}
\int_{U^{\vartheta}}\frac{\partial^2 \vartheta }{\partial \boldsymbol{\zeta}^2}\left(\phi+\frac{1}{2}C^\phi|\bfx'|^2\right)d\bfx' \geq 0,
\eeqs
 for any smooth function $\vartheta\in C^{\infty}_c(\mathbb{R}^{3})$ with a compact support $U^{\vartheta}$, where $\frac{\partial }{\partial \boldsymbol{\zeta}}$ denotes the directional derivative,

\end{enumerate}
the variational inequality
\beqs\label{inequality}
\prod(V_\phi)=\inf_{v\in K_\phi}\left\{\prod(v)\equiv\int_{\mathbb{R}^3}\frac{1}{2}|\nabla v|^2\right\},
\eeqs
where $K_\phi=\{\;\;v\in W^{1,2}_0(\mathbb{R}^3):\;\;\;v\geq \phi\;\;\}$, admits a unique minimizer $V_\phi\in W^{2,\infty}_{\mathrm{loc}}(\mathbb{R}^3)\cap K_\phi$ satisfying
\beas
\Delta V_\phi\leq0,\;\;\;\;\;\;V_\phi\geq\phi,\;\;\;\;\;\;(V_\phi-\phi)\Delta V_\phi=0\;\;\;\;\;\;\;\;\;\;\text{in}\;\;\;\;\;\mathbb{R}^3,
\eeas
and there exists a coincident set $\Omega'=\{\;\;\bfx'\;\;| \;V_\phi(\bfx')=\phi(\bfx'),\;\;\bfx'\in\mathbb{R}^3\}$ with $\Omega'\subseteq B_{r_0} $.

In the above expressions, $W^{1,2}_0(\mathbb{R}^3)$ denotes the {class} of functions in $L^2(\mathbb{R}^3)$ with a zero boundary value, and the first derivatives of the functions in $W^{1,2}_0(\mathbb{R}^3)$  also belong to $L^2(\mathbb{R}^3)$ in the sense of distribution.  $W^{2,\infty}_{\mathrm{loc}}(\mathbb{R}^3)$ denotes the class of functions in $L^\infty(\mathbb{R}^3)$, whose first and second derivatives also belong to $L^\infty(\mathbb{R}^3)$ with the norm
\beqs\label{Vnorm}
\begin{split}
||V_\phi(\bfx')||_{2,\infty}={\sum_{m\leq 2}\mathrm{ess}\sup_{\bfx'\in \mathbb{R}^3}|\partial^m V_\phi(\bfx')|}
\end{split}
\eeqs
where $m\geq 0$; $\partial^m V_\phi(\bfx')=\frac{\partial^m V_\phi(\bfx')}{\partial {x_1}^{k_1}\partial{x_2}^{k_2}\partial{x_3}^{k_3}} $ with $k_i\geq0,\;k_i\in\mathrm{Z}\;(i=1,2,3)$ and $\sum_{i=1}^3k_i=m$ denote weak derivatives; and `$\mathrm{ess\;sup}$' denotes the essential supremum. Besides, the subscript `loc' implies that the norm in \eqref{Vnorm} must be bounded with $\mathbb{R}^3$ replaced by any bounded strictly interior subdomain of it.

Based on the obstacle function $\phi$, the following over-determined problem
\beqs\label{Liu}
\begin{split}
\left\{ \begin{array}{*{20}{c}}
  &{\Delta v_{od}=\chi_{\Omega'}\Delta\phi \;\;\;\;\;\;\;\;\;\;\;\;\;\;\;\;\;\;\;\;\;\;\text{in}\;\;\;\;\;\mathbb{R}^3} \\
  &{\nabla\nabla v_{od}=\nabla\nabla\phi\;\;\;\;\;\;\;\;\;\;\;\;\;\;\;\;\;\;\;\;\;\;\bfx'\in \Omega'} \\
  &{|v_{od}|\leq \frac{\overline{C}}{|\bfx'|}\;\;\;\;\;\;\;\;\;\;\;\;\;\;\;\;\;\;\;\;\;\;\;\text{for}\;\bfx' \geq r_0}
\end{array}  \right.
\end{split}
\eeqs
admits a solution $v_{od}=V_\phi$. Here $\overline{C}$ is a constant. The details of the above formulation can be found in the work of \cite{Liu2008}.

We let $\Gamma(\bfx'-\bfy'){:=}-\frac{1}{4\pi|\bfx'-\bfy'|}$. Then owing to \eqref{Liu}$_{1}$ and \eqref{Liu}$_{3}$, for any $v_{od}\in W^{2,\infty}_{\mathrm{loc}}(\mathbb{R}^3)$, we can get
\beqs\label{Liu2}
\begin{split}
N_{\Omega'}[\Delta\phi](\bfx')&=\int_{\mathbb{R}^3}\chi_{\Omega'}\Delta_\bfy'\phi(\bfy')\Gamma(\bfx'-\bfy')d\bfy'
=\int_{\mathbb{R}^3}\Delta_\bfy' v_{od}(\bfy')\Gamma(\bfx'-\bfy')d\bfy'\\
&=\sum_{i=1}^{3}\int_{\partial B_{\infty}}\frac{\partial v_{od}(\bfy')}{\partial y_i} \Gamma(\bfx'-\bfy')n_id\bfy'-\sum_{i=1}^{3}\int_{\mathbb{R}^3}\frac{\partial v_{od}(\bfy')}{\partial y_i} \frac{\partial \Gamma(\bfx'-\bfy')}{\partial y_i}d\bfy'\\
&=-\sum_{i=1}^{3}\int_{\partial B_{\infty}} v_{od}(\bfy')\frac{\partial \Gamma(\bfx'-\bfy')}{\partial y_i}n_id\bfy'+\int_{\mathbb{R}^3} v_{od}(\bfy') \Delta_\bfy' \Gamma(\bfx'-\bfy')d\bfy'\\
&=\int_{\mathbb{R}^3} v_{od}(\bfy')\delta(\bfx'-\bfy')d\bfy'=v_{od}(\bfx'),
\end{split}
\eeqs
where $B_{\infty}=\lim_{r\to\infty}\{\bfx'||\bfx'|\leq r,\bfx'\in \mathbb{R}^3\}$, and $\bfn=(n_1,n_2,n_3)$ is the unit outward normal to $\partial B_{\infty}$. Hence we conclude that any solution $v_{od}\in W^{2,\infty}_{\mathrm{loc}}(\mathbb{R}^3)$ of \eqref{Liu} must be the Newtonian potential induced by $\Omega'$ with the mass density $\Delta\phi$.


To continue our analysis, we introduce a particular $\phi^*$ expressed as
\beqs\label{newphi}
\begin{split}
\phi^*(\bfx'){:=}\left\{ {\begin{array}{*{20}{c}}
  {\varphi(\bfx'),\;\;\;\;\;\;\;\;\;\;\;\;\;\;\bfx' \in U } \\
  {\;\;-3C,\;\;\;\;\;\;\;\;\;\;\bfx' \in \mathbb{R}^3\setminus U}
\end{array}\;\;\;\;\;\;\;\;\;\;\;\;\;\;\;\;\;\;\;\;\;\;\;\;\;\;\;\;\;\;}  \right.
\end{split}
\eeqs
with $U{:=}\{\;\bfx'\;|\;\;{x'_1}^4+{x'_2}^4+{x'_3}^4\leq 48C,\;\bfx'\in\mathbb{R}^3\}$.

We will show that  ${\phi}^*$ in \eqref{newphi} satisfies the four conditions of an obstacle function.
  Firstly, let $U'{:=}\{\;\bfx'\;|\;\;|x'_i|\leq (48C)^{\frac{1}{4}}\;(i=1,2,3),\;\bfx'\in\mathbb{R}^3\}$.  It is seen that $U'$ is bounded, and $U\subset U'$; hence $U$ is a bounded domain with $\partial U$ defined by the surface  $\;{x'_1}^4+{x'_2}^4+{x'_3}^4-48C=0$. It can be derived from \eqref{newphi} along with \eqref{phi001} that
\beqs\label{boundaryphi}
\phi^*|_{\partial U^-}=\phi^*|_{\partial U^+}=-3C,
\eeqs
where $\partial U^-$ means the limiting value approached from the interior of U, and $\partial U^+$ means the limiting value approached from the exterior of U. Thus $\phi^*\in C^{0}(\mathbb{R}^{3})$. Further, substituting \eqref{newphi} into \eqref{normdefinition} yields
\beqs\label{norm}
\begin{split}
||\phi^*||_{0,1}&=\max\left\{\sup_{\bfx'\in U}|\varphi(\bfx')|,-3C\right\}\\
&+\max\left\{\sup_{\bfx',
\bfy'\in U}\frac{|\varphi(\bfx')-\varphi(\bfy')|}{|\bfx'-\bfy'|},\sup_{\bfx'\in U,
\bfy'\in\mathbb{R}^{3}}\frac{|\varphi(\bfx')+3C|}{|\bfx'-\bfy'|},0\right\}\\
&=\sup_{\bfx'\in U}|\varphi(\bfx')|+\sup_{\bfx'\in U}|\nabla\varphi(\bfx')|.
\end{split}
\eeqs
$|\varphi(\bfx')|$ and $|\nabla\varphi(\bfx')|$ are bounded in $U$ owing to \eqref{phi001}, which implies that the norm $||\phi^*||_{0,1}$ of $\phi^*$ is bounded, and thus $\phi^*\in C^{0,1}(\mathbb{R}^3)$. Therefore, ${\phi}^*$ satisfies the first condition of an obstacle function.

Secondly, we let $r_0{:=}6\sqrt{C}$ and $B_{r_0}:=\{\bfx'||\bfx'|< r_0,\bfx'\in\mathbb{R}^3\}$. It can be verified that $B_{r_0}\subset U$, since
\beqs
\forall \bfx'\in B_{r_0},\;\;{x'_1}^4+{x'_2}^4+{x'_3}^4\leq \left({x'_1}^2+{x'_2}^2+{x'_3}^2\right)^2<36C< 48C.
\eeqs
 Thus $\{\bfx'|\;|\bfx'|\geq r_0,\;\bfx'\in\mathbb{R}^3\}=\mathbb{R}^3\setminus B_{r_0}=(U\setminus B_{r_0})\cup (\mathbb{R}^3\setminus U)$.
Then since
\beqs\label{evaluate1}
\phi^*(\bfx')=C-\frac{1}{12}\left({x'_1}^4+{x'_2}^4+{x'_3}^4\right)\leq C-\frac{1}{36}\left({x'_1}^2+{x'_2}^2+{x'_3}^2\right)^2\leq 0,\quad\mathrm{for}\;\; \bfx'\in U\setminus B_{r_0},
\eeqs
and
\beqs\label{evaluate2}
\phi^*(\bfx')=-3C<0,\quad\mathrm{for}\;\;\bfx'\in\mathbb{R}^3\setminus U,
\eeqs
we conclude that $\forall \;|\bfx'|\geq r_0$, $\phi^*(\bfx')\leq 0$, which proves that ${\phi}^*$ satisfies the second condition of an obstacle function.

Thirdly, owing to $B_{r_0}\subset U$ and \eqref{newphi}, we see $\phi^*\in C^{\infty}(B_{r_0})$. Hence $U^*=\emptyset$, where $U^*\subset B_{r_0}$ is defined as the set of singular points where $\nabla\otimes\nabla\phi^*$ is unbounded.
Thus we conclude that $|\Delta\phi^*|$ is bounded in $B_{r_0}\setminus U^*$,  which proves that $\hat{\phi}^*$ satisfies the third condition of an obstacle function.

Fourthly, let $U^{\vartheta}$ denote the compact support of a smooth function $\vartheta\in C^{\infty}(\mathbb{R}^3)$, on which $\vartheta\geq 0$. By definition, we know that $\vartheta=0$ in $\mathbb{R}^3\setminus U^{\vartheta}$, and
 \beqs\label{smooth}
\forall \;\;n=i+j+k\;\;\text{with}\;\;i,j,k\geq 0,\;\;\;\;\;\;\left.\frac{\partial^n\vartheta}{\partial {x'_1}^i\partial {x'_2}^j\partial {x'_3}^k}\right|_{\partial U}=0.
 \eeqs
Then it can be derived from \eqref{newphi} and \eqref{smooth} that $\forall \boldsymbol{\zeta}\in \bfR^3$ with $|\boldsymbol{\zeta}|=1$,
\beqs\label{judge}
\begin{split}
&\int_{U^{\vartheta}}\frac{\partial^2 \vartheta }{\partial \boldsymbol{\zeta}^2}\left(\phi^*+\frac{1}{2}C^\phi|\bfx'|^2\right)d\bfx'\\
=&\int_{U^{\vartheta}}\frac{\partial^2 \vartheta }{\partial \boldsymbol{\zeta}^2}\phi^* d\bfx'+\int_{U^{\vartheta}}\frac{1}{2}C^\phi|\bfx'|^2\frac{\partial^2 \vartheta }{\partial \boldsymbol{\zeta}^2}d\bfx'\\
=&\int_{U^{\vartheta}\cap U}\frac{\partial^2 \vartheta }{\partial \boldsymbol{\zeta}^2}\varphi d\bfx'+\int_{U^{\vartheta}\cap (\mathbb{R}^3\setminus U)}(-3C)\frac{\partial^2 \vartheta }{\partial \boldsymbol{\zeta}^2}d\bfx'+\int_{U^{\vartheta}}\vartheta C^\phi d\bfx'\\
=&\int_{\partial U \cap U^{\vartheta}}\frac{\partial \vartheta }{\partial \boldsymbol{\zeta}}\varphi \boldsymbol{\zeta}\cdot d\bfS-\int_{U^{\vartheta}\cap U}\frac{\partial \vartheta }{\partial \boldsymbol{\zeta}}\frac{\partial }{\partial \boldsymbol{\zeta}}\varphi d\bfx'+\int_{\partial U \cap U^{\vartheta}}\frac{\partial \vartheta }{\partial \boldsymbol{\zeta}}(-3C) \boldsymbol{\zeta}\cdot (-d\bfS)\\
&-\int_{U^{\vartheta}\cap (\mathbb{R}^3\setminus U)}\frac{\partial \vartheta }{\partial \boldsymbol{\zeta}}\frac{\partial }{\partial \boldsymbol{\zeta}}(-3C) d\bfx'+\int_{U^{\vartheta}}\vartheta C^\phi d\bfx'\\
=&-\int_{\partial U \cap U^{\vartheta}}\vartheta \frac{\partial \varphi}{\partial \boldsymbol{\zeta}} \boldsymbol{\zeta}\cdot d\bfS+\int_{U^{\vartheta}\cap U}\vartheta\frac{\partial^2 \varphi}{\partial \boldsymbol{\zeta}^2} d\bfx'+\int_{U^{\vartheta}}\vartheta C^\phi d\bfx'\\
\geq& -\int_{\partial U \cap U^{\vartheta}}\vartheta \sup_{\bfx'\in \partial U \cap U^{\vartheta}}\left|\frac{\partial \varphi}{\partial \boldsymbol{\zeta}}\right| dS-\int_{U^{\vartheta}\cap U}\vartheta\sup_{\bfx'\in U \cap U^{\vartheta}}\left|\frac{\partial^2 \varphi}{\partial \boldsymbol{\zeta}^2}\right| d\bfx'+\int_{U^{\vartheta}}\vartheta C^\phi d\bfx' \\
\geq&\int_{U^{\vartheta}}\vartheta \left(C^\phi-\sup_{\bfx'\in U}\left|\frac{\partial \varphi}{\partial \boldsymbol{\zeta}}\right|-\sup_{\bfx'\in U}\left|\frac{\partial^2 \varphi}{\partial \boldsymbol{\zeta}^2}\right|\right)d\bfx'.
\end{split}
\eeqs
Based on \eqref{judge}, we know that $\exists C^\phi\in \mathrm{R}$ satisfying
\beqs \label{judgeC}
C^\phi\geq\sup_{\bfx'\in U}\left|\frac{\partial \varphi}{\partial \boldsymbol{\zeta}}\right|+\sup_{\bfx'\in U}\left|\frac{\partial^2 \varphi}{\partial \boldsymbol{\zeta}^2}\right|,
\eeqs
such that $\forall \boldsymbol{\zeta}\in \bfR^3$ with $|\boldsymbol{\zeta}|=1$,
\beqs\label{judgeCnew}
\int\frac{\partial^2 \vartheta }{\partial \boldsymbol{\zeta}^2}\left(\phi^*+\frac{1}{2}C^\phi|\bfx'|^2\right)d\bfx'\geq \int_{U^{\vartheta}}\vartheta \left(C^\phi-\sup_{\bfx'\in U}\left|\frac{\partial \varphi}{\partial \boldsymbol{\zeta}}\right|-\sup_{\bfx'\in U}\left|\frac{\partial^2 \varphi}{\partial \boldsymbol{\zeta}^2}\right|\right)d\bfx' \geq 0,
\eeqs
which proves that ${\phi}^*$ satisfies the fourth condition of an obstacle function.

Therefore, for $\phi^*$, the over-determined problem \eqref{Liu} with $\phi$ replaced by $\phi^*$ admits a solution $v_{od}=V_{\phi^*}\in W^{2,\infty}_{\mathrm{loc}}(\mathbb{R}^3)$, and there is a coincident set  $\Omega'\subseteq B_{r_0}\subset U $, satisfying $V_{\phi^*}(\bfx')=\phi^*(\bfx')=\varphi(\bfx')$ for $\bfx' \in \Omega'$.

According to \eqref{Liu2}, $V_{\phi^*}(\bfx')$ is actually the Newtonian potential induced by $\Omega'$ with the mass density $\Delta\phi^*$. Since $\Delta\phi^*= \Delta\varphi=\rho$, we see that $V_{\phi^*}(\bfx')=N_{\Omega'}[\rho](\bfx')=\varphi(\bfx')$ for $\bfx'\in\Omega'$.
 Therefore, we have substantiated the existence of a domain $\Omega'$ that leads to $N_{\Omega'}[\rho](\bfx')=\varphi(\bfx')$ for $\bfx'\in\Omega'$, and thus the proof of the existence of an $\Omega'$ that yields \eqref{construct1} is achieved.

\subsection*{(2)\quad Part 2: the proof of the non-ellipsoidal shape of an $\Omega'$ that leads to \eqref{construct1}}
 \quad\\
As is mentioned before, the proof of the non-ellipsoidal shape of an $\Omega'$ that leads to \eqref{construct1} can be fulfilled by the substantiation of \eqref{construct2}. Hence our aim is to prove \eqref{construct2} in this part.

Firstly, we know that the Newtonian potential  of an ellipsoid should rely on the orientation of the ellipsoid, and it is also dependent on the position, since the mass density is not homogeneous, which varies with the position in the coordinates system $\bfx'=(x_1,x_2,x_3)$. Here we let $\bfz=(z_1,z_2,z_3)$ be the Cartesian coordinate whose origin is at the center of the ellipsoid with its axes being along the axes of the ellipsoid so that the ellipsoid is expressed as $E=\{\;\;\bfz\;\;|\;\;\frac{z_1^2}{a_1^2}+\frac{z_2^2}{a_2^2}+\frac{z_3^2}{a_3^2}\leq 1\;\;\;\}$, where $a_i>0\; (i=1,2,3)$ denote the semi-axis lengths of the ellipsoid.

By introducing transformations
\beqs\label{trans}
\bfx'{:=}\bfQ\cdot\bfz+\bfd, \quad \bfy'{:=}\bfQ\cdot\bfz'+\bfd,
\eeqs
and then substituting \eqref{trans} into \eqref{construct2}, we obtain the Newtonian potential $N_E[\rho](\bfz)$ induced by $E$ with the mass density $\rho$, which is expressed in the coordinates $\bfz=(z_1,z_2,z_3)$ of the body frame of the ellipsoid, i.e.,
\beqs\label{Nee}
N_E[\rho](\bfz)=-\int_E \frac{-|\bfQ\cdot\bfz'+\bfd|^2}{4\pi|\bfz-\bfz'|} d\bfz'=\int_E \frac{|\bfz'|^2+2(\bfd\cdot\bfQ)\cdot\bfz'+|\bfd|^2}{4\pi|\bfz-\bfz'|} d\bfz',
\eeqs
where $\bfQ$ is a second-order orthogonal tensor denoting rotation, and $\bfd\in\bfR^3$ denotes the translation.

Let $\bff=2(\bfd\cdot\bfQ)$. It can be derived from \eqref{Nee} that $N_E[\rho](\bfz)$ can be expressed in terms of $\bff$, $\bfd$ and $a_i$ $(i=1,2,3)$ \citep{Mura1987}, i.e.,
\beqs\label{NEz}
\begin{split}
N_E[\rho](\bfz)=&C_E+A_1z_1+A_2z_2+A_3z_3+B_1z_1^2+B_2z_2^2+B_3z_3^2+H_1z_1^3+H_2z_2^3+H_3z_3^3+H_4z_1z_2^2\\
&+H_5z_1z_3^2+H_6z_2z_1^2+H_7z_2z_3^2+H_8z_3z_1^2+H_9z_3z_2^2\\
&+J_1z_1^4+J_2z_2^4+J_3z_3^4+J_4z_1^2z_2^2+J_5z_2^2z_3^2+J_6z_3^2z_1^2,
\end{split}
\eeqs
where
\beqs\label{coe}
\begin{split}
&C_E{:=}\frac{1}{8}\left((a_1^2+a_2^2+a_3^2)I-(a_1^4I_1+a_2^4I_2+a_3^4I_3)\right)+\frac{1}{2}|\bfd|^2I;\\
&A_1{:=}\frac{1}{2}a_1^2I_1f_1;\;\;\;\;A_2=\frac{1}{2}a_2^2I_2f_2;\;\;\;\;A_3=\frac{1}{2}a_3^2I_3f_3;\\
&B_1{:=}\frac{3}{4}I_{11}a_1^4+\frac{1}{4}I_{12}a_2^4+\frac{1}{4}I_{13}a_3^4-\frac{1}{4}(a_1^2+a_2^2+a_3^2)I_1-\frac{1}{2}|\bfd|^2I_1;\\ &B_2{:=}\frac{3}{4}I_{22}a_2^4+\frac{1}{4}I_{21}a_1^4+\frac{1}{4}I_{23}a_3^4-\frac{1}{4}(a_1^2+a_2^2+a_3^2)I_2-\frac{1}{2}|\bfd|^2I_2;\\
&B_3{:=}\frac{3}{4}I_{33}a_3^4+\frac{1}{4}I_{31}a_1^4+\frac{1}{4}I_{32}a_2^4-\frac{1}{4}(a_1^2+a_2^2+a_3^2)I_3-\frac{1}{2}|\bfd|^2I_3;\\
&H_1{:=}-\frac{1}{2}a_1^2I_{11}f_1;\;\;\;\;H_2=-\frac{1}{2}a_2^2I_{22}f_2;\;\;\;\;H_3=-\frac{1}{2}a_3^2I_{33}f_3;\\
&H_4{:=}-\frac{1}{2}a_1^2I_{21}f_1;\;\;\;\;H_5=-\frac{1}{2}a_1^2I_{31}f_1;\;\;\;\;H_6=-\frac{1}{2}a_2^2I_{21}f_2;\\ &H_7{:=}-\frac{1}{2}a_2^2I_{23}f_2;\;\;\;\;H_8=-\frac{1}{2}a_3^2I_{13}f_3;\;\;\;\;H_9=-\frac{1}{2}a_3^2I_{23}f_3;\\
&J_1{:=}\frac{1}{8}I_{11}(a_1^2+a_2^2+a_3^2)-\frac{5}{8}I_{111}a_1^4-\frac{1}{8}a_2^4I_{112}-\frac{1}{8}a_3^4I_{113};\\ &J_2{:=}\frac{1}{8}I_{22}(a_1^2+a_2^2+a_3^2)-\frac{5}{8}I_{222}a_2^4-\frac{1}{8}a_1^4I_{221}-\frac{1}{8}a_3^4I_{223};\\
&J_3{:=}\frac{1}{8}I_{33}(a_1^2+a_2^2+a_3^2)-\frac{5}{8}I_{333}a_3^4-\frac{1}{8}a_1^4I_{331}-\frac{1}{8}a_2^4I_{332};\\
&J_4{:=}\frac{1}{4}(a_1^2+a_2^2+a_3^2)I_{12}-\frac{3}{4}(a_1^4I_{211}+a_2^4I_{122})-\frac{1}{4}a_3^4I_{321};\\
&J_5{:=}\frac{1}{4}(a_1^2+a_2^2+a_3^2)I_{23}-\frac{3}{4}(a_2^4I_{322}+a_3^4I_{233})-\frac{1}{4}a_1^4I_{321};\\
&J_6{:=}\frac{1}{4}(a_1^2+a_2^2+a_3^2)I_{31}-\frac{3}{4}(a_3^4I_{133}+a_1^4I_{311})-\frac{1}{4}a_2^4I_{321},\\
\end{split}
\eeqs
with
\beqs\label{I0}
\begin{split}
&I=\frac{\prod_{j=k}^3a_k}{2}\int_0^{+\infty}\frac{ds}{\sqrt{\prod_{q=1}^3(a_q^2+s)}};\quad I_i=\frac{\prod_{k=1}^3a_k}{2}\int_0^{+\infty}\frac{ds}{(a_i^2+s)\sqrt{\prod_{q=1}^3(a_q^2+s)}};\\
&I_{ij}=\frac{\prod_{k=1}^3a_k}{2}\int_0^{+\infty}\frac{ds}{(a_i^2+s)(a_j^2+s)\sqrt{\prod_{q=1}^3(a_q^2+s)}};\\ &I_{ijn}=\frac{\prod_{k=1}^3a_k}{2}\int_0^{+\infty}\frac{ds}{(a_i^2+s)(a_j^2+s)(a_n^2+s)\sqrt{\prod_{q=1}^3(a_q^2+s)}}.
\end{split}
\eeqs

Meanwhile, substituting \eqref{trans} into \eqref{phi001} yields
\beqs\label{varphiz}
\begin{split}
\varphi(\bfz)=&C-\frac{1}{12}(Q_{11}z_1+Q_{12}z_2+Q_{13}z_3+d_1)^4\\
&-\frac{1}{12}(Q_{21}z_1+Q_{22}z_2+Q_{23}z_3+d_2)^4-\frac{1}{12}(Q_{31}z_1+Q_{32}z_2+Q_{33}z_3+d_3)^4.
\end{split}
\eeqs

Then based on \eqref{NEz} and \eqref{varphiz}, we are going to prove 
\eqref{construct2}, and we will achieve the proof by contradiction.
Assume $N_E[\rho](\bfz)=\varphi(\bfz)$, and hence the right-hand side of \eqref{varphiz} equals the right-hand side of \eqref{NEz}. By comparison of the coefficients of $z_1^3z_2,z_1z_2^3,z_2^3z_3,z_2z_3^3,z_1^3z_3,z_1z_3^3$ in \eqref{varphiz} with those in \eqref{NEz}, we obtain
\beqs\label{contradiciton1}
\begin{split}
&Q_{11}^3Q_{12}+Q_{21}^3Q_{22}+Q_{31}^3Q_{32}=0,\\ \quad &Q_{11}Q_{12}^3+Q_{21}Q_{22}^3+Q_{31}Q_{32}^3=0,\\ \quad &Q_{12}^3Q_{13}+Q_{22}^3Q_{23}+Q_{32}^3Q_{33}=0,\\
 &Q_{12}Q_{13}^3+Q_{22}Q_{23}^3+Q_{32}Q_{33}^3=0,\\ \quad &Q_{13}^3Q_{11}+Q_{23}^3Q_{21}+Q_{33}^3Q_{31}=0,\quad \\&Q_{13}Q_{11}^3+Q_{23}Q_{21}^3+Q_{33}Q_{31}^3=0.\\
\end{split}
\eeqs
In addition, since $\bfQ$ is orthogonal, we see
\beqs\label{Qproperty}
\begin{split}
&Q_{11}Q_{12}+Q_{21}Q_{22}+Q_{31}Q_{32}=0,\quad \\&Q_{12}Q_{13}+Q_{22}Q_{23}+Q_{32}Q_{33}=0,\quad \\&Q_{13}Q_{11}+Q_{23}Q_{21}+Q_{33}Q_{31}=0,\\ &Q^2_{11}+Q^2_{21}+Q^2_{31}=1,\quad \\&Q^2_{12}+Q^2_{22}+Q^2_{32}=1,\quad \;\;\\&Q^2_{13}+Q^2_{23}+Q^2_{33}=1.\\
\end{split}
\eeqs
By combining $\eqref{contradiciton1}_2$ and $\eqref{contradiciton1}_5$  with $\eqref{Qproperty}_1$ and $\eqref{Qproperty}_3$, we obtain
\beqs\label{Qeq}
\begin{bmatrix}
Q_{12}&Q_{22}&Q_{32}\\
Q_{13}&Q_{23}&Q_{33}\\
Q_{12}^3&Q_{22}^3&Q_{32}^3\\
Q_{13}^3&Q_{23}^3&Q_{33}^3
\end{bmatrix}\cdot\begin{bmatrix}
Q_{11}\\
Q_{21}\\
Q_{31}
\end{bmatrix}=0.
\eeqs
We regard \eqref{Qeq} as a homogenous linear system of equations with respect to $(Q_{11},Q_{21},Q_{31})$, so  $(Q_{12},Q_{22},Q_{32})$, $(Q_{13},Q_{23},Q_{33})$, $(Q_{12}^3,Q_{22}^3,Q_{32}^3)$ and $(Q_{13}^3,Q_{23}^3,Q_{33}^3)$ denote four coefficients for four different linear equations in this system.

If there are more than 2 independent linear equations in \eqref{Qeq}, the solution will be trivial.
 However, \eqref{Qeq} only admits non-trivial solutions owing to $\eqref{Qproperty}_4$. Thus, there are at most two independent equations in the homogenous linear system shown in \eqref{Qeq}. Then, let us choose the first two equations in \eqref{Qeq} as two independent equations, and the independence between them can be proved by $\eqref{Qproperty}_2$.

 Given this, we have
\beqs\label{pppp}
\begin{split}
(Q_{12}^3,Q_{22}^3,Q_{32}^3){:=}k_1(Q_{12},Q_{22},Q_{32})+m_1(Q_{13},Q_{23},Q_{33}),\\ (Q_{13}^3,Q_{23}^3,Q_{33}^3){:=}k_2(Q_{13},Q_{23},Q_{33})+m_2(Q_{12},Q_{22},Q_{32}),
\end{split}
\eeqs
with $k_i,m_i\;(i=1,2)$ being four real constants.

Then by substituting $\eqref{pppp}_1$ and $\eqref{Qproperty}_2$ into $\eqref{contradiciton1}_2$, and substituting $\eqref{pppp}_2$ and $\eqref{Qproperty}_2$ into $\eqref{contradiciton1}_5$,  we get
\beqs
m_1=m_2=0,
\eeqs
which means
\beqs\label{constrain}
\begin{split}
(Q_{12}^3,Q_{22}^3,Q_{32}^3)=k_1(Q_{12},Q_{22},Q_{32}),\quad (Q_{13}^3,Q_{23}^3,Q_{33}^3)=k_2(Q_{13},Q_{23},Q_{33}),\\
\end{split}
\eeqs
with $k_1,k_2\neq 0$ due to $\eqref{Qproperty}_{5,6}$.

Then we take three cases concerning $(Q_{12},Q_{22},Q_{32})$ into consideration.
\begin{enumerate}
\item Only one component of $(Q_{12},Q_{22},Q_{32})$ is nonzero.

Without loss of generality, we take $Q_{12}\neq 0$ so that $Q_{22}=Q_{32}=0$.
Since  $|(Q_{12},Q_{22},Q_{32})|=1$,
we have $Q_{12}=\pm1$.
  Given that $(Q_{12},Q_{22},Q_{32})$ have been specified, $\bfQ$ can be determined based on \eqref{Qproperty} and \eqref{contradiciton1} once $(Q_{13},Q_{23},Q_{33})$ is determined. Then substituting $Q_{12}=\pm1$ and $Q_{22}=Q_{32}=0$ into $\eqref{Qproperty}_2$ yields $Q_{13}=0$. To further determine $(Q_{13},Q_{23},Q_{33})$, we consider three cases concerning $Q_{23}$ and $Q_{33}$.
\begin{enumerate}
\item $Q_{23}=0, Q_{33}\neq 0$ or $Q_{33}=0,Q_{23}\neq 0 $.

Without loss of generality, we take $Q_{23}=0, Q_{33}\neq 0$. Likewise, $|(Q_{13},Q_{23},Q_{33})|=1$ so that $Q_{33}=\pm1$. Based on \eqref{Qproperty}, since $(Q_{12},Q_{22},Q_{32})= (\pm1,0,0)$ and $(Q_{13},Q_{23},Q_{33})=(0,0,\pm1)$, we know that $(Q_{11},Q_{21},Q_{31})=(0,\pm1,0)$. Thus
\beas
\bfQ=\begin{bmatrix}
0&\pm1&0\\
\pm1&0&0\\
0&0&\pm1
\end{bmatrix}.
\eeas
By following the same procedure, we can construct more $\bfQ_s$ that only possesses three $\pm1$ components. Such $\bfQ_s$ denotes the rotation of the coordinate system $\bfz=(z_1,z_2,z_3)$ around any basis of it by $\pm\frac{\pi}{2}$ or symmetric transformations with respect to any plane spanned by two axes of the coordinate system $\bfz=(z_1,z_2,z_3)$ or the  superposition of them. There are 48 $\bfQ_s$ in total.

Let $\varphi^{(4)}(\bfz)$ denote the summation of the forth-degree terms in $\varphi(\bfz)$. In this case,
\beqs\label{forth1}
\varphi^{(4)}(\bfz)=-\frac{1}{12}(z_1^4+z_2^4+z_3^4).
\eeqs

\item $Q_{23}\neq0, Q_{33}\neq 0$.

According to \eqref{constrain} and \eqref{Qproperty}, we have four cases:
\beas
Q_{23}=\pm\sqrt{k_2},\;\;Q_{33}=\pm\sqrt{k_2},\;\;k_2=\frac{1}{2}.
\eeas
We discuss $Q_{23}=Q_{33}=\frac{\sqrt{2}}{2}$, and other cases can be discussed in the same way. When $Q_{23}=Q_{33}=\frac{\sqrt{2}}{2}$, by resorting to \eqref{Qproperty}, we gain $(Q_{11},Q_{21},Q_{31})=(0,\frac{\sqrt{2}}{2},-\frac{\sqrt{2}}{2})$ or $(Q_{11}$ $,Q_{21},Q_{31})=(0,-\frac{\sqrt{2}}{2},\frac{\sqrt{2}}{2})$;  we just consider the former case,  which results in
\beqs
\bfQ=\begin{bmatrix}
0&\pm1&0\\
\frac{\sqrt{2}}{2}&0&\frac{\sqrt{2}}{2}\\
-\frac{\sqrt{2}}{2}&0&\frac{\sqrt{2}}{2}
\end{bmatrix}.
\eeqs
Likewise, we could construct  more $\bfQ_s'$ in a similar form, which means the rotations of  the coordinate system $\bfz=(z_1,z_2,z_3)$ around any basis of it by $\pm\frac{\pi}{4}$  or further superposition of such rotations on the coordinates transformation represented by $\bfQ_s$. There are 72 $\bfQ_s'$ in total.
In this case,

\beqs\label{forth2}
\begin{split}
&\varphi^{(4)}(\bfz)=-\frac{1}{12}(\frac{1}{2}z_1^4+\frac{1}{2}z_2^4+z_3^4+3z^2_1z^2_2),\\
\text{or}\;\;&\varphi^{(4)}(\bfz)=-\frac{1}{12}(\frac{1}{2}z_1^4+\frac{1}{2}z_3^4+z_2^4+3z^2_1z^2_3),\\
\text{or}\;\;&\varphi^{(4)}(\bfz)=-\frac{1}{12}(\frac{1}{2}z_2^4+\frac{1}{2}z_3^4+z_1^4+3z^2_2z^2_3).
\end{split}
\eeqs

\end{enumerate}

\item Two components of $(Q_{12},Q_{22},Q_{32})$ are nonzero and the other is zero.

Without loss of generality, we take $Q_{12},Q_{22}\neq 0$ so that $Q_{32}=0$. Due to \eqref{constrain} and \eqref{Qproperty}, we have
\beas
Q_{12}=\pm\sqrt{k_1},\;\;Q_{22}=\pm\sqrt{k_1},\;\;k_1=\frac{1}{2}.
\eeas
Then we consider two cases concerning $(Q_{13},Q_{23},Q_{33})$.
\begin{enumerate}

\item At least one component of $(Q_{13},Q_{23},Q_{33})$ is zero.

This situation is the same as that discussed in (b) of (i), since $(Q_{13},Q_{23},Q_{33})$, $(Q_{12},Q_{22},$ $Q_{32})$ and $(Q_{11},Q_{21},$ $Q_{31})$ are equivalent, which can be replaced by each other.
For example, when $(Q_{12},Q_{22},Q_{32}\!)\!\!=\!\!(\!\!-\frac{\sqrt{2}}{2},\frac{\sqrt{2}}{2},0)$ is fixed, and if there is only one nonzero component in $(Q_{13},Q_{23},Q_{33})$, we have $(Q_{13},Q_{23},Q_{33})=(0,0,\pm1)$ and $(Q_{11},Q_{21},Q_{31})=(\frac{\sqrt{2}}{2}\frac{\sqrt{2}}{2},0)$; thus
\beas
\bfQ=\begin{bmatrix}
-\frac{\sqrt{2}}{2}&\-\frac{\sqrt{2}}{2}&0\\
\frac{\sqrt{2}}{2}&\frac{\sqrt{2}}{2}&0\\
0&0&\pm1
\end{bmatrix}.
\eeas

If there are two nonzero components in $(Q_{13},Q_{23},Q_{33})$, we have $(Q_{13},Q_{23},Q_{33})=(\frac{\sqrt{2}}{2},$ $\frac{\sqrt{2}}{2},0)$ and $(Q_{11}$ $,Q_{21},Q_{31})=(0,0,\pm1)$; thus
\beas
\bfQ=\begin{bmatrix}
0&-\frac{\sqrt{2}}{2}&\frac{\sqrt{2}}{2}\\
0&\frac{\sqrt{2}}{2}&\frac{\sqrt{2}}{2}\\
\pm1&0&0
\end{bmatrix}.
\eeas

\item Three components of $(Q_{13},Q_{23},Q_{33})$ are all nonzero.

In this case, according to \eqref{constrain}, we have
\beas
Q_{13}=\pm\sqrt{k_2},\;\;Q_{23}=\pm\sqrt{k_2},\;\;Q_{33}=\pm\sqrt{k_2},\;\;k_3=\frac{1}{3}.
\eeas
 Based on \eqref{constrain}, \eqref{Qproperty} and the above equation, we fix $(Q_{13},Q_{23},Q_{33})=(\frac{\sqrt{3}}{3},-\frac{\sqrt{3}}{3},\frac{\sqrt{3}}{3})$ and $(Q_{12},$ $Q_{22},Q_{32}) =(\frac{\sqrt{2}}{2},\frac{\sqrt{2}}{2},0)$.  Other situations when the sign of any component of $(Q_{12},Q_{22},Q_{32})$ and $(Q_{13},Q_{23}$ $,Q_{33})$ changes can be discussed via the same method.
Then we can easily calculate
\beas
\begin{split}
(Q_{11},Q_{21},Q_{31})&=\pm(Q_{12},Q_{22},Q_{32})\times(Q_{13},Q_{23},Q_{33})\\
&=\pm(\frac{\sqrt{6}}{6},-\frac{\sqrt{6}}{6},-\frac{\sqrt{6}}{3}),
\end{split}
\eeas
which contradicts $\eqref{contradiciton1}_6$. Hence this case is invalid.
\end{enumerate}

\item Three components of $(Q_{12},Q_{22},Q_{32})$ are all nonzero.

In this case, it is obvious that $(Q_{13},Q_{23},Q_{33})$ cannot have one nonzero component or three nonzero components due to $\eqref{Qproperty}_2$. Given this, there must be two nonzero components of $(Q_{13},Q_{23},Q_{33})$. However, when $(Q_{13},Q_{23},Q_{33})$ has two nonzero components, the situation is the same as that discussed in (b) of (ii), since $(Q_{13},Q_{23},Q_{33})$ and $(Q_{12},Q_{22},Q_{32})$ can be exchanged, which does not influence the discussion.
\end{enumerate}

Then we draw the conclusion that if there exists an ellipsoid whose Newtonian potential satisfies $N_E[\rho](\bfz)=\varphi(\bfz)$, then the sum of the forth-degree terms $\varphi^{(4)}(\bfz)$ in the polynomial $\varphi(\bfz)$ can only be expressed as either \eqref{forth1} or \eqref{forth2}.

However, according to \eqref{NEz}, the sum of the forth-degree terms $\varphi_E^{(4)}(\bfz)$ that results from the Newtonian potential of an ellipsoid must be expressed as
\beqs\label{varphi4}
\varphi_E^{(4)}(\bfz)=J_1z_1^4+J_2z_2^4+J_3z_3^4+J_4z_1^2z_2^2+J_5z_2^2z_3^2+J_6z_3^2z_1^2,
\eeqs
where $J_i\;(i=1,2,3,4,5,6)$ are introduced in \eqref{coe}.

Then 
we are going to prove  
$\varphi_E^{(4)}(\bfz)\neq \varphi^{(4)}(\bfz)$,
which ultimately implies \eqref{construct2}.

We will achieve the proof by determining the range of $J_i\;(i=1,2,3,4,5,6)$ that relies on the shape of the ellipsoid. 
It is known from \cite{Mura1987} that there are relationships among $I_i,I_{ij},I_{ijk}\;(i,j,k=1,2,3)$, i.e.,
\beqs\label{I}
\begin{split}
&I_1+I_2+I_3=1;\;\;\;\;\;\;\;I_{ij}=\frac{I_j-I_i}{a_i^2-a_j^2},\;\;\text{for}\;\;i\neq j;\quad I_{ii}=\frac{1}{3}(\frac{1}{a_i^2}-\sum_{q=1,q\neq i}^3 I_{iq});\\
&I_{ijk}=\frac{I_{jk}-I_{ik}}{a_i^2-a_j^2},\;\;\text{for}\;\;i\neq j\neq k;\;\;\;\;\;\;\;I_{iij}=\frac{I_{ij}-I_{ii}}{a_i^2-a_j^2},\;\;\text{for}\;\;i\neq j;\quad I_{iii}=\frac{1}{5}(\frac{1}{a_i^4}-\sum_{q=1,q\neq i}^3 I_{iiq}),
\end{split}
\eeqs
where the  summation convention is not utilized. Then substituting \eqref{I} and \eqref{I0} into \eqref{coe} yields
\beqs\label{J}
\begin{split}
&J_1=-\frac{1}{6}+\frac{1}{24}(4a_1^2+3a_2^2)I_{12}+\frac{1}{24}(4a_1^2+3a_3^2)I_{13};\\
&J_2=-\frac{1}{6}+\frac{1}{24}(4a_2^2+3a_1^2)I_{21}+\frac{1}{24}(4a_2^2+3a_3^2)I_{23};\\
&J_3=-\frac{1}{6}+\frac{1}{24}(4a_3^2+3a_1^2)I_{31}+\frac{1}{24}(4a_3^2+3a_2^2)I_{32};\\
&J_4=\frac{1}{4}(I_1+I_2)-\frac{3}{4}(a_1^2+a_2^2)I_{12}+\frac{a_1a_2a_3}{8}\int_0^{+\infty}\frac{s}{(a_1^2+s)(a_2^2+s)\sqrt{\prod_{q=1}^3(a_q^2+s)}}ds;\\
&J_5=\frac{1}{4}(I_2+I_3)-\frac{3}{4}(a_2^2+a_3^2)I_{23}+\frac{a_1a_2a_3}{8}\int_0^{+\infty}\frac{s}{(a_2^2+s)(a_3^2+s)\sqrt{\prod_{q=1}^3(a_q^2+s)}}ds;\\
&J_6=\frac{1}{4}(I_3+I_1)-\frac{3}{4}(a_3^2+a_1^2)I_{31}+\frac{a_1a_2a_3}{8}\int_0^{+\infty}\frac{s}{(a_3^2+s)(a_1^2+s)\sqrt{\prod_{q=1}^3(a_q^2+s)}}ds,\\
\end{split}
\eeqs
which are always valid even when two of $a_1,a_2,a_3$ are equal. Then we consider three cases concerning the shape of the ellipsoid.
\begin{enumerate}
\item When the ellipsoid is spherical.

In this case, $a_1=a_2=a_3$. Then it can be derived from \eqref{J} that
\beas
J_1=J_2=J_3=-\frac{1}{20},\;\;J_4=J_5=J_6=-\frac{1}{10},
\eeas
substitution of which into \eqref{varphi4} leads to the explicit expression of the summation of the forth-degree terms $\varphi^{(4)}_{[\text{sph}]}(\bfz)$ of the Newtonian potential induced by a sphere, i.e.,
\beqs\label{varhphi4sphere}
\varphi^{(4)}_{[\text{sph}]}(\bfz)=-\frac{1}{20}(z_1^4+z_2^4+z_3^4)-\frac{1}{10}(z_1^2z_2^2+z_2^2z_3^2+z_3^2z_1^2).
\eeqs
Since \eqref{varhphi4sphere} does not satisfy either \eqref{forth1} or \eqref{forth2}, we conclude that $\varphi^{(4)}_{[\text{sph}]}(\bfz)\neq \varphi^{(4)}(\bfz)$, which implies $\varphi(\bfz)$ cannot be the Newtonian potential induced by a sphere with the mass density $\rho$.

\item When the ellipsoid is spheroidal.

In this case, without loss of the generality, we take $a_1=a_2= \frac{ a_3}{e}$, and $\;e> 0$.
\begin{enumerate}
\item Oblate spheroid: $e<1$.

When $e<1$, it can be derived from \eqref{J} that
\beqs\label{obl}
\begin{split}
&J_1=J_2=-\frac{(2e^2-23)e^2\sqrt{1-e^2}+3e(3+4e^2)\arccos e}{64(1-e^2)^{\frac{5}{2}}};\\
 &J_3=-\frac{-(2+19e^2)\sqrt{1-e^2}+3e(3+4e^2)\arccos e}{24(1-e^2)^{\frac{5}{2}}};\\
&J_4=-\frac{(2e^2-23)e^2\sqrt{1-e^2}+3e(3+4e^2)\arccos e}{32(1-e^2)^{\frac{5}{2}}};\\
&J_5=J_6=-\frac{(2e^4+15e^2+4)\sqrt{1-e^2}-3e(3+4e^2)\arccos e}{8(1-e^2)^{\frac{5}{2}}},\\
\end{split}
\eeqs
where there is
\beqs\label{obl1}
J_1=J_2=\frac{J_4}{2}.
\eeqs
Let $\varphi^{(4)}_{[\text{obl}]}$ denote the summation of the forth-degree terms of the Newtonian potential induced by an oblate spheroid.
If $\varphi^{(4)}_{[\text{obl}]}$ is expressed as \eqref{forth1}, comparing \eqref{forth1} with \eqref{varphi4} leads to $J_1, J_2, J_3\neq0$ and  $J_4=J_5=J_6=0$, which contradicts \eqref{obl1}. Thus, $\varphi^{(4)}_{[\text{obl}]}$ cannot be expressed as \eqref{forth1}.

If $\varphi^{(4)}_{[\text{obl}]}$ is expressed as $\eqref{forth2}_1$, comparing $\eqref{forth2}_1$ with \eqref{varphi4} leads to $J_5=J_6=0$ and $J_1=J_2=\frac{1}{3}J_3=\frac{1}{6}J_4\neq 0$, which also contradicts \eqref{obl1}. Other situations when $\varphi^{(4)}_{[\text{obl}]}$ is expressed as $\eqref{forth2}_2$ or $\eqref{forth2}_3$ can be analysed in the same way.
Hence $\varphi^{(4)}_{[\text{obl}]}$ cannot be expressed as \eqref{forth2}, either.

Through the same method, we can discuss the case when $a=c=\frac{b}{e}$ and $b=c=\frac{a}{e}$. Therefore, we conclude that $\varphi^{(4)}_{[\text{obl}]}(\bfz)\neq \varphi^{(4)}(\bfz)$, which implies $\varphi(\bfz)$ cannot be the Newtonian potential induced by an oblate spheroid with the mass density $\rho$.

\item Prolate spheroid: $e>1$.

When $e>1$, it can be derived from \eqref{J} that
\beqs\label{pro}
\begin{split}
&J_1=J_2=-\frac{(2e^2-23)e^2\sqrt{e^2-1}+3e(3+4e^2)\cosh^{-1} e}{64(e^2-1)^{\frac{5}{2}}};\\
 &J_3=-\frac{-(2+19e^2)\sqrt{e^2-1}+3e(3+4e^2)\cosh^{-1} e}{24(e^2-1)^{\frac{5}{2}}};\\
&J_4=-\frac{(2e^2-23)e^2\sqrt{e^2-1}+3e(3+4e^2)\cosh^{-1} e}{32(e^2-1)^{\frac{5}{2}}};\\
 &J_5=J_6=-\frac{(2e^4+15e^2+4)\sqrt{e^2-1}-3e(3+4e^2)\cosh^{-1} e}{8(e^2-1)^{\frac{5}{2}}},\\
\end{split}
\eeqs
where there is also a result in \eqref{obl1}, and the only difference between \eqref{obl} and \eqref{pro} is the replacement of `$\arccos e$' with `$\cosh^{-1} e$'.
By a similar analysis to that for the oblate spheroid, we can reach the conclusion that $\varphi(\bfz)$ cannot be the Newtonian potential induced by a prolate spheroid with the mass density $\rho$.
\end{enumerate}

\item When the ellipsoid is in a general shape: $a_1\neq a_2\neq a_3$.

We divide our analysis into two parts:
\begin{enumerate}
\item Firstly, we are going to prove that \eqref{forth1} cannot be the summation of the forth-degree terms of the polynomial Newtonian potential $\varphi^{(4)}_{[\text{gen}]}$ induced by a general ellipsoid with $a_1\neq a_2\neq a_3$ and the mass density $\rho$.

    When $a_1\neq a_2\neq a_3$, we can simplify the expression of $J_i\;(i=1,2,3,4,5,6)$ in \eqref{J}, i.e.,
\beqs\label{Jsimple}
\begin{split}
&J_1=-\frac{1}{6}+\frac{(I_2-I_1)(4a_1^2+3a_2^2)}{24(a_1^2-a_2^2)}+\frac{(I_3-I_1)(4a_1^2+3a_3^2)}{24(a_1^2-a_3^2)};\\
&J_2=-\frac{1}{6}+\frac{(I_2-I_1)(4a_2^2+3a_1^2)}{24(a_1^2-a_2^2)}+\frac{(I_3-I_2)(4a_2^2+3a_3^2)}{24(a_2^2-a_3^2)};\\
&J_3=-\frac{1}{6}+\frac{(I_3-I_1)(4a_3^2+3a_1^2)}{24(a_1^2-a_3^2)}+\frac{(I_3-I_2)(4a_3^2+3a_2^2)}{24(a_2^2-a_3^2)};\\
&J_4=\frac{(5a_1^2+2a_2^2)I_1-(5a_2^2+2a_1^2)I_2}{4(a_1^2-a_2^2)};\\
& J_5=\frac{(5a_2^2+2a_3^2)I_2-(5a_3^2+2a_2^2)I_3}{4(a_2^2-a_3^2)};\\
& J_6=\frac{(5a_3^2+2a_1^2)I_3-(5a_1^2+2a_3^2)I_1}{4(a_3^2-a_1^2)}.
\end{split}
\eeqs
If $\varphi^{(4)}_{[\text{gen}]}$ is given as \eqref{forth1}, comparing \eqref{forth1} with \eqref{varphi4} yields $J_4=J_5=J_6=0$. Then substituting $J_4=J_5=J_6=0$ into \eqref{Jsimple} leads to
\beqs\label{gen1}
\begin{bmatrix}
5a_1^2+2a_2^2&-(5a_2^2+2a_1^2)&0\\
0&5a_1^2+2a_3^2&-(5a_3^2+2a_2^2)\\
5a_1^2+2a_3^2&0&-(5a_3^2+2a_1^2)
\end{bmatrix}\cdot\begin{bmatrix}
I_1\\
I_2\\
I_3
\end{bmatrix}=
\begin{bmatrix}
0\\
0\\
0
\end{bmatrix}.
\eeqs
Since
\beqs
\begin{split}
&\left|\begin{matrix}
5a_1^2+2a_2^2&-(5a_2^2+2a_1^2)&0\\
0&5a_1^2+2a_3^2&-(5a_3^2+2a_2^2)\\
5a_1^2+2a_3^2&0&-(5a_3^2+2a_1^2)
\end{matrix}\right|=30(a_1^2-a_2^2)(a_2^2-a_3^2)(a_3^2-a_1^2)\neq 0,
\end{split}
\eeqs
we know that the linear system of equations with respect $(I_1,I_2,I_3)$ in \eqref{gen1} only admits a trivial solution. However, according to \eqref{I0}, we know that $I_i>0\;(i=1,2,3)$, which forms a contradiction. Hence we conclude that  $\varphi^{(4)}_{[\text{gen}]}$ cannot be in the form of \eqref{forth1}.

\item Secondly, we are going to prove that \eqref{forth2} cannot be the summation of the forth-degree terms of the polynomial Newtonian potential $\varphi^{(4)}_{[\text{gen}]}$ induced by a general ellipsoid with $a_1\neq a_2\neq a_3$ and the mass density $\rho$.

    Based on \eqref{forth2}, we take $J_4=J_5=0,J_6\neq 0$, and the situations when $J_4=J_6=0,J_5\neq 0$ and $J_5=J_6=0,J_4\neq 0$ can be analysed through the same procedure.
When $J_4=J_5=0,J_6\neq 0$, comparing\eqref{forth2} with \eqref{varphi4} and \eqref{Jsimple} yields
\beqs\label{Jsimple2}
\begin{split}
&I_1=\frac{5a_2^2+2a_1^2}{5a_1^2+2a_2^2}\frac{5a_3^2+2a_2^2}{5a_2^2+2a_3^2}I_3;\quad I_2=\frac{5a_3^2+2a_2^2}{5a_2^2+2a_3^2}I_3;\\
&J_1=-\frac{1}{6}+\frac{2 a_1^2 (9 a_2^2 + 5 a_3^2) + 3 (a_2^4 + 6 a_2^2 a_3^2)}{4(5 a_1^2 +
   2 a_2^2) (5 a_2^2 + 2 a_3^2)}I_3=-\frac{1}{24};\\
&J_2=-\frac{1}{6}+\frac{48 a_2^4 + 78 a_2^2 a_3^2 + a_1^2 (78 a_2^2 + 90 a_3^2)}{24(5 a_1^2 +
   2 a_2^2) (5 a_2^2 + 2 a_3^2)}I_3=-\frac{1}{12};\\
&J_3=-\frac{1}{6}+\frac{2 a_1^2 (9 a_2^2 + 5 a_3^2) + 3 (a_2^4 + 6 a_2^2 a_3^2)}{4(5 a_1^2 + 2 a_2^2) (5 a_2^2 + 2 a_3^2)}I_3=-\frac{1}{24};\\
&J_4=J_5=0;\;\;\;\;J_6=-\frac{15 (a_1^2 - a_2^2) (a_2^2 - a_3^2) }{2(5 a_1^2 + 2 a_2^2) (5 a_2^2 + 2 a_3^2)}I_3=-\frac{1}{4},\\
\end{split}
\eeqs
which are valid only when
\beqs\label{haha}
\begin{split}
&\frac{2 a_1^2 (9 a_2^2 + 5 a_3^2) + 3 (a_2^4 + 6 a_2^2 a_3^2)}{(5 a_1^2 + 2 a_2^2) (5 a_2^2 + 2 a_3^2)}I_3=\frac{15 (a_1^2 - a_2^2) (a_2^2 - a_3^2) }{(5 a_1^2 + 2 a_2^2) (5 a_2^2 + 2 a_3^2)}I_3=\frac{1}{2},\\
 &\frac{48 a_2^4 + 78 a_2^2 a_3^2 + a_1^2 (78 a_2^2 + 90 a_3^2)}{(5 a_1^2 + 2 a_2^2) (5 a_2^2 + 2 a_3^2)}I_3=2.
\end{split}
\eeqs
It follows from $\eqref{haha}_1$ that
\beqs
\label{hahaha}
\begin{split}
&2 a_1^2 (9 a_2^2 + 5 a_3^2) + 3 (a_2^4 + 6 a_2^2 a_3^2)=15 (a_1^2 - a_2^2) (a_2^2 - a_3^2)\\
\;\;\Rightarrow\; \;&a_2^2 (6 a_2^2 + a_3^2) + a_1^2 (3 a_2^2 + 25 a_3^2)=0.
\end{split}
\eeqs
However, $a_2^2 (6 a_2^2 + a_3^2) + a_1^2 (3 a_2^2 + 25 a_3^2)>0$, which indicates that \eqref{hahaha} is impossible. Hence we conclude that $\varphi$ cannot be the Newtonian potential induced by a general ellipsoid with $a_1\neq a_2\neq a_3$ and the mass density $\rho$.
\end{enumerate}
\end{enumerate}
In conclusion,   by reductio, we have proved that $\varphi$ introduced in \eqref{phi001} is not equal to the Newtonian potential induced by spheres, spheroids, and general ellipsoids with {any} orientation, {any} position, and the fixed mass density $\rho$, which implies the completion of the proof that $\varphi$ cannot be the Newtonian potential induced by any ellipsoidal inclusion which possesses the mass density $\rho$. Therefore, the proof of \eqref{construct2} is completed, and thus an $\Omega'$ that leads to \eqref{construct1} must be non-ellipsoidal,  combining which with the proof of the existence of an $\Omega'$ that generates \eqref{construct1} in the first part leads to the proof of Lemma \ref{keylemma}.

\setcounter{equation}{0}
\renewcommand\theequation{D.\arabic{equation}}
\renewcommand\thetheorem{D.\arabic{theorem}}
\renewcommand\thefigure{D.\arabic{figure}}
\section*{D. \quad The shapes of counter-example non-ellipsoidal inclusions}

\subsection*{(1)\quad Construction of $\Omega^{(1)}$}

The counter-example inclusion $\Omega^{(1)}$ will be constructed by searching for an $\Omega'$ that generates
\beqs\label{constructforomega1}
N_{\Omega'}[\rho](\bfx'){=}-\int_{\Omega'} \frac{\rho(\bfy')}{4\pi|\bfx'-\bfy'|} d\bfy'=C-\frac{1}{12}\left({x'_1}^4+{x'_2}^4+{x'_3}^4\right), \;\;\bfx'\in \Omega',
\eeqs
where $C$ is a positive real constant and $\rho(\bfy')=-|\bfy'|^2$, and then transforming $\Omega'$  into $\Omega^{(1)}$ via the transformation
\beqs\label{shapeforomega1}
\Omega^{(1)} = \left\{ {\bfx'\left| {\left( {\begin{array}{*{20}{c}}
  1&0&0 \\
  0&1&0 \\
  0&0&{s}
\end{array}} \right)\cdot\bfx'} \right. \in \Omega' } \right\}
\eeqs
with $s$ a positive real constant.
Substituting \eqref{constructforomega1} into \eqref{keyequation} results in the quadratic elastic strain $\boldsymbol{\varepsilon}$ induced by $\Omega^{(1)}$ when subjected to the quadratic eigenstress $\boldsymbol{\sigma}^*$ defined by \eqref{eigenstresscubic} and \eqref{rhoquadratic}, which verifies $\Omega^{(1)}$ is a counter-example inclusion for a quadratic eigenstrain.

Since $\Omega^{(1)}$ is constructed due to \eqref{shapeforomega1}, we then describe the scheme for the determination of the shape of $\Omega'$.
According to Appendix C, we know that \eqref{constructforomega1} is the solution of the over-determined problem \eqref{Liu} for $\bfx'\in\Omega'$, when the corresponding obstacle function is taken as
\beqs\label{phicounter1}
\phi(\bfx') = C-\frac{1}{12}\left({x'_1}^4+{x'_2}^4+{x'_3}^4\right),
\eeqs
which is consistent with the right-hand side of \eqref{constructforomega1}.
Let $V_\phi(\bfx')$ denote the solution of the over-determined problem \eqref{Liu} under the condition \eqref{phicounter1}. It is known in Appendix C that $V_\phi(\bfx')$ can be directly solved by the variational inequality \eqref{inequality}. Once $V_\phi(\bfx')$ is obtained, the shape of $\Omega'$ can be constructed by assembling the points where $V_\phi(\bfx')=\phi(\bfx')$, since by definition,
\beqs\label{specificomega}
\Omega'=\{\;\;\bfx'\;\;| \;V_\phi(\bfx')=\phi(\bfx'),\;\;\bfx'\in\mathbb{R}^3\}.
\eeqs

According to the work of \cite{Liu2008},  the variational inequality \eqref{inequality} can be discretized into
\beqs\label{discretized}
\min\left\{\hat{\prod}(\hat{\bfv})=-\frac{1}{2}\hat{\bfV}_\phi\cdot\hat{\bfK}\cdot\hat{\bfV}_\phi,\;\hat{\bfV}_\phi\geq \hat{\boldsymbol{\phi}}\right\},
\eeqs
where $\hat{\bfV}_\phi$ denotes the vector whose components are the values of $V_\phi$ at the nodal points when using discretization of the finite element method; $\hat{\boldsymbol{\phi}}$  denotes the vector whose components are the values of $\phi$ at the nodal points; and $\hat{\bfK}$ is the stiffness tensor corresponding to the Laplacian equation $\Delta V_\phi=0$ discretized via the finite element method. Note that \eqref{discretized} is a standard quadratic programming problem that can be easily solved.

Therefore, the construction of $\Omega^{(1)}$ is  achieved by two steps. Firstly, we let $C=\frac{1}{36}$ in \eqref{phicounter1} and then numerically solve \eqref{discretized} to obtain $V_\phi(\bfx')$ and thus the shape of $\Omega'$ via \eqref{specificomega}. Secondly, based on the shape of $\Omega'$, we let $s=0.5$ in \eqref{shapeforomega1} and then construct $\Omega^{(1)}$ via \eqref{shapeforomega1}. The shape of $\Omega^{(1)}$ is shown in Figure~\ref{counterexample1} below. The notation $\{x,y,z\}$ denoting the axes of the coordinates in the figure corresponds to $\{x_1,x_2,x_3\}\ (\mathrm{or}\;\{x_1',x_2',x_3'\})$ in the main text. More counter-example inclusions can be constructed by choosing other $C$ and $s$.

 \begin{figure}[H]
\centering
\subfigure[Nodal points where $|\{\hat{V}_\phi\}_k- \{\hat{\phi}\}_k|\leq 1\times 10^{-4}$ with  $k$ the sequence of the nodes. ]{
\begin{minipage}[t]{0.4\linewidth}
\center
\includegraphics[width=1.1in]{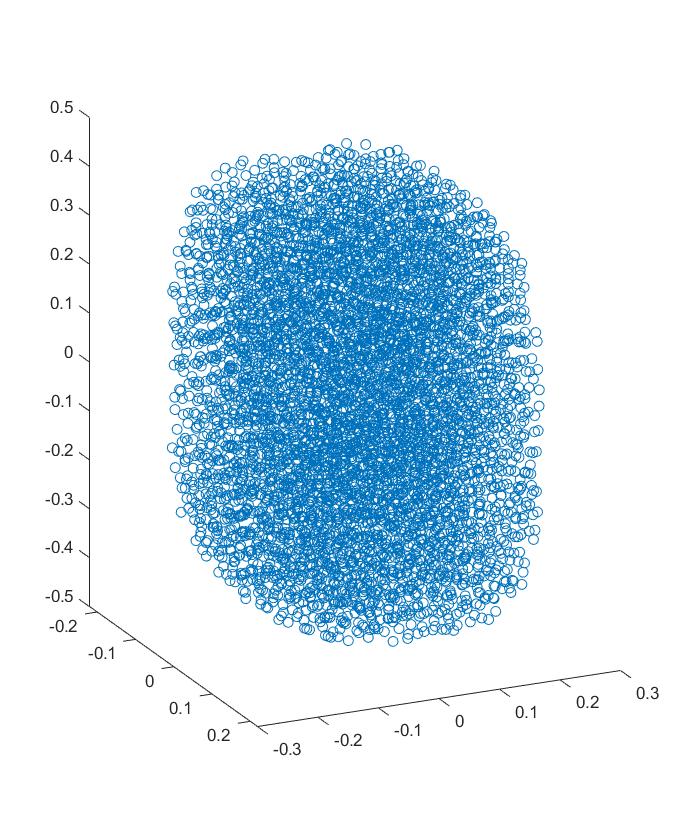}
\end{minipage}%
}%
\subfigure[The configuration of $\Omega^{(1)}$ assembled by the nodal points in (a).]{
\begin{minipage}[t]{0.5\linewidth}
\center
\includegraphics[width=1.1in]{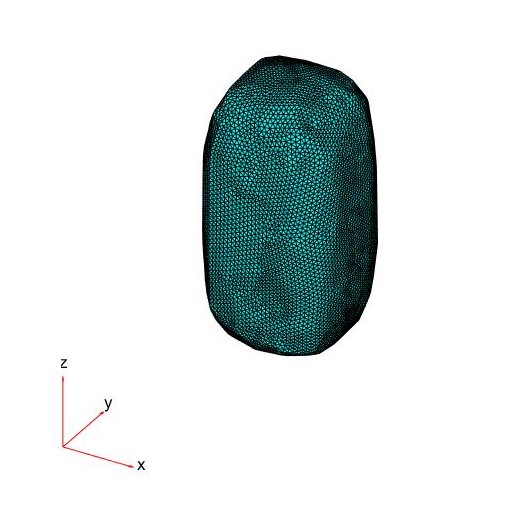}
\end{minipage}%
}%

\subfigure[The view of $\Omega^{(1)}$ in x-y plane.]{
\begin{minipage}[t]{0.33\linewidth}
\center
\includegraphics[width=1.3in]{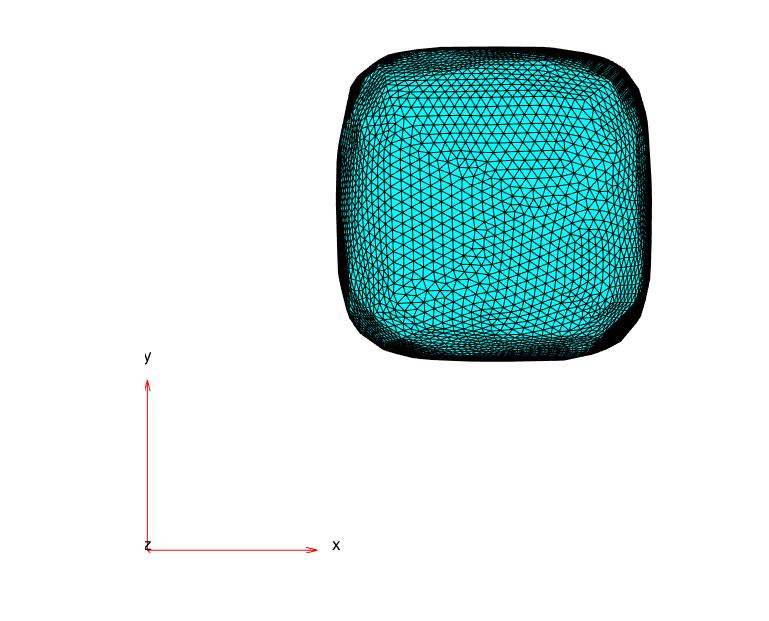}
\end{minipage}
}%
\subfigure[The view of $\Omega^{(1)}$ in x-z plane.]{
\begin{minipage}[t]{0.33\linewidth}
\center
\includegraphics[width=1.3in]{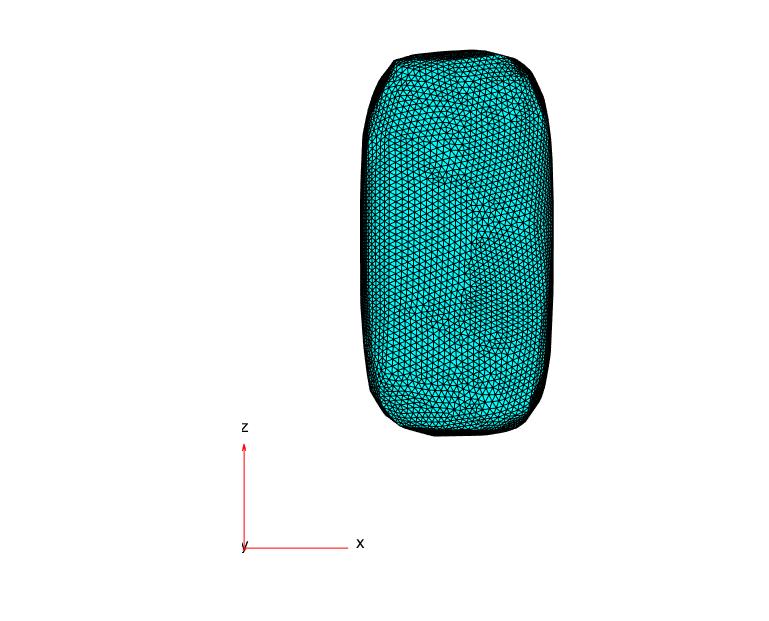}
\end{minipage}
}%
\subfigure[The view of $\Omega^{(1)}$ in y-z plane.]{
\begin{minipage}[t]{0.33\linewidth}
\center
\includegraphics[width=1.3in]{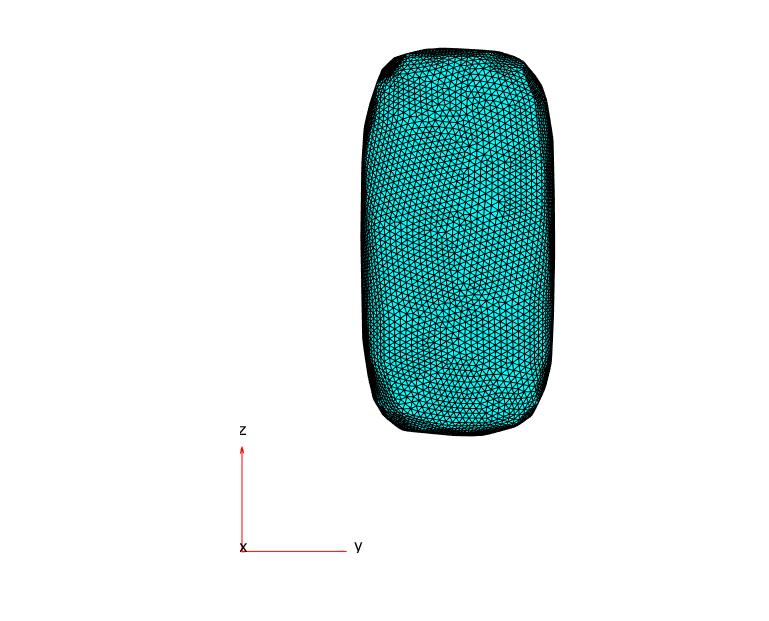}
\end{minipage}
}%
\centering
\caption{ The counter-example $\Omega^{(1)}$ for a quadratic eigenstrain  when $C=\frac{1}{36}$ in \eqref{phicounter1}, and $s=0.5$ in \eqref{shapeforomega1}.}\label{counterexample1}
\end{figure}

\subsection*{(2)\quad  Construction of $\Omega^{(2)}$}

The counter-example inclusion $\Omega^{(2)}$ will be constructed by searching for an $\bar{\Omega}$ that generates
\beqs\label{constructforomega2}
\begin{split}
N_{\bar{\Omega}}[\rho](\bfx')&=-\!\!\!\int_{\bar{\Omega}}\! \frac{\rho(\bfy')}{4\pi|\bfx'-\bfy'|} d\bfy'\\&=\hat{C}-\frac{1}{30}\left({x'_1}^{6}+{x'_2}^{6}+{x'_3}^{6}\right)-\beta\log \frac{(x_1'-12\sqrt{C})^2+(x_2'-12\sqrt{C})^2}{36C}, \;\;\bfx'\in \bar{\Omega}.
\end{split}
\eeqs
where $\beta$, $C$ and $\hat{C}$ are positive real constants and $\rho(\bfy')=-({y'_1}^4+{y'_2}^4+{y'_3}^4)$, and then transforming $\bar{\Omega}$  into $\Omega^{(2)}$ via the transformation
\beqs\label{shapeforomega2}
\Omega^{(2)} = \left\{ {\bfx'\left| {\left( {\begin{array}{*{20}{c}}
  s_1&0&0 \\
  0&s_2&0 \\
  0&0&{1}
\end{array}} \right)\cdot\bfx'} \right. \in \bar{\Omega} } \right\}
\eeqs
with $s_1$ and $s_2$ two positive real constants. Here $s_1=\sqrt{\frac{a}{c}}$ and $s_2=\sqrt{\frac{b}{c}}$, where $a,b,c$ are defined in \eqref{abc}.
Substituting \eqref{constructforomega2} into \eqref{keyequationforanydegree} results in the quartic elastic strain induced by $\Omega^{(2)}$ when subjected to the quartic eigenstress $\boldsymbol{\sigma}^*$ defined by \eqref{eigenstresscubic} and \eqref{rhohighorder}, which verifies $\Omega^{(2)}$ is a counter-example inclusion for a quartic eigenstrain.

In this case, we take an obstacle function $\phi(\bfx')$ expressed as
\beqs\label{phicounter2}
\phi(\bfx')=\hat{C}-\frac{1}{30}\left({x'_1}^{6}+{x'_2}^{6}+{x'_3}^{6}\right)-\beta\log \frac{(x_1'-12\sqrt{C})^2+(x_2'-12\sqrt{C})^2}{36C},
\eeqs
which is consistent with the right-hand side of \eqref{constructforomega2}. Then based on \eqref{phicounter2}, $\bar{\Omega}$ can be determined by
\beqs\label{specificomega2}
\bar{\Omega}=\{\;\;\bfx'\;\;| \;V_\phi(\bfx')=\phi(\bfx'),\;\;\bfx'\in\mathbb{R}^3\}
\eeqs
once $V_\phi(\bfx')$ is numerically solved by \eqref{discretized}. Further, $\Omega^{(2)}$ is constructed by \eqref{shapeforomega2}.

Specifically, the construction of $\Omega^{(2)}$ is also achieved by two steps. Firstly, we let $C=\hat{C}=\frac{1}{36}$ and $\beta=\frac{1}{600}$ in \eqref{phicounter2} and then numerically solve \eqref{discretized} to obtain $V_\phi(\bfx')$ and thus the shape  of $\bar{\Omega}$ via \eqref{specificomega2}. Secondly, based on the shape of $\bar{\Omega}$, we let $s_1=0.5$ and $s_2=1.5$ in \eqref{shapeforomega2} and then construct $\Omega^{(2)}$ via \eqref{shapeforomega2}. The shape of $\Omega^{(2)}$ is shown in Figure~\ref{counterexample2} below. The notation $\{x,y,z\}$ denoting the axes of the coordinates in the figure is corresponding to $\{x_1,x_2,x_3\}(\mathrm{or}\;\{x_1',x_2',x_3'\})$ in the main text. And more counter-example inclusions can be constructed by choosing other $C$, $\hat{C}$, $\beta$, $s_1$ and $s_2$.



 \begin{figure}[H]
\centering
\subfigure[Nodal points where $|\{\hat{V}_\phi\}_k- \{\hat{\phi}\}_k|\leq 1\times 10^{-4}$ with  $k$ the sequence of the nodes.]{
\begin{minipage}[t]{0.4\linewidth}
\center
\includegraphics[width=1.5in]{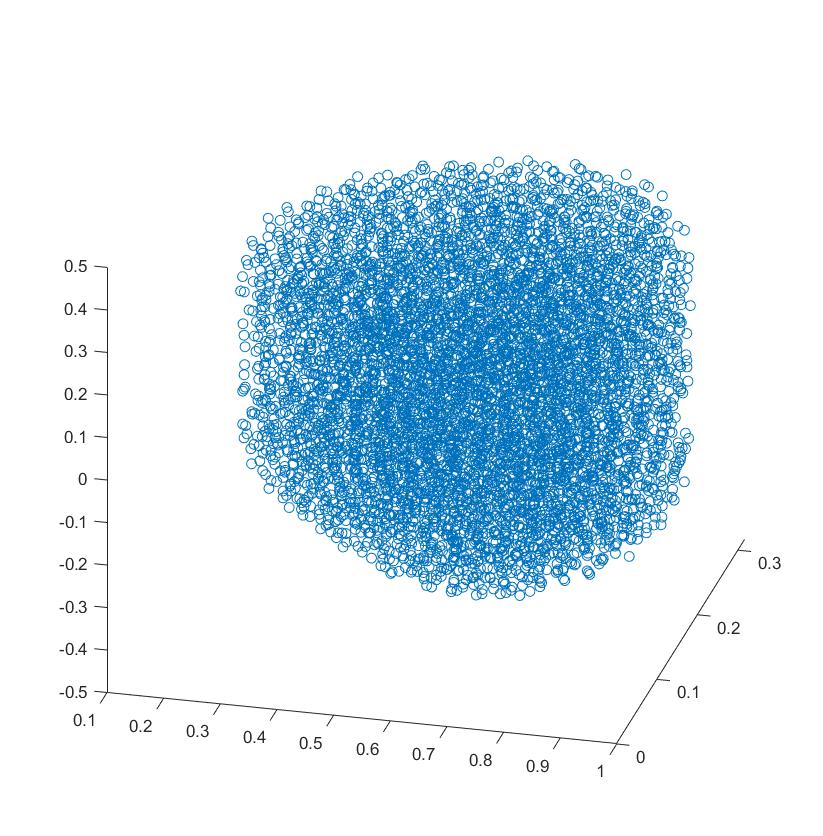}
\end{minipage}%
}%
\subfigure[The configuration of $\Omega^{(2)}$ assembled by the nodal points in (a).]{
\begin{minipage}[t]{0.5\linewidth}
\center
\includegraphics[width=1.7in]{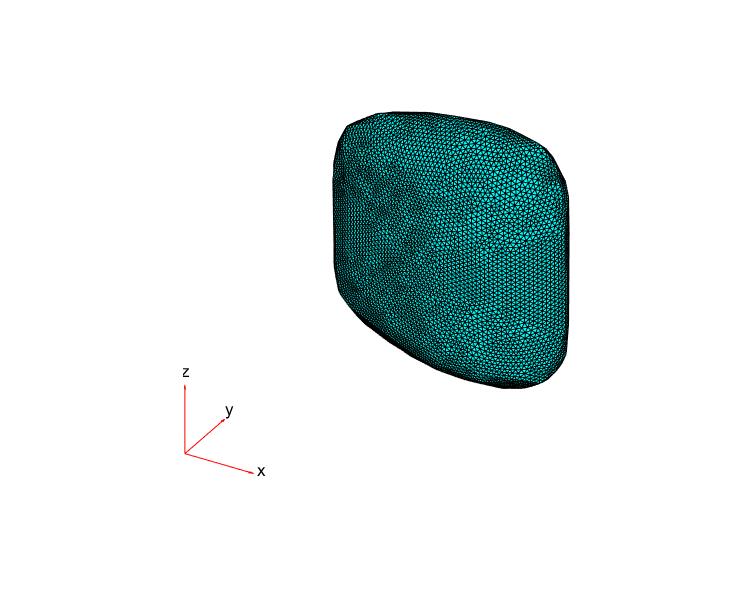}
\end{minipage}%
}%

\subfigure[The view of $\Omega^{(2)}$ in x-y plane.]{
\begin{minipage}[t]{0.33\linewidth}
\center
\includegraphics[width=1.2in]{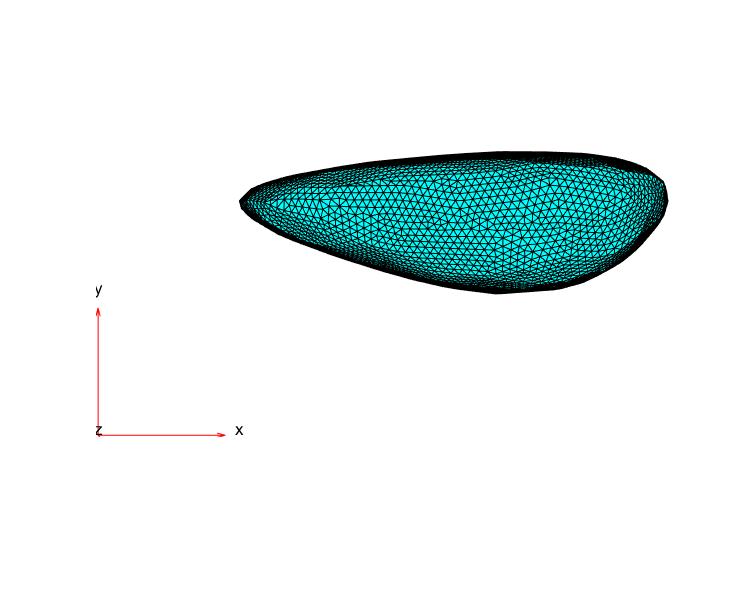}
\end{minipage}
}%
\subfigure[The view of $\Omega^{(2)}$ in x-z plane.]{
\begin{minipage}[t]{0.33\linewidth}
\center
\includegraphics[width=1.2in]{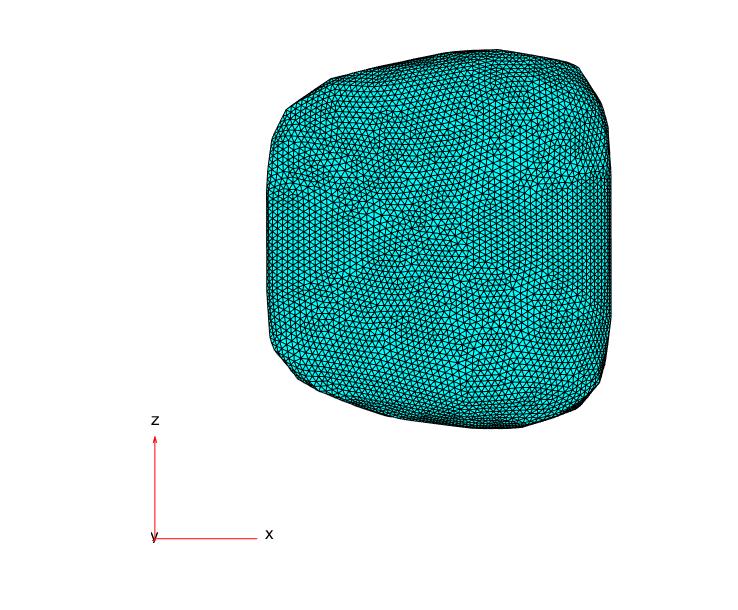}
\end{minipage}
}%
\subfigure[The view of $\Omega^{(2)}$ in y-z plane.]{
\begin{minipage}[t]{0.33\linewidth}
\center
\includegraphics[width=1.2in]{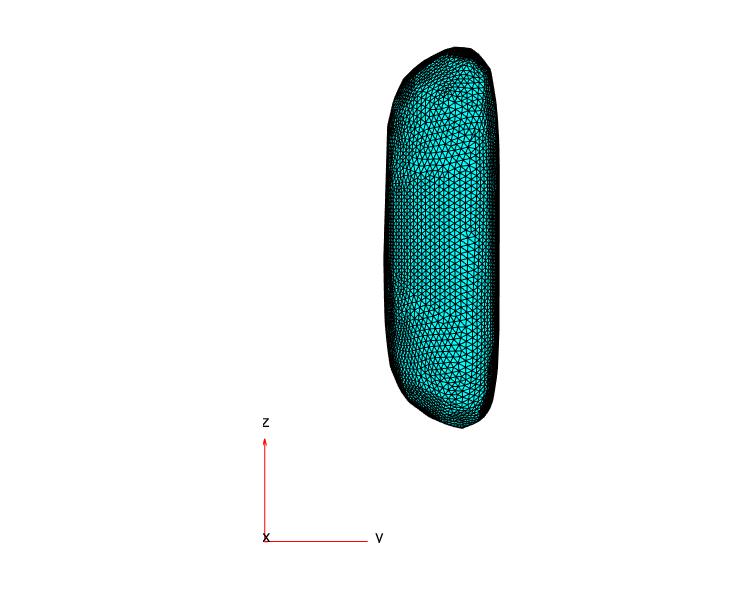}
\end{minipage}
}%
\centering
\caption{ The counter-example $\Omega^{(2)}$ for a quartic eigenstrain  when $C=\hat{C}=\frac{1}{36}$ and $\beta=\frac{1}{600}$ in \eqref{phicounter2}, and $s_1=0.5$ and $s_2=1.5$ in \eqref{shapeforomega2}.}\label{counterexample2}
\end{figure}

\setcounter{equation}{0}
\renewcommand\theequation{E.\arabic{equation}}
\renewcommand\thetheorem{E.\arabic{theorem}}
\renewcommand\thefigure{E.\arabic{figure}}

\section*{E.\quad The proof of the existence of an $\Omega'$ that yields \eqref{NewformnNewtonian}}


We introduce
\beqs\label{phi02}
\phi'{:=}\phi^*+\omega^*,
\eeqs
where $\phi^*$ is introduced in \eqref{newphi},  and
\beqs\label{Omegastar}
\omega^*(x_1',x_2'){:=}\left\{ \begin{array}{*{20}{c}}
  &{0\;\;\;\;\;\;\;\;\;\;\;\;\;\;\;\;\;\;\;\;\;\;\bfx' \in U^{(1)}, } \\
  &{-\beta\log9\;\;\;\;\;\;\;\;\;\bfx' \in U^{(2)}, }  \\
  &{\omega\;\;\;\;\;\;\;\;\;\;\;\;\;\;\;\;\;\;\;\;\bfx' \in U^\omega, }
\end{array}  \right.\;
\eeqs
with $\beta$ being a positive real constant, $\omega$ being defined in \eqref{omegaform}, $U^{(1)}{:=}\{\bfx'|(x_1'-12\sqrt{C})^2+(x_2'-12\sqrt{C})^2\leq 36C,\;\bfx'\in \mathbb{R}^3\}$, $U^{(2)}{:=}\{\bfx'|(x_1'-12\sqrt{C})^2+(x_2'-12\sqrt{C})^2\geq 324C,\;\bfx'\in \mathbb{R}^3\}$ and $U^\omega{:=}\{\bfx'|$ $ 324C\geq(x_1'-12\sqrt{C})^2+(x_2'-12\sqrt{C})^2\geq 36C,\;\bfx'\in \mathbb{R}^3\}$.

Then we will verify that $\phi'$ is an obstacle function defined in Appendix C,
 by demonstrating that  $\phi'$  satisfies all of the four conditions.

Firstly, it has been proved in Appendix C that $\phi^*\in C^{0}(\mathbb{R}^{3})$. Since $\omega^*\in C^{0}(\mathbb{R}^{3})$ due to \eqref{Omegastar} and \eqref{omegaform}, we see $\phi'\in C^{0}(\mathbb{R}^{3})$.  Further, by substituting \eqref{phi02} into \eqref{normdefinition}, we obtain
\beqs\label{phi01}
||\phi'(\bfx')||_{0,1}\leq\sup_{\bfx'\in {U}}|{\varphi}(\bfx')|+\sup_{\bfx'\in {U}}|\nabla_{x'}{\varphi}(\bfx')|+\sup_{\bfx'\in U^\omega}|\omega(x_1',x_2')|+\sup_{\bfx'\in  U^\omega}|\nabla_{x'}\omega(x_1',x_2')|,
\eeqs
where $\varphi$ is given in \eqref{phi001}, and $U\in \mathbb{R}^3$ is a bounded domain defined in \eqref{newphi}.
$|\varphi(\bfx')|$ and $|\nabla_{x'}\varphi(\bfx')|$ are bounded in $U$ owing to \eqref{newphi}, and $|\omega(x_1',x_2')|$ and $|\nabla_{x'}\omega(x_1',x_2')|$ are bounded in $U^\omega$ owing to \eqref{omegaform} , which means that
 the norm $||\phi'(\bfx')||_{0,1}$ of $\phi'$ is bounded, and thus $\phi'\in C^{0,1}(\mathbb{R}^3)$. Hence, $\phi'$ satisfies the first condition of an obstacle function.


 Secondly, we let $r_0:=6\sqrt{C}$. It has been proved in Appendix C that $\forall \;|\bfx'|\geq r_0$, $\phi^*(\bfx')\leq 0$; in addition, \eqref{Omegastar} and \eqref{omegaform} imply $\omega^*(x'_1, x_2')\leq 0$ for any $\bfx'$. We conclude that $\forall \;|\bfx'|\geq r_0$, $\phi'(\bfx')=\phi^*(\bfx')+\omega^*(x'_1, x_2')\leq 0$, which proves that $\phi'$ satisfies the second condition of an obstacle function.

 Thirdly, based on \eqref{newphi} and \eqref{Omegastar}, it can be verified that $\phi'\in C^{\infty}(B_{r_0})$. Hence $U^*=\emptyset$, where $U^*\subset B_{r_0}$ is defined as the set of singular points where $\nabla\otimes\nabla\phi'$ is unbounded.
Thus we conclude that $|\Delta\phi'|$ is bounded in $B_{r_0}\setminus U^*$, which proves that $\phi'$ satisfies the third condition of an obstacle function.

Fourthly, by introducing $U^{\vartheta}$ that denotes the compact support of the smooth function $\vartheta$, on which $\vartheta\geq 0$, and resorting to \eqref{judgeCnew},   it can be derived from \eqref{phi02} that $\forall \boldsymbol{\zeta}\in \bfR^3$ with $|\boldsymbol{\zeta}|=1$,
\beqs\label{judgenew1}
\begin{split}
&\int_{U^{\vartheta}}\frac{\partial^2 \vartheta }{\partial \boldsymbol{\zeta}^2}\left(\phi'+\frac{1}{2}C^{\phi'}|\bfx'|^2\right)d\bfx'\geq\int_{U^{\vartheta}}\vartheta \left(C^{\phi'}-\sup_{\bfx'\in U}\left|\frac{\partial \varphi}{\partial \boldsymbol{\zeta}}\right|-\sup_{\bfx'\in U}\left|\frac{\partial^2 \varphi}{\partial \boldsymbol{\zeta}^2}\right|\right)d\bfx'+\int_{U^{\vartheta}}\frac{\partial^2 \vartheta }{\partial \boldsymbol{\zeta}^2}\omega^* d\bfx',
\end{split}
\eeqs
whose second term satisfies
\beqs\label{judgenew2}
\begin{split}
\int_{U^{\vartheta}}\frac{\partial^2 \vartheta }{\partial \boldsymbol{\zeta}^2}\omega^* d\bfx'=&\int_{U^{\vartheta}\cap U^\omega}\frac{\partial^2 \vartheta }{\partial \boldsymbol{\zeta}^2}\omega d\bfx'+\int_{U^{\vartheta}\cap U^{(2)}}(-\beta\log9)\frac{\partial^2 \vartheta }{\partial \boldsymbol{\zeta}^2}d\bfx'\\
=&\int_{U^{\vartheta}\cap \partial U^\omega}\frac{\partial \vartheta }{\partial \boldsymbol{\zeta}}\omega \boldsymbol{\zeta}\cdot d\bfS-\int_{U^{\vartheta}\cap U^\omega}\frac{\partial \vartheta }{\partial \boldsymbol{\zeta}}\frac{\partial \omega }{\partial \boldsymbol{\zeta}} d\bfx'\\
&-\int_{U^{\vartheta}\cap \partial U^{(2)}}(\beta\log9)\frac{\partial \vartheta }{\partial \boldsymbol{\zeta}}\boldsymbol{\zeta}\cdot d\bfS+\int_{U^{\vartheta}\cap U^{(2)}}\frac{\partial \vartheta }{\partial \boldsymbol{\zeta}}\boldsymbol{\zeta}\frac{\partial (\beta\log9) }{\partial \boldsymbol{\zeta}}\boldsymbol{\zeta}d\bfx'\\
=&-\int_{U^{\vartheta}\cap \partial U^\omega}\boldsymbol{\vartheta}\frac{\partial \omega }{\partial \boldsymbol{\zeta}}\boldsymbol{\zeta}\cdot d\bfS+\int_{U^{\vartheta}\cap U^\omega}\boldsymbol{\vartheta}\frac{\partial^2 \omega }{\partial \boldsymbol{\zeta}^2} d\bfx'\\
\geq& -\int_{U^{\vartheta}}\sup_{\bfx'\in U^\omega}\left|\frac{\partial \omega}{\partial \boldsymbol{\zeta}}\right|-\sup_{\bfx'\in U^\omega}\left|\frac{\partial^2 \omega}{\partial \boldsymbol{\zeta}^2}\right|d\bfx'.
\end{split}
\eeqs
Then substitution of \eqref{judgenew2} back into \eqref{judgenew1} indicates that $\exists C^{\phi'}\in\bfR$ satisfying
\beas
\begin{split}
&C^{\phi'}\geq \sup_{\bfx'\in U}\left|\frac{\partial \varphi}{\partial \boldsymbol{\zeta}}\right|+\sup_{\bfx'\in U}\left|\frac{\partial^2 \varphi}{\partial \boldsymbol{\zeta}^2}\right|+\sup_{\bfx'\in U^\omega}\left|\frac{\partial \omega}{\partial \boldsymbol{\zeta}}\right|+\sup_{\bfx'\in U^\omega}\left|\frac{\partial^2 \omega}{\partial \boldsymbol{\zeta}^2}\right|,
\end{split}
\eeas
such that $\forall \boldsymbol{\zeta}\in \bfR^3$ with $|\boldsymbol{\zeta}|=1$,
\beas
\int_{U^{\vartheta}}\frac{\partial^2 \vartheta }{\partial \boldsymbol{\zeta}^2}\left(\phi'+\frac{1}{2}C^{\phi'}|\bfx'|^2\right)d\bfx'\geq 0,
\eeas
which proves that $\phi'$ satisfies the fourth condition of an obstacle function.
Therefore, we have shown that $\phi'$ is  an obstacle function for the over-determined problem \eqref{Liu}, which results in the existence of a coincident set $\Omega'\subseteq B_{r_0}$, where $N_{\Omega'}[\rho](\bfx')=\phi'(\bfx')$ for $\bfx' \in \Omega'$. Further, since $\Omega'\subseteq B_{r_0}\subset (U\cap U^\omega)$, we conclude that $N_{\Omega'}[\rho](\bfx')=\varphi(\bfx')+\omega(x_1',x_2')$ for $\bfx' \in \Omega'$ due to \eqref{phi02}. Hence we have proved the existence of an $\bar{\Omega}$ that makes \eqref{NewformnNewtonian} hold.

\quad\\
\quad\\
\quad\\



\begin{thebibliography}{71}
\expandafter\ifx\csname natexlab\endcsname\relax\def\natexlab#1{#1}\fi
\providecommand{\url}[1]{\texttt{#1}}
\providecommand{\href}[2]{#2}
\providecommand{\path}[1]{#1}
\providecommand{\DOIprefix}{doi:}
\providecommand{\ArXivprefix}{arXiv:}
\providecommand{\URLprefix}{URL: }
\providecommand{\Pubmedprefix}{pmid:}
\providecommand{\doi}[1]{\href{http://dx.doi.org/#1}{\path{#1}}}
\providecommand{\Pubmed}[1]{\href{pmid:#1}{\path{#1}}}
\providecommand{\bibinfo}[2]{#2}
\ifx\xfnm\relax \def\xfnm[#1]{\unskip,\space#1}\fi
\bibitem[{Ammari et~al.(2010)Ammari, Capdeboscq, Kang, Lee, Milton and
  Zribi}]{Ammari2010}
\bibinfo{author}{Ammari, H.}, \bibinfo{author}{Capdeboscq, Y.},
  \bibinfo{author}{Kang, H.}, \bibinfo{author}{Lee, H.},
  \bibinfo{author}{Milton, G.W.}, \bibinfo{author}{Zribi, H.},
  \bibinfo{year}{2010}.
\newblock \bibinfo{title}{Progress on the strong {Eshelby's} conjecture and
  extremal structures for the elastic moment tensor}.
\newblock \bibinfo{journal}{Journal De Mathematiques Pures Et Appliquees}
  \bibinfo{volume}{94}, \bibinfo{pages}{93--106}.

\bibitem[{Asaro and Barnett(1975)}]{Asaro1975}
\bibinfo{author}{Asaro, R.}, \bibinfo{author}{Barnett, D.},
  \bibinfo{year}{1975}.
\newblock \bibinfo{title}{The non-uniform transformation strain problem for an
  anisotropic ellipsoidal inclusion}.
\newblock \bibinfo{journal}{Journal of the Mechanics and Physics of Solids}
  \bibinfo{volume}{23}, \bibinfo{pages}{77 -- 83}.

\bibitem[{Calvo-Jurado and Parnell(2020)}]{CarmenandParnell2020}
\bibinfo{author}{Calvo-Jurado, C.}, \bibinfo{author}{Parnell, W.J.},
  \bibinfo{year}{2020}.
\newblock \bibinfo{title}{Induced fields in isolated elliptical inhomogeneities
  due to imposed polynomial fields at infinity}.
\newblock \bibinfo{journal}{International Journal of Computer Mathematics}
  \bibinfo{volume}{97}, \bibinfo{pages}{18--29}.

\bibitem[{Cassette et~al.(2012)Cassette, Mahler, Guigner, Patriarche, Dubertret
  and Pons}]{Cassette2012}
\bibinfo{author}{Cassette, E.}, \bibinfo{author}{Mahler, B.},
  \bibinfo{author}{Guigner, J.M.}, \bibinfo{author}{Patriarche, G.},
  \bibinfo{author}{Dubertret, B.}, \bibinfo{author}{Pons, T.},
  \bibinfo{year}{2012}.
\newblock \bibinfo{title}{Colloidal {CdSe/CdS} dot-in-plate nanocrystals with
  {2D-polarized} emission}.
\newblock \bibinfo{journal}{Acs Nano} \bibinfo{volume}{6},
  \bibinfo{pages}{6741--6750}.

\bibitem[{Chen et~al.(2016)Chen, Zhang, Zopf, Jung, Zhang, Keil, Ding and
  Schmidt}]{Chen2016}
\bibinfo{author}{Chen, Y.}, \bibinfo{author}{Zhang, J.}, \bibinfo{author}{Zopf,
  M.}, \bibinfo{author}{Jung, K.}, \bibinfo{author}{Zhang, Y.},
  \bibinfo{author}{Keil, R.}, \bibinfo{author}{Ding, F.},
  \bibinfo{author}{Schmidt, O.G.}, \bibinfo{year}{2016}.
\newblock \bibinfo{title}{Wavelength-tunable entangled photons from
  silicon-integrated {III-V} quantum dots}.
\newblock \bibinfo{journal}{Nature Communications} \bibinfo{volume}{7},
  \bibinfo{pages}{10387}.

\bibitem[{Choi et~al.(2009)Choi, Koski, Sivasankar and Alivisatos}]{Choi2009}
\bibinfo{author}{Choi, C.L.}, \bibinfo{author}{Koski, K.J.},
  \bibinfo{author}{Sivasankar, S.}, \bibinfo{author}{Alivisatos, A.P.},
  \bibinfo{year}{2009}.
\newblock \bibinfo{title}{Strain-dependent photoluminescence behavior of
  {CdSe/CdS} nanocrystals with spherical, linear, and branched topologies}.
\newblock \bibinfo{journal}{Nano Letters} \bibinfo{volume}{9},
  \bibinfo{pages}{3544--3549}.

\bibitem[{Ding et~al.(2006)}]{Ding2006}
\bibinfo{author}{Ding, H.J.}, \bibinfo{author}{Chen, W.Q.}, \bibinfo{author}{Zhang, L.C.}, \bibinfo{year}{2006}.
\newblock \bibinfo{title}{Elasticity of Transversely Isotropic Materials}.
\newblock \bibinfo{publisher}{Springer}, \bibinfo{address}{Dordrecht}.

{ \bibitem[{Dive(1931)}]{Dive1931}
\bibinfo{author}{Dive, P.},
  \bibinfo{year}{1931}.
\newblock \bibinfo{title}{Attraction des ellipso{\"\i}des homog\`enes et r\'eciproques d'un th\'eor\`eme de Newton}.
\newblock \bibinfo{journal}{Bulletin de la Soci\'et\'e Math\'ematique de France} \bibinfo{volume}{59},
  \bibinfo{pages}{128--140}.}

\bibitem[{Downes et~al.(1997)Downes, Faux and O'reilly}]{Downes1997}
\bibinfo{author}{Downes, J.R.}, \bibinfo{author}{Faux, D.A.},
  \bibinfo{author}{O'reilly, E.P.}, \bibinfo{year}{1997}.
\newblock \bibinfo{title}{A simple method for calculating strain distributions
  in quantum dot structures}.
\newblock \bibinfo{journal}{Journal of Applied Physics} \bibinfo{volume}{81},
  \bibinfo{pages}{6700--6702}.
\bibitem[{Duan et~al.(2005)Duan, Wang, Huang and Karihaloo}]{Nanoinc2005}
\bibinfo{author}{Duan, H.L.}, \bibinfo{author}{Wang, J.},
  \bibinfo{author}{Huang, Z.P.}, \bibinfo{author}{Karihaloo, B.L.},
  \bibinfo{year}{2005}.
\newblock \bibinfo{title}{{Eshelby} formalism for nano-inhomogeneities}.
\newblock \bibinfo{journal}{Proceedings of the Royal Society a-Mathematical
  Physical and Engineering Sciences} \bibinfo{volume}{461},
  \bibinfo{pages}{3335--3353}.

\bibitem[{Duong et~al.(2001)Duong, Wang and Yu}]{Duong2001}
\bibinfo{author}{Duong, C.N.}, \bibinfo{author}{Wang, J.J.},
  \bibinfo{author}{Yu, J.}, \bibinfo{year}{2001}.
\newblock \bibinfo{title}{An approximate algorithmic solution for the elastic
  fields in bonded patched sheets}.
\newblock \bibinfo{journal}{International Journal of Solids and Structures}
  \bibinfo{volume}{38}, \bibinfo{pages}{4685--4699}.

\bibitem[{Dyson(1891)}]{Dyson}
\bibinfo{author}{Dyson, F.}, \bibinfo{year}{1891}.
\newblock \bibinfo{title}{The potentials of ellipsoids of variable densities}.
\newblock \bibinfo{journal}{The Quarterly Journal of Pure and Applied
  Mathematics} \bibinfo{volume}{25}, \bibinfo{pages}{259--288}.

\bibitem[{Efros et~al.(1996)Efros, Rosen, Kuno, Nirmal, Norris and
  Bawendi}]{Efros1996}
\bibinfo{author}{Efros, A.L.}, \bibinfo{author}{Rosen, M.},
  \bibinfo{author}{Kuno, M.}, \bibinfo{author}{Nirmal, M.},
  \bibinfo{author}{Norris, D.J.}, \bibinfo{author}{Bawendi, M.},
  \bibinfo{year}{1996}.
\newblock \bibinfo{title}{Band-edge exciton in quantum dots of semiconductors
  with a degenerate valence band: dark and bright exciton states}.
\newblock \bibinfo{journal}{Physical Review B} \bibinfo{volume}{54},
  \bibinfo{pages}{4843--4856}.

\bibitem[{Eshelby(1957)}]{Eshelby1957}
\bibinfo{author}{Eshelby, J.D.}, \bibinfo{year}{1957}.
\newblock \bibinfo{title}{The determination of the elastic field of an
  ellipsoidal inclusion, and related problems}.
\newblock \bibinfo{journal}{Proceedings of the Royal Society of London. Series
  A. Mathematical and Physical Sciences} \bibinfo{volume}{241},
  \bibinfo{pages}{376--396}.

\bibitem[{Eshelby(1959)}]{Eshelby1959}
\bibinfo{author}{Eshelby, J.D.}, \bibinfo{year}{1959}.
\newblock \bibinfo{title}{The elastic field outside an ellipsoidal inclusion}.
\newblock \bibinfo{journal}{Proceedings of the Royal Society of London. Series
  A. Mathematical and Physical Sciences} \bibinfo{volume}{252},
  \bibinfo{pages}{561--569}.

\bibitem[{Eshelby(1961)}]{Eshelby1961}
\bibinfo{author}{Eshelby, J.D.}, \bibinfo{year}{1961}.
\newblock \bibinfo{title}{Elastic inclusions and inhomogeneities}, in:
  \bibinfo{booktitle}{Progress in solid mechanics II (editors: Sneddon, I. N., Hill, R.)}, \bibinfo{publisher}{North-Holland Publishing Company},
  \bibinfo{address}{Amsterdam}. pp. \bibinfo{pages}{89--140}.

\bibitem[{Fan et~al.(2017)}]{Fan2017}
\bibinfo{author}{Fan, F.}, \bibinfo{author}{Voznyy, O.},
  \bibinfo{author}{Sabatini, R.P.}, \bibinfo{author}{Bicanic, K.T.},
  \bibinfo{author}{Adachi, M.M.}, \bibinfo{author}{McBride, J.R.},
  \bibinfo{author}{Reid, K.R.}, \bibinfo{author}{Park, Y.S.},
  \bibinfo{author}{Li, X.}, \bibinfo{author}{Jain, A.},
  \bibinfo{author}{Quintero-Bermudez, R.},
  \bibinfo{author}{Saravanapavanantham, M.}, \bibinfo{author}{Liu, M.},
  \bibinfo{author}{Korkusinski, M.}, \bibinfo{author}{Hawrylak, P.},
  \bibinfo{author}{Klimov, V.I.}, \bibinfo{author}{Rosenthal, S.J.},
  \bibinfo{author}{Hoogland, S.}, \bibinfo{author}{Sargent, E.H.},
  \bibinfo{year}{2017}.
\newblock \bibinfo{title}{Continuous-wave lasing in colloidal quantum dot
  solids enabled by facet-selective epitaxy}.
\newblock \bibinfo{journal}{Nature} \bibinfo{volume}{544},
  \bibinfo{pages}{75--79}.

\bibitem[{Ferrers(1877)}]{Ferrers}
\bibinfo{author}{Ferrers, N.M.}, \bibinfo{year}{1877}.
\newblock \bibinfo{title}{On the potentials of ellipsoids, ellipsoidal shells,
  elliptic laminae and elliptic rings of variable densities}.
\newblock \bibinfo{journal}{The Quarterly Journal of Pure and Applied
  Mathematics} \bibinfo{volume}{14}, \bibinfo{pages}{1--22}.

\bibitem[{Friedman(1982)}]{Friedman1982}
\bibinfo{author}{Friedman, A.}, \bibinfo{year}{1982}.
\newblock \bibinfo{title}{Variational Principles and Free Boundary Problems}.
\newblock \bibinfo{publisher}{Wiley}, \bibinfo{address}{New York}.

\bibitem[{Gosling and Willis(1995)}]{Gosling1995}
\bibinfo{author}{Gosling, T.J.}, \bibinfo{author}{Willis, J.R.},
  \bibinfo{year}{1995}.
\newblock \bibinfo{title}{Mechanical stability and electronic properties of
  buried strained quantum wire arrays}.
\newblock \bibinfo{journal}{Journal of Applied Physics} \bibinfo{volume}{77},
  \bibinfo{pages}{5601--5610}.

\bibitem[{Guo et~al.(2011)Guo, Nie and Chan}]{NIE2011}
\bibinfo{author}{Guo, L.}, \bibinfo{author}{Nie, G.H.}, \bibinfo{author}{Chan,
  C.K.}, \bibinfo{year}{2011}.
\newblock \bibinfo{title}{Elliptical inhomogeneity with polynomial eigenstrains
  embedded in orthotropic materials}.
\newblock \bibinfo{journal}{Archive of Applied Mechanics} \bibinfo{volume}{81},
  \bibinfo{pages}{157--170}.

\bibitem[{Han(2012)}]{Han2012}
\bibinfo{author}{Han, Q.}, \bibinfo{year}{2012}.
\newblock \bibinfo{title}{A Basic Course in Partial Differential Equations}.
\newblock \bibinfo{publisher}{American Mathematical Society},
  \bibinfo{address}{Providence}.

\bibitem[{Jing et~al.(2016)Jing, Kershaw, Li, Huang, Li, Rogach and
  Gao}]{Jing2016}
\bibinfo{author}{Jing, L.}, \bibinfo{author}{Kershaw, S.V.},
  \bibinfo{author}{Li, Y.}, \bibinfo{author}{Huang, X.}, \bibinfo{author}{Li,
  Y.}, \bibinfo{author}{Rogach, A.L.}, \bibinfo{author}{Gao, M.},
  \bibinfo{year}{2016}.
\newblock \bibinfo{title}{Aqueous based semiconductor nanocrystals}.
\newblock \bibinfo{journal}{Chemical Reviews} \bibinfo{volume}{116},
  \bibinfo{pages}{10623--10730}.

\bibitem[{Joyce and Parnell(2017)}]{Joyce2017}
\bibinfo{author}{Joyce, D.}, \bibinfo{author}{Parnell, W.J.},
  \bibinfo{year}{2017}.
\newblock \bibinfo{title}{The {Newtonian} potential inhomogeneity problem:
  non-uniform eigenstrains in cylinders of non-elliptical cross section}.
\newblock \bibinfo{journal}{Journal of Engineering Mathematics}
  \bibinfo{volume}{107}, \bibinfo{pages}{283--303}.

\bibitem[{Kang and Milton(2008)}]{Kang2008}
\bibinfo{author}{Kang, H.}, \bibinfo{author}{Milton, G.W.},
  \bibinfo{year}{2008}.
\newblock \bibinfo{title}{Solutions to the p\'{o}lya-szeg\"{o} conjecture and
  the weak {Eshelby} conjecture}.
\newblock \bibinfo{journal}{Archive for Rational Mechanics and Analysis}
  \bibinfo{volume}{188}, \bibinfo{pages}{93--116}.

\bibitem[{Kawashita and Nozaki(2001)}]{Kawashita2001}
\bibinfo{author}{Kawashita, M.}, \bibinfo{author}{Nozaki, H.},
  \bibinfo{year}{2001}.
\newblock \bibinfo{title}{Eshelby tensor of a polygonal inclusion and its
  special properties}.
\newblock \bibinfo{journal}{Journal of Elasticity} \bibinfo{volume}{64},
  \bibinfo{pages}{71--84}.

\bibitem[{Lee et~al.(2015)Lee, Zou and Pan}]{Lee2015}
\bibinfo{author}{Lee, Y.G.}, \bibinfo{author}{Zou, W.N.}, \bibinfo{author}{Pan, E.}, \bibinfo{year}{2015}.
\newblock\bibinfo{title}{Eshelby's problem of polygonal inclusions with
  polynomial eigenstrains in an anisotropic magneto-electro-elastic full plane}.
\newblock \bibinfo{journal}{Proceedings of the Royal Society a-Mathematical
  Physical and Engineering Sciences} \bibinfo{volume}{471},
  \bibinfo{pages}{20140827}.

\bibitem[{Lim et~al.(2006)Lim, Li and He}]{Lim2006}
\bibinfo{author}{Lim, C.}, \bibinfo{author}{Li, Z.}, \bibinfo{author}{He, L.},
  \bibinfo{year}{2006}.
\newblock \bibinfo{title}{Size dependent, non-uniform elastic field inside a
  nano-scale spherical inclusion due to interface stress}.
\newblock \bibinfo{journal}{International Journal of Solids and Structures}
  \bibinfo{volume}{43}, \bibinfo{pages}{5055--5065}.

{ \bibitem[{Liu et~al.(2007)Liu, James and Leo}]{Liu2007}
\bibinfo{author}{Liu, L.P.}, \bibinfo{author}{James, R. D.}, \bibinfo{author}{Leo, P.H.},
\bibinfo{year}{2007}.
\newblock \bibinfo{title}{Periodic inclusion-matrix microstructures with constant field inclusions}.
\newblock \bibinfo{journal}{Metallurgical and Materials Transactions A} \bibinfo{volume}{38},
  \bibinfo{pages}{1543--1940}.}

\bibitem[{Liu(2008)}]{Liu2008}
\bibinfo{author}{Liu, L.P.}, \bibinfo{year}{2008}.
\newblock \bibinfo{title}{Solutions to the {Eshelby} conjectures}.
\newblock \bibinfo{journal}{Proceedings of the Royal Society a-Mathematical
  Physical and Engineering Sciences} \bibinfo{volume}{464},
  \bibinfo{pages}{573--594}.

\bibitem[{Liu(2013)}]{Liu2013}
\bibinfo{author}{Liu, L.P.}, \bibinfo{year}{2013}.
\newblock \bibinfo{title}{Polynomial eigenstress inducing polynomial strain of
  the same degree in an ellipsoidal inclusion and its applications}.
\newblock \bibinfo{journal}{Mathematics and Mechanics of Solids}
  \bibinfo{volume}{18}, \bibinfo{pages}{168--180}.

  { \bibitem[{Liu et~al.(2021)Liu, James and Leo}]{Liu2021}
\bibinfo{author}{Liu, L.P.}, \bibinfo{author}{James, R.D.}, \bibinfo{author}{Leo,
  P.H.}, \bibinfo{year}{2021}.
\newblock \bibinfo{title}{New extremal inclusions and their applications to
  two-phase composites}.
\newblock \href{http://arxiv.org/abs/2107.04088}{{\tt arXiv:2107.04088}}.}

\bibitem[{Lu and Pan(2014)}]{Lu2014}
\bibinfo{author}{Lu, C.Y.}, \bibinfo{author}{Pan, J.W.}, \bibinfo{year}{2014}.
\newblock \bibinfo{title}{Quantum optics push-button photon entanglement}.
\newblock \bibinfo{journal}{Nature Photonics} \bibinfo{volume}{8},
  \bibinfo{pages}{174--176}.

\bibitem[{Lubarda and Markenscoff(1998)}]{Lubarda1998}
\bibinfo{author}{Lubarda, V.A.}, \bibinfo{author}{Markenscoff, X.},
  \bibinfo{year}{1998}.
\newblock \bibinfo{title}{On the absence of {Eshelby} property for
  non-ellipsoidal inclusions}.
\newblock \bibinfo{journal}{International Journal of Solids and Structures}
  \bibinfo{volume}{35}, \bibinfo{pages}{3405--3411}.

\bibitem[{Ma et~al.(2018)Ma, Hu, Wei and Liang}]{Ma2018}
\bibinfo{author}{Ma, H.}, \bibinfo{author}{Hu, G.}, \bibinfo{author}{Wei, Y.},
  \bibinfo{author}{Liang, L.}, \bibinfo{year}{2018}.
\newblock \bibinfo{title}{Inclusion problem in second gradient elasticity}.
\newblock \bibinfo{journal}{International Journal of Engineering Science}
  \bibinfo{volume}{132}, \bibinfo{pages}{60--78}.

\bibitem[{Markenscoff(1997)}]{Markenscoff1997}
\bibinfo{author}{Markenscoff, X.}, \bibinfo{year}{1997}.
\newblock \bibinfo{title}{On the shape of the {Eshelby} inclusions}.
\newblock \bibinfo{journal}{Journal of Elasticity} \bibinfo{volume}{49},
  \bibinfo{pages}{163--166}.

\bibitem[{Markenscoff(1998)}]{Markenscoff1998}
\bibinfo{author}{Markenscoff, X.}, \bibinfo{year}{1998}.
\newblock \bibinfo{title}{Inclusions with constant eigenstress}.
\newblock \bibinfo{journal}{Journal of the Mechanics and Physics of Solids}
  \bibinfo{volume}{46}, \bibinfo{pages}{2297--2301}.

\bibitem[{Monchiet and Bonnet(2011)}]{Monchiet2011}
\bibinfo{author}{Monchiet, V.}, \bibinfo{author}{Bonnet, G.},
  \bibinfo{year}{2011}.
\newblock \bibinfo{title}{Inversion of higher order isotropic tensors with
  minor symmetries and solution of higher order heterogeneity problems}.
\newblock \bibinfo{journal}{Proceedings of the Royal Society a-Mathematical
  Physical and Engineering Sciences} \bibinfo{volume}{467},
  \bibinfo{pages}{314--332}.

\bibitem[{Monchiet and Bonnet(2013)}]{MonchietanndBonnet2013}
\bibinfo{author}{Monchiet, V.}, \bibinfo{author}{Bonnet, G.},
  \bibinfo{year}{2013}.
\newblock \bibinfo{title}{Algebra of transversely isotropic sixth order tensors
  and solution to higher order inhomogeneity problems}.
\newblock \bibinfo{journal}{Journal of Elasticity} \bibinfo{volume}{110},
  \bibinfo{pages}{159--183}.

\bibitem[{Mura(1987)}]{Mura1987}
\bibinfo{author}{Mura, T.}, \bibinfo{year}{1987}.
\newblock \bibinfo{title}{Micromechanics of Defects in Solids}.
\newblock \bibinfo{publisher}{Springer Netherlands}, \bibinfo{address}{Leiden}.

\bibitem[{Mura(1997)}]{Mura1997}
\bibinfo{author}{Mura, T.}, \bibinfo{year}{1997}.
\newblock \bibinfo{title}{The determination of the elastic field of a polygonal
  star shaped inclusion}.
\newblock \bibinfo{journal}{Mechanics Research Communications}
  \bibinfo{volume}{24}, \bibinfo{pages}{473 -- 482}.

\bibitem[{Mura and Kinoshita(1978)}]{Mura1978}
\bibinfo{author}{Mura, T.}, \bibinfo{author}{Kinoshita, N.},
  \bibinfo{year}{1978}.
\newblock \bibinfo{title}{The polynomial eigenstrain problem for an anisotropic
  ellipsoidal inclusion}.
\newblock \bibinfo{journal}{Physica Status Solidi (a)} \bibinfo{volume}{48},
  \bibinfo{pages}{447--450}.

\bibitem[{Mura et~al.(1994)Mura, Shodja, Lin, Safadi and Makkawy}]{Mura1994}
\bibinfo{author}{Mura, T.}, \bibinfo{author}{Shodja, H.M.}, \bibinfo{author}{Lin,
  T.}, \bibinfo{author}{Safadi, A.}, \bibinfo{author}{Makkawy, A.},
  \bibinfo{year}{1994}.
\newblock \bibinfo{title}{The determination of the elastic field of a
  pentagonal star shaped inclusion}.
\newblock \bibinfo{journal}{Bulletin of the Technical University of Istanbul}
  \bibinfo{volume}{47}, \bibinfo{pages}{267--280}.

\bibitem[{Nie et~al.(2007)Nie, Guo, Chan and Shin}]{NIE2007}
\bibinfo{author}{Nie, G.H.}, \bibinfo{author}{Guo, L.}, \bibinfo{author}{Chan,
  C.K.}, \bibinfo{author}{Shin, F.}, \bibinfo{year}{2007}.
\newblock \bibinfo{title}{Non-uniform eigenstrain induced stress field in an
  elliptic inhomogeneity embedded in orthotropic media with complex roots}.
\newblock \bibinfo{journal}{International Journal of Solids and Structures}
  \bibinfo{volume}{44}, \bibinfo{pages}{3575 -- 3593}.


{ \bibitem[{Nikliborc(1932)}]{Nikliborc1932}
\bibinfo{author}{Nikliborc, W.},
  \bibinfo{year}{1932}.
\newblock \bibinfo{title}{Eine Bemerkung {\"u}ber die Volumpotentiale}.
\newblock \bibinfo{journal}{Mathematische Zeitschrift} \bibinfo{volume}{35},
  \bibinfo{pages}{625--631}.}


\bibitem[{Nozaki and Taya(1997)}]{Nozaki1997}
\bibinfo{author}{Nozaki, H.}, \bibinfo{author}{Taya, M.}, \bibinfo{year}{1997}.
\newblock \bibinfo{title}{Elastic fields in a polygon-shaped inclusion with
  uniform eigenstrains}.
\newblock \bibinfo{journal}{Journal of Applied Mechanics-Transactions of the
  ASME} \bibinfo{volume}{64}, \bibinfo{pages}{495--502}.

\bibitem[{Pan and Chou(1976)}]{Pan1976}
\bibinfo{author}{Pan, Y.C.}, \bibinfo{author}{Chou, T.W.},
  \bibinfo{year}{1976}.
\newblock \bibinfo{title}{Point force solution for an infinite transversely
  isotropic solid}.
\newblock \bibinfo{journal}{Journal of Applied Mechanics-Transactions of the
  ASME} \bibinfo{volume}{43}, \bibinfo{pages}{608--612}.

\bibitem[{Parnell(2016)}]{Parnell2016}
\bibinfo{author}{Parnell, W.J.}, \bibinfo{year}{2016}.
\newblock \bibinfo{title}{The {Eshelby}, {Hill}, {Moment} and {Concentration}
  tensors for ellipsoidal inhomogeneities in the {Newtonian} potential problem
  and linear elastostatics}.
\newblock \bibinfo{journal}{Journal of Elasticity} \bibinfo{volume}{125},
  \bibinfo{pages}{231--294}.

\bibitem[{Rahman(2002)}]{Rahman2002}
\bibinfo{author}{Rahman, M.}, \bibinfo{year}{2002}.
\newblock \bibinfo{title}{{ The isotropic ellipsoidal inclusion with a
  polynomial distribution of eigenstrain}}.
\newblock \bibinfo{journal}{Journal of Applied Mechanics-Transactions of the
  ASME} \bibinfo{volume}{69}, \bibinfo{pages}{593--601}.

\bibitem[{Rashidinejad and Shodja(2019)}]{Rashidinejad2019}
\bibinfo{author}{Rashidinejad, E.}, \bibinfo{author}{Shodja, H.M.},
  \bibinfo{year}{2019}.
\newblock \bibinfo{title}{On the exact nature of the coupled-fields of
  magneto-electro-elastic ellipsoidal inclusions with non-uniform eigenfields
  and general anisotropy}.
\newblock \bibinfo{journal}{Mechanics of Materials} \bibinfo{volume}{128},
  \bibinfo{pages}{89--104}.

\bibitem[{Rodin(1998)}]{Rodin1998}
\bibinfo{author}{Rodin, G.J.}, \bibinfo{year}{1998}.
\newblock \bibinfo{title}{Elastic fields in a polygon-shaped inclusion with
  uniform eigenstrains}.
\newblock \bibinfo{journal}{Journal of Applied Mechanics-Transactions of the
  ASME} \bibinfo{volume}{65}, \bibinfo{pages}{278--278}.

\bibitem[{Ru and Schiavone(1996)}]{Ru1996}
\bibinfo{author}{Ru, C.Q.}, \bibinfo{author}{Schiavone, P.},
  \bibinfo{year}{1996}.
\newblock \bibinfo{title}{On the elliptic inclusion in anti-plane shear}.
\newblock \bibinfo{journal}{Mathematics and Mechanics of Solids}
  \bibinfo{volume}{1}, \bibinfo{pages}{327--333}.

\bibitem[{Sendeckyj(1967)}]{Sendeckyj1967}
\bibinfo{author}{Sendeckyj, G.P.}, \bibinfo{year}{1967}.
\newblock \bibinfo{title}{Ellipsoidal inhomogeneity problem (Ph.D.
  Dissertation)}.
\newblock \bibinfo{publisher}{Northwestern University},
  \bibinfo{address}{Evanston}.

\bibitem[{Sendeckyj(1970)}]{Sendeckyj1970}
\bibinfo{author}{Sendeckyj, G.P.}, \bibinfo{year}{1970}.
\newblock \bibinfo{title}{Elastic inclusion problems in plane elastostatics}.
\newblock \bibinfo{journal}{International Journal of Solids and Structures}
  \bibinfo{volume}{6}, \bibinfo{pages}{1535--1543}.

\bibitem[{Sharma et al.(2003)}]{Sharma2003}
\bibinfo{author}{Sharma, P.}, \bibinfo{author}{Ganti, S.}, \bibinfo{author}{Bhate, N.}
  \bibinfo{year}{2003}.
\newblock \bibinfo{title}{{Effect of surfaces on the size-dependent elastic state of nano-inhomogeneities}}.
\newblock \bibinfo{journal}{Applied Physics Letters} \bibinfo{volume}{82}, \bibinfo{pages}{535--537}.

\bibitem[{Sharma and Ganti(2004)}]{Sharma2004}
\bibinfo{author}{Sharma, P.}, \bibinfo{author}{Ganti, S.},
  \bibinfo{year}{2004}.
\newblock \bibinfo{title}{{Size-dependent {Eshelby's} tensor for embedded
  nano-inclusions incorporating surface/interface energies }}.
\newblock \bibinfo{journal}{Journal of Applied Mechanics-Transactions of the
  ASME} \bibinfo{volume}{71}, \bibinfo{pages}{663--671}.

\bibitem[{Sharma and Wheeler(2007)}]{Sharma2007}
\bibinfo{author}{Sharma, P.}, \bibinfo{author}{Wheeler, L.},
  \bibinfo{year}{2007}.
\newblock \bibinfo{title}{{Size-dependent elastic state of ellipsoidal
  nano-inclusions incorporating surface/interface tension}}.
\newblock \bibinfo{journal}{Journal of Applied Mechanics-Transactions of the
  ASME} \bibinfo{volume}{74}, \bibinfo{pages}{447--454}.

\bibitem[{Smith et~al.(2009)Smith, Mohs and Nie}]{Smith2009}
\bibinfo{author}{Smith, A.M.}, \bibinfo{author}{Mohs, A.M.},
  \bibinfo{author}{Nie, S.}, \bibinfo{year}{2009}.
\newblock \bibinfo{title}{Tuning the optical and electronic properties of
  colloidal nanocrystals by lattice strain}.
\newblock \bibinfo{journal}{Nature Nanotechnology} \bibinfo{volume}{4},
  \bibinfo{pages}{56--63}.

\bibitem[{Steindl et~al.(2019)}]{Steindl2019}
\bibinfo{author}{Steindl, P.}, \bibinfo{author}{Sala, EM.}, \bibinfo{author}{Al\'{e}n, B.},
  \bibinfo{author}{Marr\'{o}n, DF.}, \bibinfo{author}{Bimberg, D.}, \bibinfo{author}{Klenovsk\'{y}, P.},
  \bibinfo{year}{2019}.
\newblock \bibinfo{title}{Optical response of {(InGa)(AsSb)/GaAs} quantum dots
  embedded in a {GaP} matrix}.
\newblock \bibinfo{journal}{Physical Review B} \bibinfo{volume}{100},\bibinfo{pages}{195407}.

\bibitem[{Stepanov et~al.(2016)Stepanov, Elzo-Aizarna, Bleuse, Malik, Cur\'{e},
  Gautier, Favre-Nicolin, G\'{e}rard and Claudon}]{Stepanov2016}
\bibinfo{author}{Stepanov, P.}, \bibinfo{author}{Elzo-Aizarna, M.},
  \bibinfo{author}{Bleuse, J.}, \bibinfo{author}{Malik, N.S.},
  \bibinfo{author}{Cur\'{e}, Y.}, \bibinfo{author}{Gautier, E.},
  \bibinfo{author}{Favre-Nicolin, V.}, \bibinfo{author}{G\'{e}rard, J.M.},
  \bibinfo{author}{Claudon, J.}, \bibinfo{year}{2016}.
\newblock \bibinfo{title}{Large and uniform optical emission shifts in quantum
  dots strained along their growth axis}.
\newblock \bibinfo{journal}{Nano Letters} \bibinfo{volume}{16},
  \bibinfo{pages}{3215--3220}.

\bibitem[{Tanuma(2007)}]{Tanuma2007}
\bibinfo{author}{Tanuma, K.}, \bibinfo{year}{2007}.
\newblock \bibinfo{title}{Stroh formalism and rayleigh waves}.
\newblock \bibinfo{journal}{Journal of Elasticity} \bibinfo{volume}{89},
  \bibinfo{pages}{5--154}.


\bibitem[{Tian and Rajapakse(2007)}]{Tian2007}
\bibinfo{author}{Tian, L.}, \bibinfo{author}{Rajapakse, R.},
  \bibinfo{year}{2007}.
\newblock \bibinfo{title}{Elastic field of an isotropic matrix with a nanoscale
  elliptical inhomogeneity}.
\newblock \bibinfo{journal}{International Journal of Solids and Structures}
  \bibinfo{volume}{44}, \bibinfo{pages}{7988--8005}.

\bibitem[{Trotta et~al.(2015)Trotta, Mart\'{i}n-S\'{a}nchez, Daruka, Ortix and
  Rastelli}]{Trotta2015}
\bibinfo{author}{Trotta, R.}, \bibinfo{author}{Mart\'{i}n-S\'{a}nchez, J.},
  \bibinfo{author}{Daruka, I.}, \bibinfo{author}{Ortix, C.},
  \bibinfo{author}{Rastelli, A.}, \bibinfo{year}{2015}.
\newblock \bibinfo{title}{Energy-tunable sources of entangled photons: a viable
  concept for solid-state-based quantum relays}.
\newblock \bibinfo{journal}{Physical Review Letters} \bibinfo{volume}{114},
  \bibinfo{pages}{150502}.

\bibitem[{Trotta et~al.(2016)Trotta, Mart\'{i}n-S\'{a}nchez, Wildmann, Piredda,
  Reindl, Schimpf, Zallo, Stroj, Edlinger and Rastelli}]{Trotta2016}
\bibinfo{author}{Trotta, R.}, \bibinfo{author}{Mart\'{i}n-S\'{a}nchez, J.},
  \bibinfo{author}{Wildmann, J.S.}, \bibinfo{author}{Piredda, G.},
  \bibinfo{author}{Reindl, M.}, \bibinfo{author}{Schimpf, C.},
  \bibinfo{author}{Zallo, E.}, \bibinfo{author}{Stroj, S.},
  \bibinfo{author}{Edlinger, J.}, \bibinfo{author}{Rastelli, A.},
  \bibinfo{year}{2016}.
\newblock \bibinfo{title}{Wavelength-tunable sources of entangled photons
  interfaced with atomic vapours}.
\newblock \bibinfo{journal}{Nature Communications} \bibinfo{volume}{7},
  \bibinfo{pages}{10375}.

\bibitem[{Veilleux et~al.(2010)Veilleux, Lachance-Quirion, Dore, Landry,
  Charette and Allen}]{Veilleux2010}
\bibinfo{author}{Veilleux, V.}, \bibinfo{author}{Lachance-Quirion, D.},
  \bibinfo{author}{Dore, K.}, \bibinfo{author}{Landry, D.B.},
  \bibinfo{author}{Charette, P.G.}, \bibinfo{author}{Allen, C.N.},
  \bibinfo{year}{2010}.
\newblock \bibinfo{title}{Strain-induced effects in colloidal quantum dots:
  lifetime measurements and blinking statistics}.
\newblock \bibinfo{journal}{Nanotechnology} \bibinfo{volume}{21},
  \bibinfo{pages}{134024}.

\bibitem[{Vigdergauz(2000)}]{Vigdergauz2000}
\bibinfo{author}{Vigdergauz, S.}, \bibinfo{year}{2000}.
\newblock \bibinfo{title}{Constant-stress inclusions in an elastic plate}.
\newblock \bibinfo{journal}{Mathematics and Mechanics of Solids}
  \bibinfo{volume}{5}, \bibinfo{pages}{265--279}.

\bibitem[{Walpole(1981)}]{Walpole1981}
\bibinfo{author}{Walpole, L.J.}, \bibinfo{year}{1981}.
\newblock \bibinfo{title}{Elastic behavior of composite materials: theoretical
  foundations}.
\newblock \bibinfo{journal}{Advances in Applied Mechanics}
  \bibinfo{volume}{21}, \bibinfo{pages}{169--242}.

\bibitem[{Wang(1992)}]{Wang1992}
\bibinfo{author}{Wang, B.}, \bibinfo{year}{1992}.
\newblock \bibinfo{title}{Three-dimensional analysis of an ellipsoidal
  inclusion in a piezoelectric material}.
\newblock \bibinfo{journal}{International Journal of Solids and Structures}
  \bibinfo{volume}{29}, \bibinfo{pages}{293--308}.

\bibitem[{Wang et~al.(2012)Wang, Gong, Guo and He}]{Wang2012}
\bibinfo{author}{Wang, J.}, \bibinfo{author}{Gong, M.}, \bibinfo{author}{Guo,
  G.C.}, \bibinfo{author}{He, L.}, \bibinfo{year}{2012}.
\newblock \bibinfo{title}{Eliminating the fine structure splitting of excitons
  in self-assembled {InAs/GaAs} quantum dots via combined stresses}.
\newblock \bibinfo{journal}{Applied Physics Letters} \bibinfo{volume}{101},
  \bibinfo{pages}{063114}.

\bibitem[{Withers(1989)}]{Withers1989}
\bibinfo{author}{Withers, P.J.}, \bibinfo{year}{1989}.
\newblock \bibinfo{title}{The determination of the elastic field of an
  ellipsoidal inclusion in a transversely isotropic medium, and its relevance
  to composite materials}.
\newblock \bibinfo{journal}{Philosophical Magazine a-Physics of Condensed
  Matter Structure Defects and Mechanical Properties} \bibinfo{volume}{59},
  \bibinfo{pages}{759--781}.

\bibitem[{Xu and Wang(2005)}]{Xu2005}
\bibinfo{author}{Xu, B.X.}, \bibinfo{author}{Wang, M.Z.}, \bibinfo{year}{2005}.
\newblock \bibinfo{title}{The quasi {Eshelby} property for rotational
  symmetrical inclusions of uniform eigencurvatures within an infinite plate}.
\newblock \bibinfo{journal}{Proceedings of the Royal Society a-Mathematical
  Physical and Engineering Sciences} \bibinfo{volume}{461},
  \bibinfo{pages}{2899--2910}.

\bibitem[{Xu et~al.(2009)Xu, Zhao, Gross and Wang}]{Xu2009}
\bibinfo{author}{Xu, B.X.}, \bibinfo{author}{Zhao, Y.T.},
  \bibinfo{author}{Gross, D.}, \bibinfo{author}{Wang, M.Z.},
  \bibinfo{year}{2009}.
\newblock \bibinfo{title}{Proof of the strong {Eshelby} conjecture for plane
  and anti-plane anisotropic inclusion problems}.
\newblock \bibinfo{journal}{Journal of Elasticity} \bibinfo{volume}{97},
  \bibinfo{pages}{173--188}.

\bibitem[{Yue et~al.(2015)Yue, Xu, Chen and Pan}]{Yue2015}
\bibinfo{author}{Yue, Y.M.}, \bibinfo{author}{Xu, K.Y.}, \bibinfo{author}{Chen,
  Q.D.}, \bibinfo{author}{Pan, E.}, \bibinfo{year}{2015}.
\newblock \bibinfo{title}{Eshelby problem of an arbitrary polygonal inclusion
  in anisotropic piezoelectric media with quadratic eigenstrains}.
\newblock \bibinfo{journal}{Acta Mechanica} \bibinfo{volume}{226},
  \bibinfo{pages}{2365--2378}.

\bibitem[{Zhang et~al.(2010)Zhang, Tang, Lee and Min}]{Zhang2010}
\bibinfo{author}{Zhang, J.}, \bibinfo{author}{Tang, Y.}, \bibinfo{author}{Lee,
  K.}, \bibinfo{author}{Min, O.}, \bibinfo{year}{2010}.
\newblock \bibinfo{title}{Nonepitaxial growth of hybrid core-shell
  nanostructures with large lattice mismatches}.
\newblock \bibinfo{journal}{Science} \bibinfo{volume}{327},
  \bibinfo{pages}{1634--1638}.

\bibitem[{Zhao et~al.(2019)Zhao, Chen, Chen, Luo, Sun, Zhang, Wang, Lin, Su,
  Qiao and Wang}]{Zhao2019}
\bibinfo{author}{Zhao, J.}, \bibinfo{author}{Chen, X.}, \bibinfo{author}{Chen,
  B.}, \bibinfo{author}{Luo, X.}, \bibinfo{author}{Sun, T.},
  \bibinfo{author}{Zhang, W.}, \bibinfo{author}{Wang, C.},
  \bibinfo{author}{Lin, J.}, \bibinfo{author}{Su, D.}, \bibinfo{author}{Qiao,
  X.}, \bibinfo{author}{Wang, F.}, \bibinfo{year}{2019}.
\newblock \bibinfo{title}{Accurate control of core-shell upconversion
  nanoparticles through anisotropic strain engineering}.
\newblock \bibinfo{journal}{Advanced Functional Materials}
  \bibinfo{volume}{29}, \bibinfo{pages}{1903295}.

\bibitem[{Zheng et~al.(2006)Zheng, Zhao and Du}]{Zheng2006}
\bibinfo{author}{Zheng, Q.S.}, \bibinfo{author}{Zhao, Z.H.},
  \bibinfo{author}{Du, D.X.}, \bibinfo{year}{2006}.
\newblock \bibinfo{title}{Irreducible structure, symmetry and average of
  {Eshelby's} tensor fields in isotropic elasticity}.
\newblock \bibinfo{journal}{Journal of the Mechanics and Physics of Solids}
  \bibinfo{volume}{54}, \bibinfo{pages}{368-383}.


\end{thebibliography}
\end{document}